\documentstyle[iopconf1,epsf]{article}

\newcommand{\be}[1]{\begin{equation} \label{(#1)}}
\newcommand{\ee}{\end{equation}}
\newcommand{\ba}[1]{\begin{eqnarray} \label{(#1)}}
\newcommand{\ea}{\end{eqnarray}}

\newcommand{\nn}{\nonumber}

\def \znbb {$0\nu\beta\beta$}

\def \Rpv{R_{P} \hspace{-0.9em}/\;\:}
\def\rp{$R_p \hspace{-1em}/\;\:$}
\def \emass {\langle m_{\nu} \rangle}

\def \onbb {$0\nu\beta\beta$ }

\def \be {\begin{equation}}
\def \ee {\end{equation}}

\def\be{\begin{equation}}
\def\ee{\end{equation}}
\def\bea{\begin{eqnarray}}
\def\eea{\end{eqnarray}}

\begin{document}

\begin{center}
\renewcommand{\baselinestretch}{1.5}
\Large\bf

\begin{center}
{\Huge {\bf GENIUS}}\\
{\Large {\bf  a Supersensitive Germanium Detector System for Rare Events}}\\
\vspace*{4cm}

\normalsize
{H.V.~Klapdor--Kleingrothaus, L.~Baudis, G.~Heusser,
  B.~Majorovits, H.~P\"as}\\
Max--Planck--Institut f\"ur Kernphysik, Heidelberg, Germany\\
\end{center}

\vskip 2cm
\Large
\begin{center}
{\bf August 1999}\\
\end{center}

\end{center}
\vspace*{2cm}

\small
\noindent
The GENIUS collaboration includes groups from:\\
MPIK Heidelberg, Germany; Inst. of Nucl. Research, Kiev, Ukraine;
Physik. Techn. Bundesanstalt, Braunschweig, Germany;
Internat. Center for Theoret. Physics, Trieste, Italy;
JINR Dubna, Russia; Inst. of Radiophys. Research, Nishnij Novgorod,
Russia; Dep. of Physics, Northeastern Univ. Boston, USA;
Dep. of Physics, Univ. of Maryland, USA;
Los Alamos Nat. Lab., USA; Dep. de Fisica Teorica, Univ. of Valencia, Spain;
Dep. of Physics, Texas A\&M Univ., USA \\

\noindent
{\normalsize Spokesman of the Collaboration: H. V. Klapdor--Kleingrothaus }

\begin{flushright}
{\bf MPI-Report MPI-H-V26-1999}\\
\end{flushright}

\newpage

\renewcommand{\baselinestretch}{1}


\renewcommand{\baselinestretch}{1}

\vspace*{-3cm}
\begin{center}
{\Large {\bf  The GENIUS Collaboration}}\\
{\large Status September 1999}
\vspace*{1cm}
\end{center}
\normalsize


\noindent
{\bf  H. V. Klapdor--Kleingrothaus$^*$,L. Baudis, A. Dietz, G. Heusser,
St. Kolb, B. Majorovits, H. P\"as, F. Schwamm, H. Strecker}\\
Max--Planck--Institut f\"ur Kernphysik, Heidelberg, Germany\\

\noindent
{\bf O.A. Ponkratenko, V.I. Tretyak, Yu.G. Zdesenko}\\
Institute of Nuclear research, Kiev, Ukraine\\

\noindent
{\bf V. Alexeev, A. Balysh, A. Bakalyarov, S. T. Belyaev, 
V. I. Lebedev, S. Zhukov}\\
Kurchatov Institute, Moscow, Russia\\

\noindent
{\bf U. Keyser, A.~Paul, S.~R\"ottger, A.~Zimbal}\\
Physikalisch-Technische Bundesanstalt, Braunschweig, Germany\\

\noindent
{\bf A. Yu. Smirnov}\\
Internat. Center for Theoretical Physics, Trieste, Italy\\

\noindent
{\bf V. Bednyakov}\\
Joint Institute for Nuclear Research, Dubna, Russia\\

\noindent
{\bf I.V. Krivosheina, V. Melnikov}\\
Institute of Radiophysical Research, Nishnij Novgorod, Russia\\

\noindent
{\bf P. Nath}\\
Department of Physics, Northeastern University, Boston, USA\\

\noindent
{\bf R.N. Mohapatra}\\
Department of Physics, University of Maryland, USA\\

\noindent
{\bf J.W.F Valle}\\
Departamento de Fisica Teorica, University of Valencia, Spain\\

\noindent
{\bf R. Arnowitt}\\
Department of Physics, Texas A\&M University, USA\\

\noindent
{\small $^*$ Spokesman of the Collaboration}\\

\renewcommand{\baselinestretch}{1}


\pagenumbering{arabic}


\clearpage

\newpage

\section{The New Physics Potential of GENIUS}

\subsection{Introduction}

Two outstanding problems in contemporary astro- and particle physics
are the nature of the dark matter in the Universe and the question for 
the neutrino mass. There is compelling evidence on all cosmological
scales that the dominant form of matter is nonbaryonic
\cite{kolb94}. Attractive   
candidates for nonbaryonic dark matter are relic elementary particles
left over from the big bang, the three most promising being neutrinos, 
axions and neutralinos \cite{jkg96}.

There is significant evidence from theories of structure
formation against neutrinos as the bulk of dark matter. A mixed hot
plus cold dark matter scenario still gives the better fit to the CMB and 
LSS-data \cite{eric98}. Recently $\Lambda$CDM scenarios have become
the most attractive ones \cite{turner99}.

To adress both issues we propose the GENIUS experiment
\cite{Kla98,KK2}.
The optimal locations would be 
the Gran Sasso or WIPP underground laboratories. 
GENIUS, using
ionization in a Ge detector as detection technique, would operate naked Ge
crystals in ultrapure liquid nitrogen.
The aim is to reach the background level of 10$^{-3}$ events/kg y keV
in the low energy region, thus to cover most of the  parameter space
predicted for neutralinos in the MSSM, and  
to be sensitive to the low-energy pp and $^7$Be solar neutrino flux. In the energy region of the
\onbb{}-decay of $^{76}$Ge the goal is to reach a count rate of 0.3
events/t y keV, thus testing the effective Majorana neutrino mass down 
to 0.01 eV for one ton of enriched $^{76}$Ge (0.001 eV for ten
tons). While for dark matter search only 100 kg of natural Ge are needed as 
detectors, an amount of the order of one ton of (natural
 or enriched) Ge would allow to observe for the first time pp
 neutrinos in a real-time measurement.

In addition to the unique information on neutrino masses and mixings 
obtainable, GENIUS would allow also a breakthrough into the multi-TeV range 
for many other beyond standard models of particle physics, such as
supersymmetry (R-parity breaking, sneutrino mass), compositeness, right-handed
W boson mass, test of special relativity and equivalence principle in the 
neutrino sector, and others, competitive to corresponding research at future
high-energy colliders.   

\subsection{Direct Dark Matter Detection}

\subsubsection{Three dark matter problems}

There is evidence on all cosmological scales that most of the matter
in our Universe is dark. The quantity and composition of dark matter
is of fundamental importance in cosmology. 

We know of three so-called dark matter problems at present:
the bulk of baryonic matter is dark; the dominant form of matter
is nonbaryonic; an additional dark, exotic form of energy contributes
about 60\% to the critical density $\Omega_0$.

A precise determination of the universal baryon density is provided by
the big-bang nucleosynthesis (BBN). 
Comparison of the measured primeval
deuterium abundance with its big-bang prediction yields a baryon density
of $\Omega_B$h$^2$ = 0.019 $\pm$ 0.0012 \cite {Bur99}. 
However, clusters of galaxies account only for about 10\% of the baryons
 \cite{per92}.
Promising candidates for the dark baryons are so-called MACHOS.
The search for MACHOs in the halo of our own galaxy, in form of
planets, white and brown dwarfs or primordial black holes, exploits the
gravitational microlensing effect, the temporary brightening
of a background star as an unseen object passes close to the line of sight.  
For several years two groups are monitoring the brightness of millions
of stars in the Magellanic clouds, the MACHO \cite {macho} and the EROS 
\cite {eros98} collaborations.
 Several candidates have already been
detected. If interpreted as dark matter they would make up half  the
amount needed in the galactic halo. The most probable mass of
these candidates, which can be inferred from the duration of a stars
brightening together with the lense distance, is about half the solar
mass. However, no stellar candidate seems able to explain the observations.
Measurements of carbon abundances in Lyman $\alpha$ forest lines speak 
against white dwarfs, even if their masses lie in the
expected region \cite{free99}. The possibility remains, that MACHOs are
an exotic form of baryonic matter, like primordial
black holes, or that they are not located in the halo of our galaxy 
\cite{griest99}.
Two events which have been discovered in the direction of the Small
Magellanic Cloud underline this hypothesis; both
lenses are stars within the satellite galaxy
itself \cite{moniez}.

Clusters of galaxies provide  very reliable methods of estimating
the total matter density \cite{white}.
They are the largest observed structures, which in part already attained 
hydrostatic equilibrium. The
relative amount of baryons and dark matter within their hydrostatic
region provides a measure of the cosmic mix of these components.
The baryonic component of rich clusters is dominated by the X-ray
emitting intracluster gas. Using the cluster baryon fraction
determined from X-ray measurements and assuming that clusters provide
a fair sample of matter in the Universe, a matter density of about a
third of the critical density (i.e. a flat Universe) is inferred \cite{gus}. 

There are many other methods of inferring the total matter density, 
involving different physics.

For example, compelling evidence for both baryonic and nonbaryonic dark
matter comes from observation of the rotation curves of galaxies. In
particular the rotation curves of dwarf spirals are completely dark
matter dominated \cite{burkert}.
Also distant field galaxies, which are much
younger than nearby galaxies are  entirely embedded in dark halos 
\cite{fuchs}. 
This is indeed expected from
theories of galaxy cosmogony, where the dark
matter haloes themselves are thought to be the sites of galaxy formation.      

On the other side, there is  strong evidence  from structure formation 
that the total matter density, $\Omega_M$ is significantly greater then the
density of baryons, $\Omega_B$ \cite{dodelson}.

Evidence for an additional dark, smoothly distributed form of energy
 comes from observation of distant supernovae of type Ia.
By measuring the deviation of the Hubble law from linearity at high
redshifts, the acceleration or deceleration of the expansion can be
determined. Objects with well understood properties, 
which can be observed up to very large distance (standard
candles) 
are type Ia supernovae, thermonuclear explosions of white dwarfs in 
binary systems. Two groups succeeded in measuring distances
to some 50 supernovae type Ia \cite{perl,riess}. 
The measurements indicate that the
Universe is speeding up, the simplest explanation being a so-called
cosmological constant, a smooth contribution to the energy density of
the Universe, which cannot clump. Such a contribution is also
supported by inflation, since the dynamically determined 
matter density ($\Omega_M$ $\simeq$ 0.4) is too low to yield the predicted 
flat Universe ($\Omega_0$ = 1).

\begin{figure}[h!]
\epsfysize=8cm
\hspace*{1.8cm}
\epsfbox{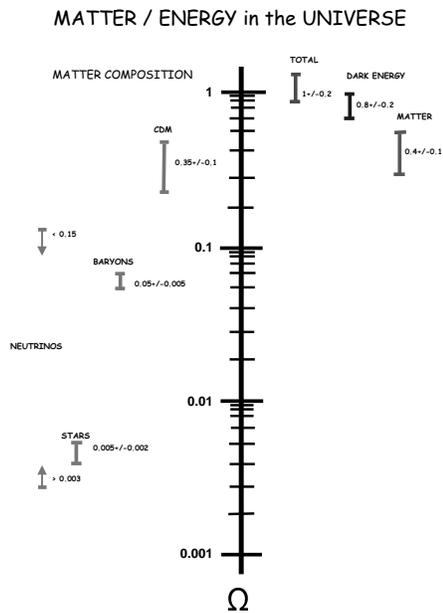}
\caption{Summary of an overall accounting of matter and energy in the 
Universe, from \cite{turner99}}
\label{turner}
\end{figure}

A summary of the matter/energy composition of the Universe is shown
in Fig. \ref{turner} (from \cite{turner99}).

\subsubsection{Nonbaryonic dark matter candidates}

From the nonbaryonic dark matter candidates proposed in the 1980s,
only WIMPs (weakly interacting massive particles) and axions survived
\cite{kamion}. 
WIMPs, which arise in supersymmetric  or
other theories beyond the standard model of particle physics, were in
thermal equilibrium with other particles during the early phase of the
Universe.
Their abundance depends only on their annihilation
cross section, but not on the WIMP mass.
Their annihilation cross section must be in the
order of a typical weak interaction should their abundance be of
cosmological relevance today. 
If supersymmetry is realized in
nature on the required energy scale, then the lightest supersymmetric
particle (LSP), the neutralino, shows exactly the properties of a WIMP 
\cite{kamion}. 

Axions, the other leading dark matter particle candidates, were
postulated two decades ago to explain why the strong interaction
conserves the CP-symmetry, which is violated in the standard model.
Axions could have been produced during the QCD phase
transition; their masses are constrained by experimental searches,
astrophysical and cosmological arguments to the order of 10$^{-5}$ eV.
Since axions couple to two photons, they can be detected by stimulating 
their conversion to
photons in a cavity permeated by a strong static magnetic field
\cite{sikivie}.
 In order to cover 
a wide mass range the cavity must be tuned to different frequencies.
Two pilot experiments in Brookhaven \cite{rbf} and Florida \cite{uf}
already demonstrated the
feasibility of the cavity detection method; the second generation
experiments presently under way at the Lawrence Livermore National
Laboratory \cite{lnll} and at the Kyoto University \cite{ku} 
will have a sensitivity which
is sufficient to discover axions if they populate the galactic halo.

Besides these dark matter candidates, another class of candidates, 
the so-called superheavy dark matter, emerged. If one gives up the
assumption, that the particle was in thermal equilibrium in the early
universe, then its present abundance is
no longer  determined by the annihilation cross section and much
heavier particles, so-called WIMPZILLAs, are allowed \cite{rockyI}.
There are two necessary conditions for WIMPZILLAs, they
must be stable, or at least have a lifetime much greater than the age
of the universe and their interaction rate must be sufficiently weak
such that thermal equilibrium with the primordial plasma was never
obtained. For this, the particle must be extremely massive, of the
order of the Hubble parameter at the end of inflation and the
annihilation rate per particle must be smaller than the expansion rate 
of the Universe \cite{rockyI}.

Another kind of heavy particles, which could form a natural dark matter
candidate, are stable baryonic Q-balls \cite{kusenko}. 
They are predicted by supersymmetric models and could
have been produced during the baryogenesis epoch.
In this scenario, the baryonic matter and the dark matter are
produced in the same process, therefore it is easy to understand why the
observed abundances of these are in the same order of magnitude.
The way to detect Q-balls depends on their ability
to pick up electric charge as they travel through ordinary matter. 
Electrically charged superballs would loose their energy in atomic
collisions, their expected signature is similar to those of 
nuclearites, which are searched for in the MACRO experiment.
Non observation of these gives a lower limit on the baryon number of
dark matter Q-balls of 10$^{21}$. The signature of electrically
neutral Q-balls  are similar to those expected from GUT-monopoles, 
the lower limit from the Baikal experiment on their baryonic charge is 
10$^{23}$. These limits will be improved by future experiments,
like AMANDA or ANTARES. However for covering the entire 
cosmologically interesting range a detector with an are of several
square kilometers would be needed \cite{kusenko}.

\subsubsection{Status of the direct search for WIMPs}

If WIMPs populate the halo of our galaxy they could be detected directly
in low background laboratory experiments or indirectly through
their annihilation products in the halo, the centre of the Sun or
Earth. 
The goal of the direct detection experiments is to look for  
the elastic scattering
of a WIMP off nuclei in a low-background detector. The recoil nucleus looses
its energy through ionization and thermal processes. The methods to
detect this energy loss span from scintillators, ionization
detectors, bolometers to superheated droplet or superconducting granular 
detectors.

The deposited energy for neutralinos
with masses between 10 GeV and 1 TeV is below 100 keV \cite{jkg96}.
For a standard halo comprised of WIMPs with a Maxwellian velocity
distribution characterized by v$_{rms}$ = 270 km/s and a mass density
of 0.4 GeV/cm$^2$, the expected  event
rates are well below 1 event per kilogramm target material and day 
\cite{jkg96}.
This makes any experimental attempt to directly detect neutralinos a great
technical challenge, requiring a large detector mass, a low energy
threshold, a low background and/or an effective background discrimination
technique.    

Direct detection experiments operating so far have reached
sensitivities low enough to enter the parameter space predicted for
neutralinos in the MSSM.
The best current limits on WIMP-nucleon cross section come from the
DAMA NaI experiment \cite{DAMA}, from the Heidelberg-Moscow 
experiment \cite{HM98} and from CDMS \cite{cdms98}
  (see Figure \ref{limits}).

\begin{figure}[h]
\epsfysize=8cm
\epsfbox{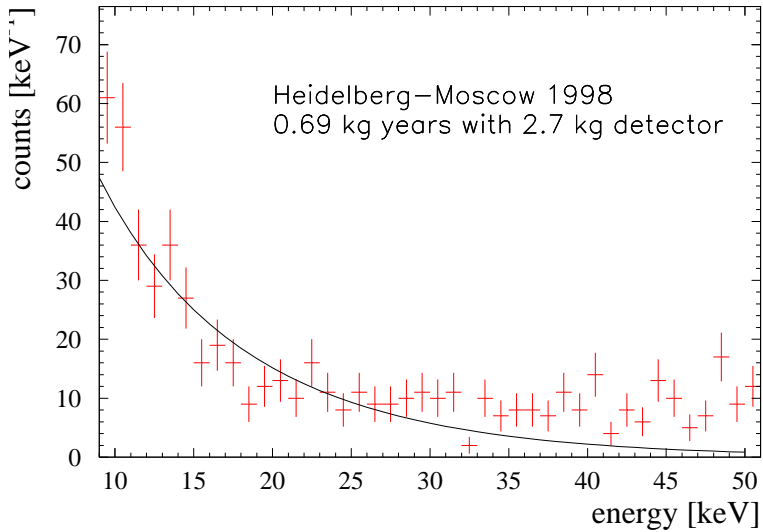}
\caption{Total measured spectrum with one enriched $^{76}$Ge detector
of the Heidelberg-Moscow experiment after an exposure of 0.69 kg yr
and
a theoretical spectrum for a 100 GeV WIMP.}
\label{ang2}
\end{figure}

The Stanford CDMS (Cold Dark Matter Search) experiment \cite{rick97}
uses thermal detectors of
ultrapure germanium and silicon which are operated at a temperature of 20 
mK. The simultaneous measurement of both ionization and phonon signals
allows the discrimination of a nuclear recoil event from an electron
interaction. This represents a very effective background suppression method. 
For the moment the experiment is located at the Stanford Underground Facility,
10.6 m below ground, where it is planned to obtain an exposure of 100 kg d
with two silicon and four germanium devices. For the future 
it is planned to operate the detector in the Soudan 
Mine  with 2000 mwe overburden, which will reduce the muon flux by
five orders of magnitude  and thus reduce the cosmogenic activities and
the neutron background.
 Fig. \ref{limits} shows the
expected sensitivity at the Stanford site and at the Soudan site.

The DAMA experiment is running 115.5 kg NaI detectors in the Gran
Sasso Underground Laboratory \cite{Ber97}. The high obtainable statistic
opens the possibility to look for a  WIMP signature, as a
variation of the event rate due to the movement of the Sun in the
galactic halo and the Earth rotation around the Sun. The analysis  of
about 54 kg yr in terms of a WIMP annual modulation signature
favours a positive signal, the allowed region of WIMP masses and cross sections
is well embedded in the minimal supersymmetric parameter space
predicted for neutralinos \cite{damaevid,Bot98}. 
However, a further confirmation by DAMA and by other experiments must
be awaited.

\begin{figure}[h]
\epsfysize=9cm
\epsfbox{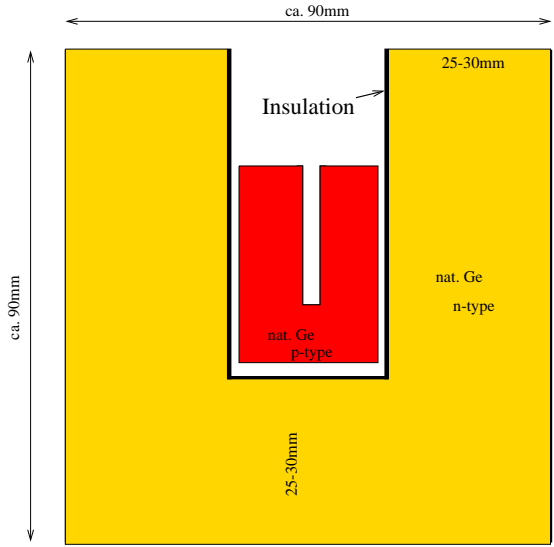}
\caption{Schematic figure of the HDMS detector.}
\label{detindet}
\end{figure}

The Hei\-del\-berg-Moscow \cite{HM97} 
and HDMS (Heidelberg Dark Matter Search) \cite{Bau97}  
experiments are both located in the Gran Sasso Laboratory,
where the muon component of the cosmic rays is reduced to one part in
a million. 
The Heidelberg-Moscow experiment, which also searches
for the neutrinoless double beta decay in enriched $^{76}$Ge, gives
at present the most stringent limits on WIMP-nucleon cross section on
spin-independent interactions 
for using raw data without pulse shape analysis \cite{HM98}.
Fig. \ref{ang2} shows 
the measured spectrum with one enriched $^{76}$Ge detector
 after an exposure of 0.69 kg yr and a calculated WIMP spectrum for a 100 GeV
WIMP.  

\begin{figure}[h]
\epsfysize=7cm
\epsfbox{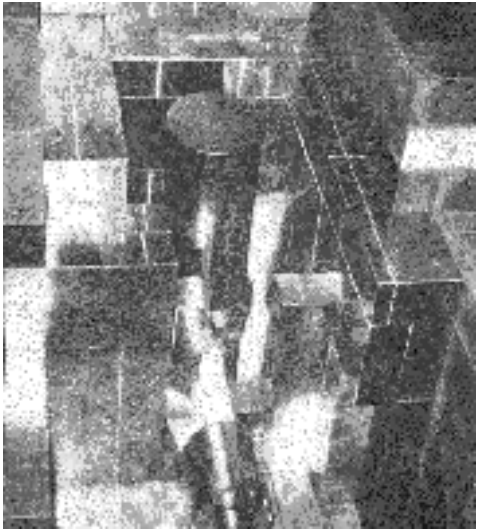}
\caption{The HDMS detector during its installation in the Gran Sasso
  Laboratory.}
\label{hdms}
\end{figure}

HDMS, which is a
dedicated dark matter experiment \cite{Bau97}, 
aims to improve this limit by one
order of magnitude. As in the Heidelberg-Moscow experiment, the
aim is to look for a small ionization signal inside a high purity Ge crystal.  
The actual dark matter target, a 200 g crystal made of natural Ge, 
is surrounded by a
well-type Ge crystal (see Fig. \ref{detindet}). 

The outer detector acts as an effective veto against
multiple scattered photons,  allowing to suppress the background
originating  from the latter by a factor 6-10. 
The HDMS prototype (see Fig. \ref{hdms}) started to measure in April 1998, 
while the full HDMS experiment (inner crystal made of enriched $^{73}$Ge 
and a new cryostat system of selected copper) will start measuring in
the course of the year 2000. 
With the expected sensitivity (see Fig. \ref{limits}) it will be able 
to test, like CDMS, the complete DAMA evidence region.

\subsubsection{GENIUS as a dark matter detector}

For an almost complete covering of the MSSM parameter space, an
increase in sensitivity by more than three orders of magnitude relative
to running experiments is required.

The GENIUS experiment could accomplish this task by operating about 40
'naked' natural Ge detectors (100 kg) in a tank of ultrapure liquid nitrogen.
The idea is to increase the target mass to 100 kg while decreasing 
at the same time the absolute background by a considerable amount.

The final goal would be to reach a background level of 
10$^{-3}$ events/kg y keV in the energy region relevant for dark matter
searches.

The energy threshold would be about 11 keV, the energy resolution
being better than 0.3 \%. 
With 100 kg of target material, GENIUS could
also look for a WIMP signature in form of an annual modulation of
the WIMP-signal.

A comparable sensitivity can be reached in principle 
only by LHC. An advantage of GENIUS is that it will be particulary
sensitive in regions of large tan$\beta$ in the minimal SUGRA space,
where conventional signals for supersymmetry in collider experiments
are difficult to detect. Thus, if the parameter tan$\beta$ is large,
then the first direct evidence for supersymmetry could come from
GENIUS, rather than from collider searches for sparticles
\cite{Bae97}. But also if SUSY will be detected by collider
experiments, it would still be fascinating - and necessary - to verify 
the existence and properties of neutralino dark matter.

\begin{figure}[h]
\epsfysize=8cm
\epsfbox{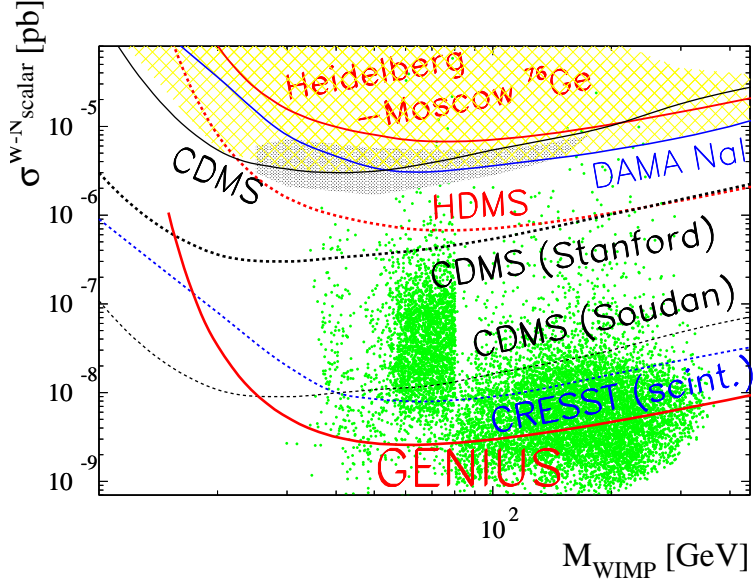}
\caption{WIMP--nucleon cross section limits as a function of the WIMP
  mass. The hatched region is excluded by the Heidelberg--Moscow
  \cite{HM98} and the DAMA experiment \cite{DAMA}, the dashed lines are 
  expectations for recently started or future experiments, like HDMS
  \cite{Bau97}, CDMS \cite{cdms98} and CRESST \cite{cresst96}. The
  filled contour represents  the 2$\sigma$ evidence region of the DAMA
  experiment \cite{damaevid}.    
The solid fat
  line denotes the expectation for the GENIUS project with a background 
  level of 0.01 counts/(keV kg y), an energy threshold of 11 keV
and an exposure of 300 kg yr.
The experimental limits are compared to
expectations (scatter plot) for WIMP--neutralinos calculated in the
MSSM framework with non--universal scalar mass unification \cite{Bed97c}.}
\label{limits}
\end{figure}

Fig. \ref{limits} shows a comparison of existing constraints and future 
sensitivities of cold dark matter experiments, together with the 
theoretical expectations for neutralino scattering rates 
\cite{Bed97b}.
Obviously, GENIUS could easily cover the range of positive 
evidence for dark matter
recently claimed by DAMA \cite{Ber97a,Bot97}. 
It would also be by far more
sensitive than all other dark matter experiments at present under construction
or proposed, like the cryogenic experiment CDMS. Furthermore,
obviously GENIUS will be the only experiment, 
which could seriously test the MSSM predictions over a large part of the 
SUSY parameter space. In this way, GENIUS could compete even 
with LHC in the search for SUSY, see for example the discussion 
in \cite{Bae97}. It is important to note, that GENIUS could reach the 
sensitivity shown in Fig. \ref{limits} with only 100 kg of {\it natural} Ge
detectors in a measuring time of three years \cite{Kla98d}.  

\subsection{Double Beta Decay}

Strong hints for neutrino masses are given by the atmospheric and
solar neutrino data, in particular by the SuperKamiokande confirmation
of the atmospheric neutrino deficit \cite{SuperK}. However, neutrino oscillation
experiments can measure only neutrino mass differences. In view of the 
SuperKamiokande data and of the dark matter problem, a determination
of the absolute neutrino mass scale should become a high priority.
A method for measuring the Majorana neutrino mass is provided by
neutrinoless double beta decay, at the same time a unique method of
discerning between a Majorana and a Dirac neutrino. 
The current most stringent experimental limit on the effective  
Majorana neutrino mass, $\langle {\rm m} \rangle < $0.2 eV, comes from the
Heidelberg-Moscow experiment \cite{Bau99a}. Future planned experiments, like NEMO or 
Kamland will, like the Heidelberg-Moscow experiment, improve this
limit at best by a factor of two. 
Fig. \ref{mass_time} gives the present status of the at  present most sensitive double beta experiments and of future plans.

\begin{figure}[h]
\epsfysize=10cm
\hspace*{1.8cm}
\epsfbox{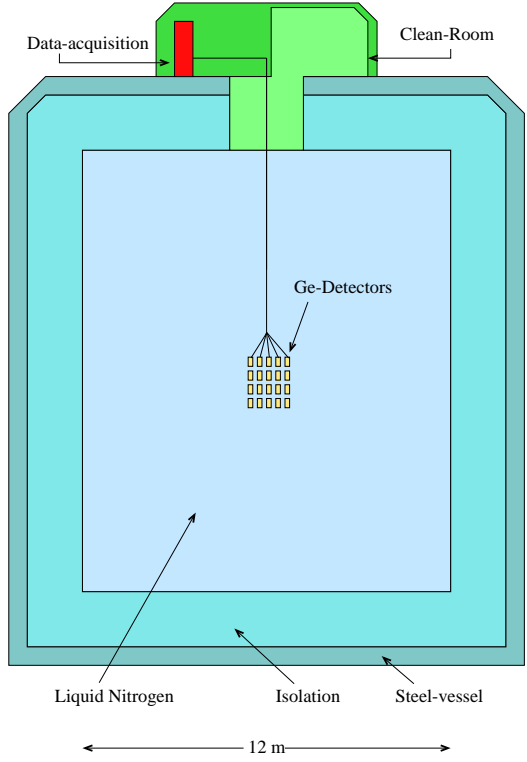}
\caption{Schematic view of the GENIUS experiment.}
\label{tank_sch}
\end{figure}

For a significant step beyond this
limit, much higher source strenghts and lower background levels are
needed.
This goal could be accomplished by the GENIUS experiment operating 300
detectors made of enriched $^{76}$Ge, (1 ton) in a liquid nitrogen
shielding
 (see Fig. \ref{tank_sch}).
 
GENIUS would search for the \onbb{}-decay in $^{76}$Ge, at the
Q-value of 2038.56 $\pm$ 0.32 keV \cite{hyka91}. The aim is to reach the sensitivity of
$\langle {\rm m} \rangle < $0.01 eV after one year of measuring
time. In an extended version using 10 tons of $^{76}$Ge, 
 a sensitivity of 0.001 eV could be reached. Already the first step will have
striking influence on presently discussed neutrino mass scenarios. The 
potential of GENIUS would also allow a breakthrough into the multi TeV 
range for many beyond standard models. It will give information on
supersymmetry (R-parity breaking, sneutrino mass), leptoquarks
(leptoquark -Higgs coupling or leptoquark mass), compositeness,
right-handed W boson mass, test of special relativity and equivalence
principle in the neutrino
sector and others, competitive to
corresponding results from  future high-energy colliders.
 The sensitivity of GENIUS in the neutrino sector would be larger than of many 
present terrestrial neutrino oscillation
 experiments and would provide complementary informations
to those planned for the future.
GENIUS with one ton would be able to check the LSND indication for
neutrino oscillations and GENIUS with ten tons could probe directly  
the large angle solution of the solar neutrino problem. For an almost
degenerate  neutrino mass scenario it could even probe the small angle 
solution of the solar neutrino problem.
This potential has been descibed recently in various papers \cite{KK2,Kla97d,Pan99,KPS}.
In the following subsubsections we give a short background of the general potential of double beta decay, a report of the status of double beta experiments, proposals and results, and an outline of the potential of GENIUS for investigation of neutrino masses and mixings and of other beyond standard model physics.

\subsubsection{General new physics potential of double beta decay}

\begin{figure}
\epsfxsize=90mm
\epsfbox{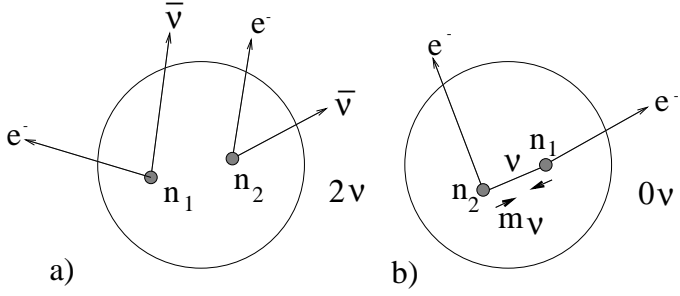}
\caption{Schematic representation of $2\nu$ and $0\nu$ double beta decay.}
\label{fig1-paes}
\end{figure}

\begin{figure}
\epsfxsize=90mm
\epsfbox{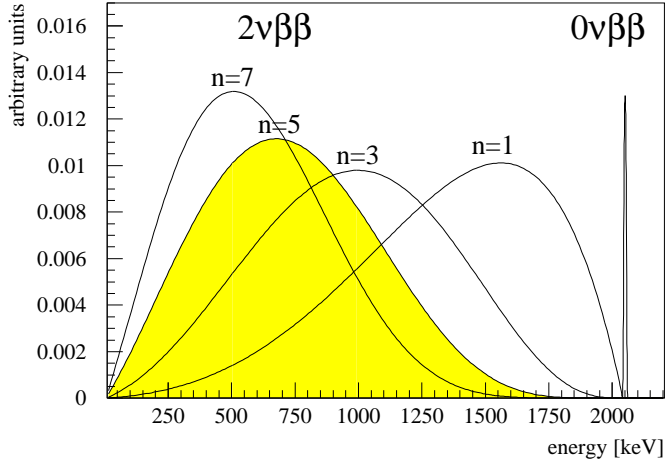}
\caption{Spectral shapes of the different modes of double beta decay,
  denotes the spectral index, n=5 for 2$\nu\beta\beta$ decay}
\label{fig2-paes}
\end{figure}

Double beta decay can occur in several decay modes (Fig. \ref{fig1-paes}):

\be
^{A}_{Z}X \rightarrow ^A_{Z+2}X + 2 e^- + 2 {\overline \nu_e}
\ee
\be        
^{A}_{Z}X \rightarrow ^A_{Z+2}X + 2 e^- 
\ee
\be
^{A}_{Z}X \rightarrow ^A_{Z+2}X + 2 e^- + \phi
\ee
\be
^{A}_{Z}X \rightarrow ^A_{Z+2}X + 2 e^- + 2\phi
\ee
the last three of them violating lepton number conservation by $\Delta L=2$.
For the neutrinoless mode
(2) we expect a sharp line at $E=Q_{\beta\beta}$, for the two--neutrino mode
and the various Majoron--accompanied modes classified by their spectral index,
continuous spectra (see Fig. \ref{fig2-paes}).
Important for particle physics are the decay modes (2)--(4).

 
\begin{figure}
\vspace*{1cm}
\hspace*{10mm}
\epsfxsize=60mm
\epsfysize=50mm
\epsfbox{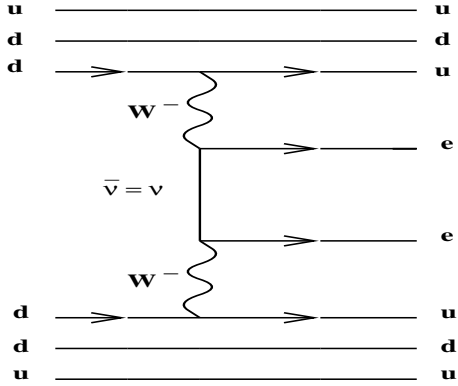}
\caption{Feynman graph for neutrinoless double beta decay 
 triggered  by exchange of a left--handed light or heavy neutrino}
\label{fig3-paes}
\end{figure}

\begin{figure}
\vspace*{1cm}
\hspace*{10mm}
\epsfxsize=100mm
\epsfbox{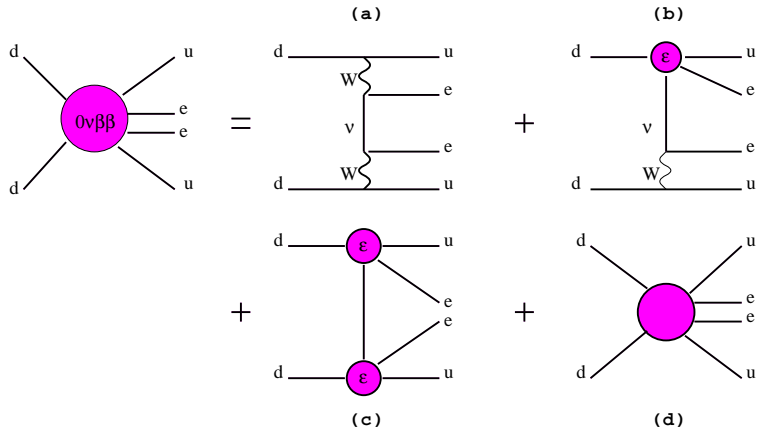}
\caption{Feynman graphs of the general double beta decay rate: 
The contributions (a)-(c)correspond to the long range part, 
the contribution d) is the short range part.}
\label{fig4-paes}
\end{figure}

Figure \ref{fig3-paes} gives the Feynman graphs of the neutrinoless 
double beta decay mode triggered by the exchange of a neutrino.
The neutrinoless mode (2) needs not be necessarily connected with the 
exchange of a virtual neutrino or sneutrino. {\it Any} process violating 
lepton number can
in principle lead to a process with the same signature as usual 
$0\nu\beta\beta$
decay. It may be triggered by exchange of neutralinos, gluinos, squarks,
sleptons, leptoquarks,... (see below and \cite{KK2,Paes97,Paes99}). 
Fig. \ref{fig4-paes} gives the graph of the general neutrinoless double
beta decay mode.

This gives rise
to the broad potential of double beta decay for testing or yielding 
restrictions on
quantities of beyond standard model physics (Table \ref{rah}).

\begin{table}
{\footnotesize
{  
\begin{tabular}[!h]{|lll|}
\hline
Observ. & Restrictions & Topics investigated\\
\hline
\hline
$0\nu$: &\underline{via $\nu$ exchange:} & 
Beyond the standard model and SU(5)\\
&Neutrino mass & model; early universe, matter--antimatter\\
& \hskip 3mm Light Neutrino & asymmetry, Dark matter\\
& \hskip 3mm Heavy Neutrino & L--R --symmetric models (e.g. SO(10)),\\ 
&&compositeness\\
&Test of Lorentz invariance & \\
&and equivalence principle & \\
&Right handed weak currents & $ V+ A$ interaction, $W^{\pm}_{R}$ masses \\
&\underline{via photino, gluino, zino} & SUSY models: Bounds for parameter \\
&\underline{(gaugino)  or sneutrino} & space beyond the range of accelerators\\
&\underline{exchange:}& \\
&R-parity breaking, & \\
&sneutrino mass & \\
&\underline{via leptoquark exchange} & leptoquark masses and models\\
& leptoquark-Higgs interaction & \\
\hline
$0\nu\chi$: &existence of the Majoron & Mechanism of (B-L) breaking\\
&& 
-explicit\\ 
&& -spontaneous breaking of the\\ 
&& local/global B-L symmetry\\
&& new Majoron models\\
\hline
\end{tabular}
}}
\caption {$\beta\beta$ decay and particle physics}
\label{rah}
\end{table}

There is, however, a generic relation between the amplitude of $0\nu\beta\beta$
decay and the $(B-L)$ violating Majorana mass of the neutrino. It has been 
recognized about 15 years ago \cite{Sch81} that if any of these two quantities
vanishes, the other one vanishes, too, and vice versa, if one of them is
non--zero, the other one also differs from zero. This Schechter-Valle-theorem 
is valid for
any gauge model with spontaneously broken symmetry at the weak scale,
independent of the mechanism of $0\nu\beta\beta$ decay. A generalisation
of this theorem to supersymmetry has been given recently \cite{Hir97,Hir97a}.
This theorem claims for the neutrino 
Majorana mass, the $B-L$ violating mass of the
sneutrino and neutrinoless double beta decay amplitude:
If one of them is non--zero, also the others are non--zero and vice versa,
independent of the mechanisms of $0\nu\beta\beta$ decay and (s-)neutrino
mass generation. This theorem connects double beta research with new processes
potentially observable at future colliders like NLC (next linear collider)
\cite{Hir97,Kolb1}.

\subsubsection*{1.3.1.1 Neutrino mass}


Neutrino physics has entered an era of new actuality in connection
with several possible indications of physics beyond the standard model
(SM) of particle physics: A lack of solar
($^7Be$) neutrinos, an atmospheric $\nu_{\mu}$ deficit and mixed dark matter 
models could all be explained simultaneously by non--vanishing neutrino masses.
Recent GUT models, for example an extended SO(10) scenario with $S_4$
horizontal symmetry could explain these observations by requiring 
degenerate neutrino masses of the order of 1 eV 
\cite{19,Moh94,20,21,22,23} \cite{12,13}.
For an overview see \cite{Smi96a,Mohneu}.
More recent theoretical discussions are given in 
\cite{Kla99b,Adh98,Min97,Giu99,Ma99,Vis99,Bil99}.

From all this work it is clear that double beta decay experiments have come 
into some key
position, since 
the predictions of or assumptions in such
scenarios start now to become testable partly already by the most advanced present experiments like the Heidelberg-Moscow experiment.

Neutrinoless double beta decay can be triggered by exchange of a
light or heavy left-handed Majorana neutrino (see Fig. \ref{fig3-paes}).
For exchange of a heavy {\it right}--handed neutrino see below.
      The propagators in the first and second case show a different $m_{\nu}$
dependence: Fermion propagator $\sim \frac {m}{q^2-m^2} \Rightarrow$
\be
a)\hskip5mm m\ll q \rightarrow \sim m \hskip5mm 'light' \hskip2mm neutrino
\ee
\be
b)\hskip5mm m\gg q \rightarrow \sim \frac{1}{m} \hskip5mm 'heavy' \hskip2mm
neutrino
\ee   
The half--life for $0\nu\beta\beta$ decay induced by exchange of a light 
neutrino is given by \cite{27}
\ba{71}
[T^{0\nu}_{1/2}(0^+_i \rightarrow 0^+_f)]^{-1}= C_{mm} 
\frac{\langle m_{\nu} \rangle^2}{m_{e}^2}
+C_{\eta\eta} \langle \eta \rangle^2 + C_{\lambda\lambda} 
\langle \lambda \rangle^2 +C_{m\eta} \frac{m_{\nu}}{m_e} 
\nn
\ea
\be
+ C_{m\lambda}
\langle \lambda \rangle \frac{\langle m_{\nu} \rangle}{m_e} 
+C_{\eta\lambda} 
\langle \eta \rangle \langle \lambda \rangle
\ee
or, when neglecting the effect of right--handed weak currents, by
\be
[T^{0\nu}_{1/2}(0^+_i \rightarrow 0^+_f)]^{-1}=C_{mm} 
\frac{\langle m_{\nu} \rangle^2}{m_{e}^2}
=(M^{0\nu}_{GT}-M^{0\nu}_{F})^2 G_1 
\frac{\langle m_{\nu} \rangle^2}{m_e^2}
\ee
where $G_1$ denotes the phase space integral, $ \langle m_{\nu} \rangle$
denotes an effective neutrino mass
\be
\langle m_{\nu} \rangle = \sum_i m_i U_{ei}^2,
\label{obs}
\ee
respecting the possibility of the electron neutrino to be a mixed state
(mass matrix not diagonal in the flavor space)
\be
|\nu_e \rangle = \sum_i U_{ei} |\nu_{i}\rangle
\ee
For Majorana neutrinos, $U$ is given by
 \be
\footnotesize
{\tiny{
\left(
     \begin{array}{ccc}
  c_{12}c_{13} & s_{12}c_{13}e^{-i\delta_{12}} & s_{13}e^{-i\delta_{13}}\\
 -s_{12}c_{23}e^{i\delta_{12}}
 -c_{12}s_{23}s_{13}e^{i(\delta_{13}+\delta_{23})} &
  c_{12}c_{23}-s_{12}s_{23}s_{13}e^{i(\delta_{23}+\delta_{13}-\delta_{12})}& 
  s_{23}c_{13}e^{i\delta_{23}} \\
  s_{12}s_{23}e^{i(\delta_{23}+\delta_{13})}
 -c_{12}c_{23}s_{13}e^{i(\delta_{23}+\delta_{13})} &
 -c_{12}s_{23}e^{i\delta_{23}}
 -s_{12}c_{23}s_{13}e^{i(\delta_{13}-\delta_{12})} & c_{23}c_{13} \\
     \end{array}
     \right),}}
\ee
\normalsize
where $s_{ij}=sin \theta_{ij}, c_{ij}=cos \theta_{ij}$ and $\delta$ is a 
CP violating phase. For a given neutrino oscillation pattern the absolute
$\nu$ masses are not fixed, adding an arbitrary $m_0$
\be
m_i \rightarrow m_i + m_0
\ee
does not change the oscillation probabilities, but the $0\nu\beta\beta$ 
rate. This means the oscillation pattern does not restrict the effective 
Majorana 
mass. Thus neutrinoless double beta decay is a indispensable crucial check of 
neutrino mass 
models. 

The effective mass $\langle m_{\nu} \rangle$ could be smaller than $m_i$
for all i for appropriate CP phases of the mixing coefficiants $U_{ei}$
\cite{Wol81}.
In general not too pathological GUT models yield 
$m_{\nu_e}=\langle m_{\nu_e}
\rangle$ (see \cite{15}).

$\eta$,$\lambda$ describe an admixture of right--handed weak currents, and
$M^{0\nu}\equiv M_{GT}^{0\nu}-M_{F}^{0\nu}$ denote nuclear matrix elements.

\subsubsection*{Nuclear matrix elements:}

A detailed discussion of $\beta\beta$ matrix elements for neutrino induced
transitions including the substantial (well--understood) differences
in the precision with which $2\nu$ and $0\nu\beta\beta$ rates can be 
calculated, can be found in \cite{16,27,28} \cite{29,KK1,KK2}.

\subsubsection*{1.3.1.2  Supersymmetry}

\begin{figure}
\epsfxsize=50mm
\epsfbox{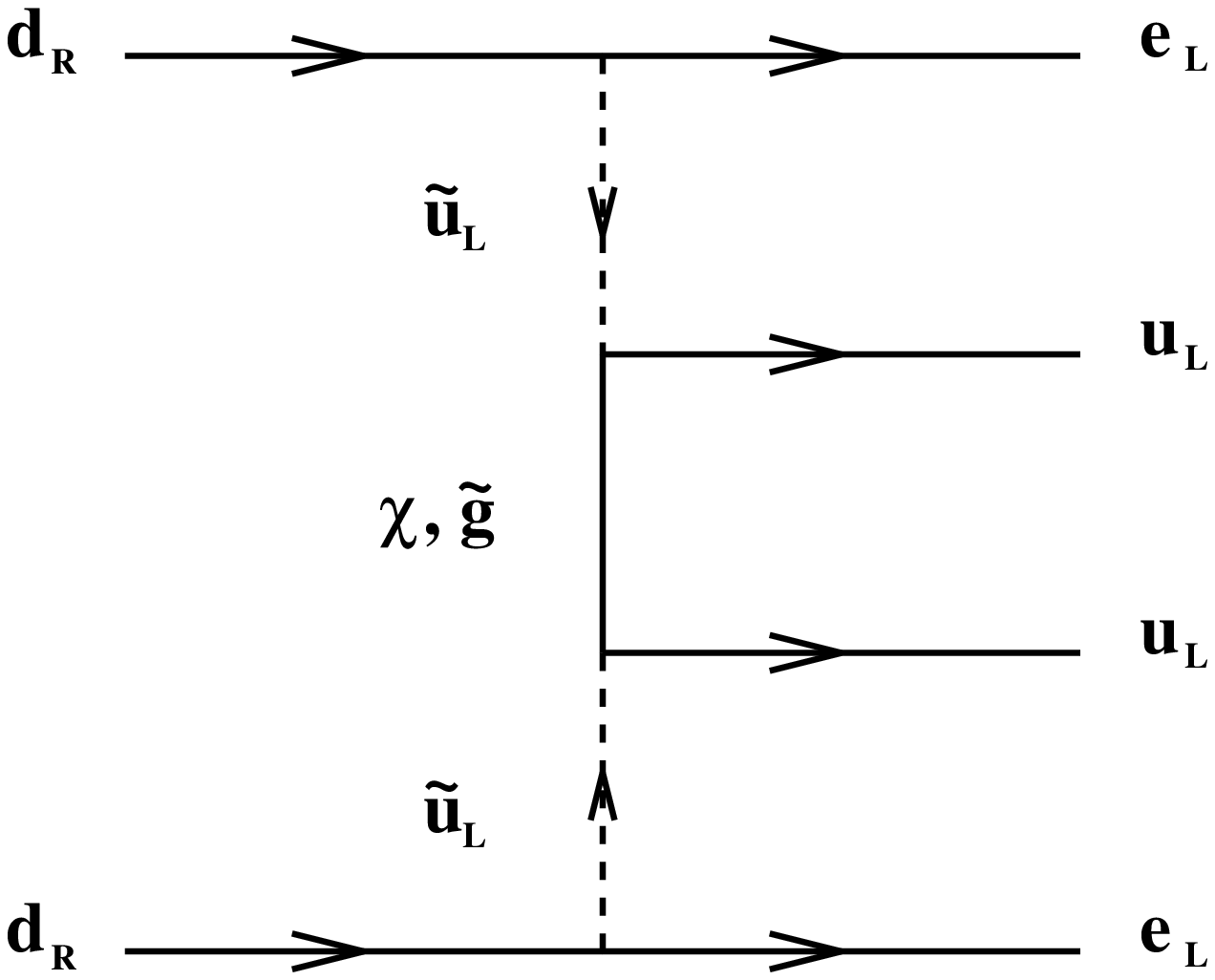}
\vspace*{-45mm}
\hspace*{60mm}
\epsfxsize=50mm
\epsfbox{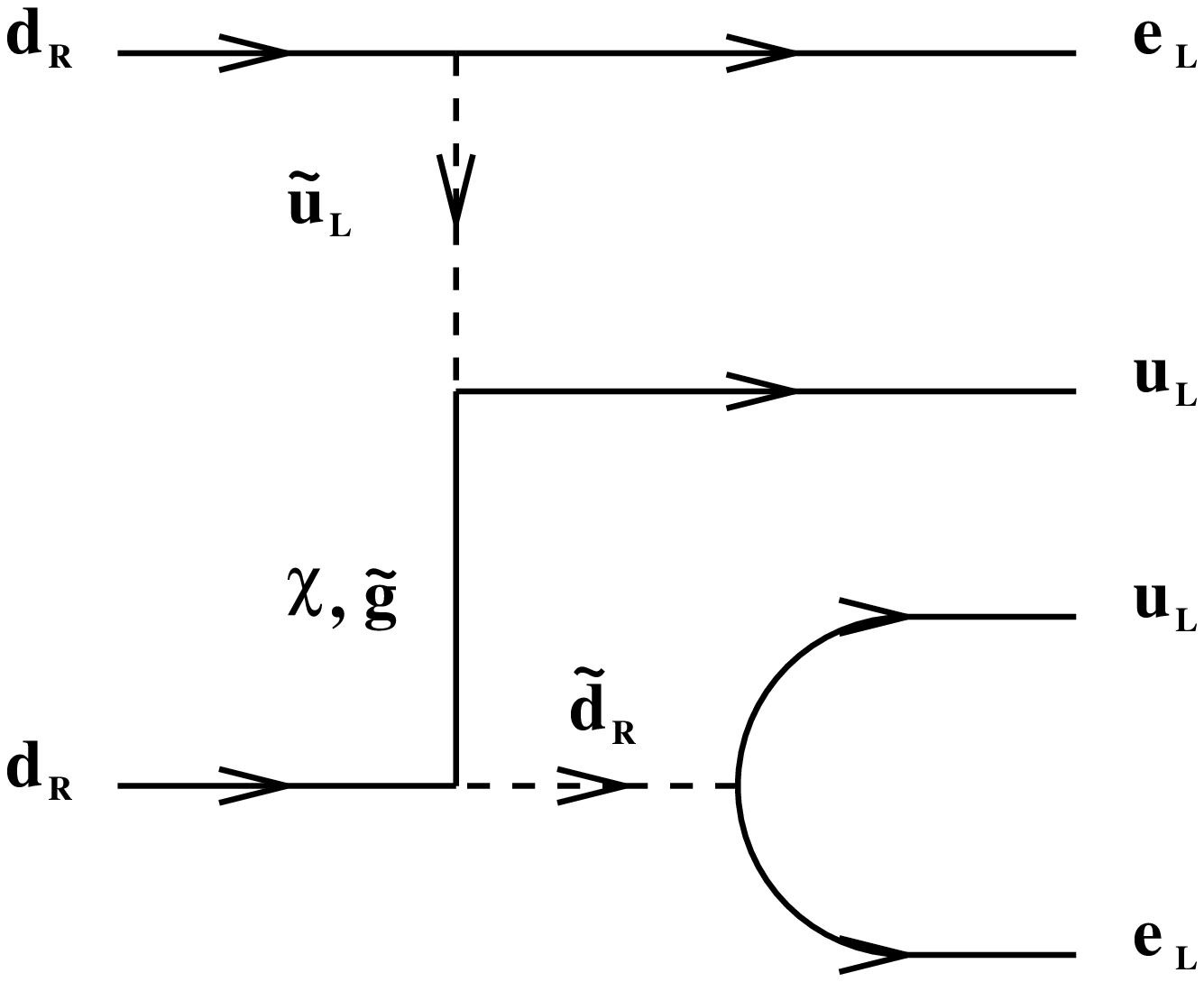}
\caption{ Examples of Feynman graphs for $0\nu\beta\beta$ decay within
  R--parity violating supersymmetric models (from [Hir95a]).}
\label{fig5}
\end{figure}

Supersymmetry (SUSY) is considered as prime candidate for a theory beyond the
standard model, which could overcome some of the most puzzling questions of
today's particle physics (see, e.g. \cite{44,45,Kan97}). 
Generally one can add the following R--parity violating terms  
to the usual superpotential \cite{hal84}.
\be
W_{\Rpv}=\lambda_{ijk}L_{i}L_{j}\overline{E}_{k}+\lambda^{'}_{ijk}
L_i Q_j \overline{D}_k + \lambda^{''}\overline{U}_i \overline{D}_j 
\overline{D}_k,
\ee
where indices $i,j,k$ denote generations. $L$,$Q$ denote lepton and quark 
doublet superfields and $\overline{E}, \overline{U}, \overline{D}$ lepton and
up, down quark singlet superfields. Terms proportional to $\lambda$, 
$\lambda^{'}$
violate lepton number, those proportional to $\lambda^{''}$ violate baryon 
number. From proton decay limits it is clear that both types of terms cannot 
be present at the same time in the superpotential. On the other hand, once the
$\lambda^{''}$ terms being assumed to be zero, $\lambda$ and $\lambda^{'}$
terms are not limited. $0\nu\beta\beta$ decay can occur within the 
\rp MSSM through Feynman graphs  such as those of Fig. \ref{fig5} 
In lowest order 
there are alltogether six different graphs of this kind. \cite{6,47,75}.    
Thus  $0\nu\beta\beta$ decay can be used to restrict R--parity violating
SUSY models \cite{6,hir96c,17,47,48}. From these graphs one derives \cite{6}
under some assumptions 
\be
[T^{0\nu}_{1/2}(0^+ \rightarrow 0^+)]^{-1} \sim G_{01}
(\frac{\lambda_{111}'^2}{m^4_{{\tilde q},{\tilde e}}m_{{\tilde g}\chi}}M)^2
\ee
where $G_{01}$ is a phase space factor, 
$m_{{\tilde q}{\tilde e}{\tilde g}\chi}$
are the masses of supersymmetric particles involved: squarks, selectrons,
gluinos, or neutralinos. $\lambda'_{111}$ is the strength of an R--parity
breaking interaction (eq. 11), and $M$ is a nuclear matrix element. For the
matrix elements and their calculation (see \cite{hir96c}).

\begin{figure}
\epsfxsize=50mm
\epsfbox{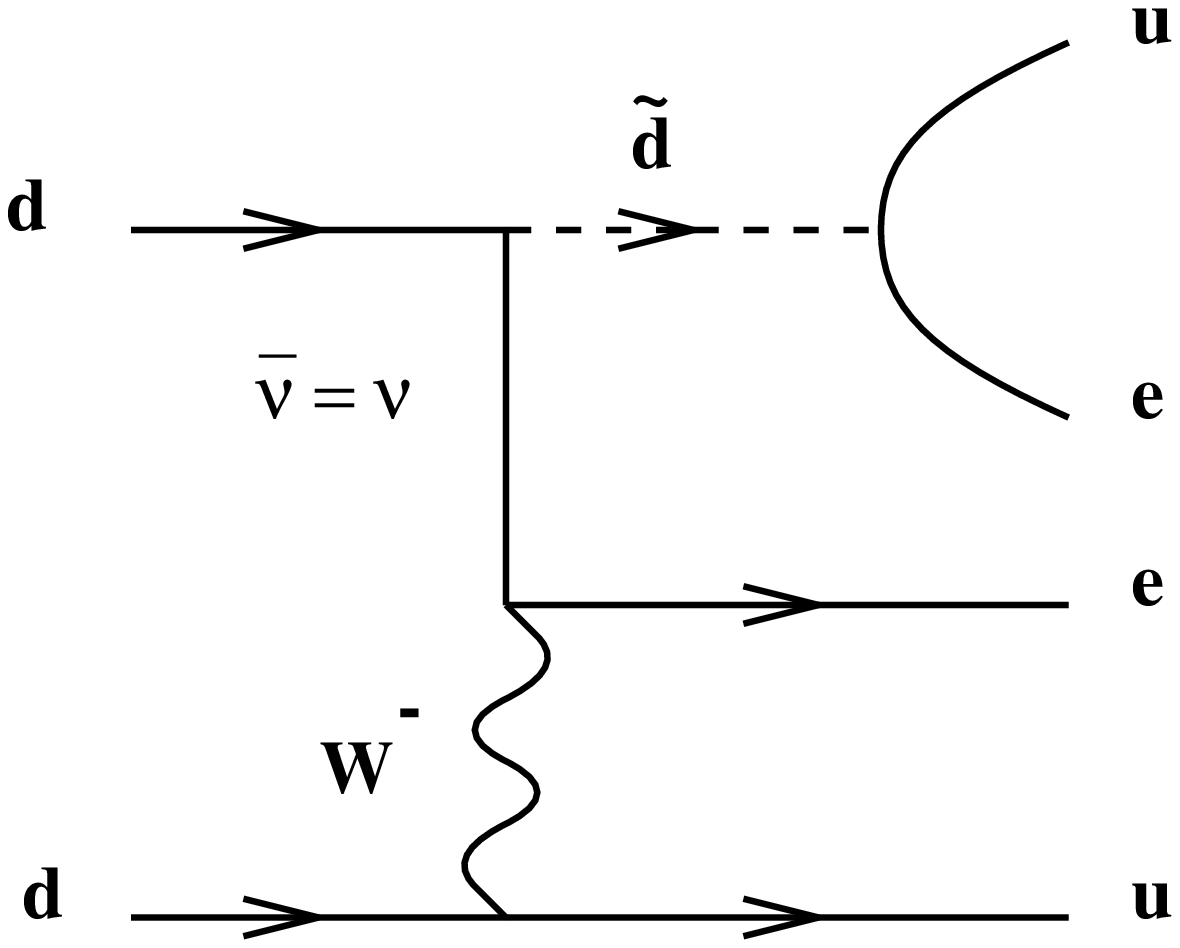}
\vspace*{-45mm}
\hspace*{60mm}
\epsfxsize=50mm
\epsfbox{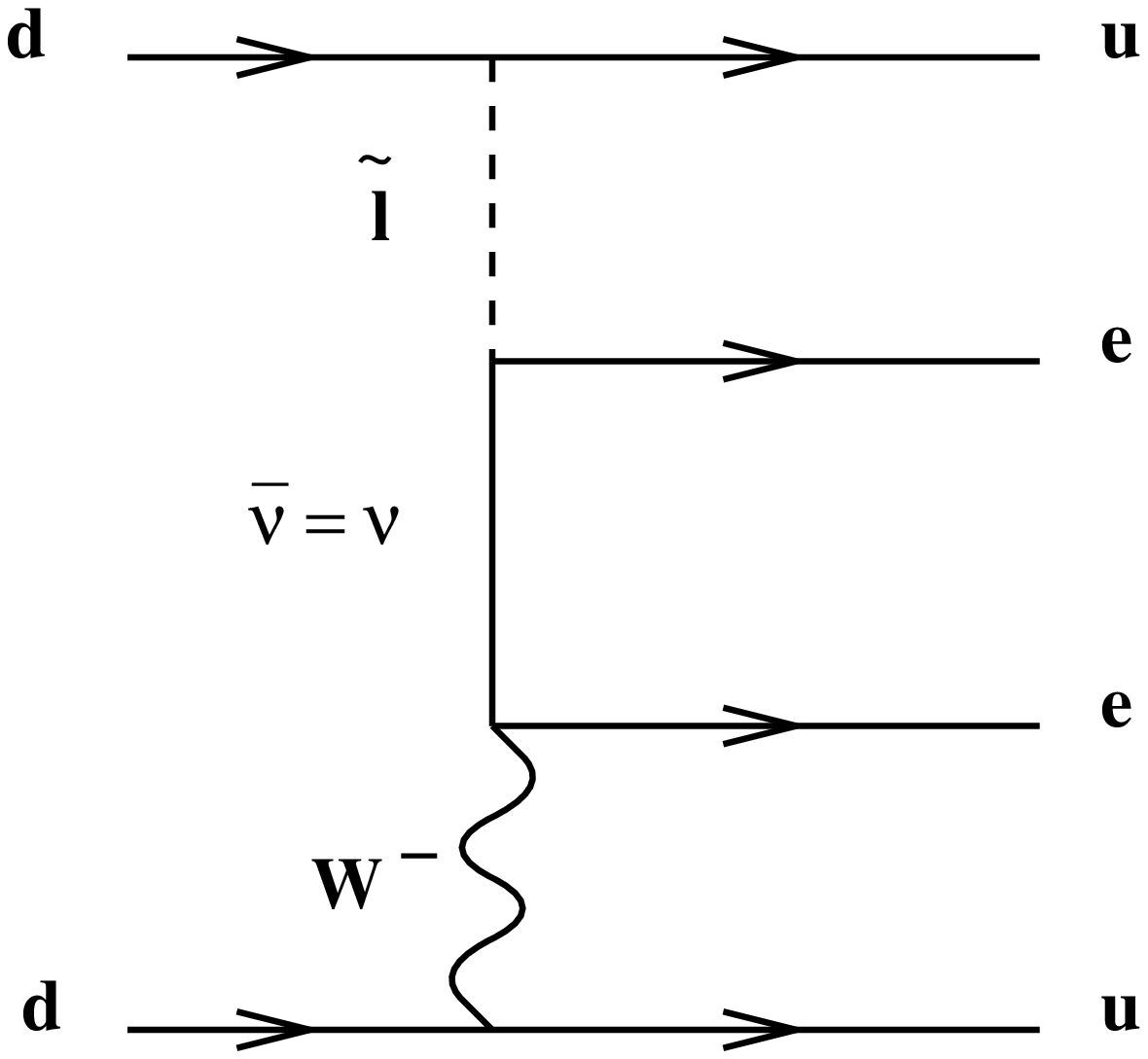}
\caption{ (left) Feynman graph for the mixed SUSY-neutrino exchange mechanism 
  of 0$\nu\beta\beta$ decay. R-parity violation occurs through scalar
  quark exchange. (right) As left figure, 
but for scalar lepton exchange (from [Hir96]).}
\label{fig6}
\end{figure}

\begin{figure}
\parbox{14cm}{
\vspace*{6mm}
\epsfxsize=50mm
\epsfbox{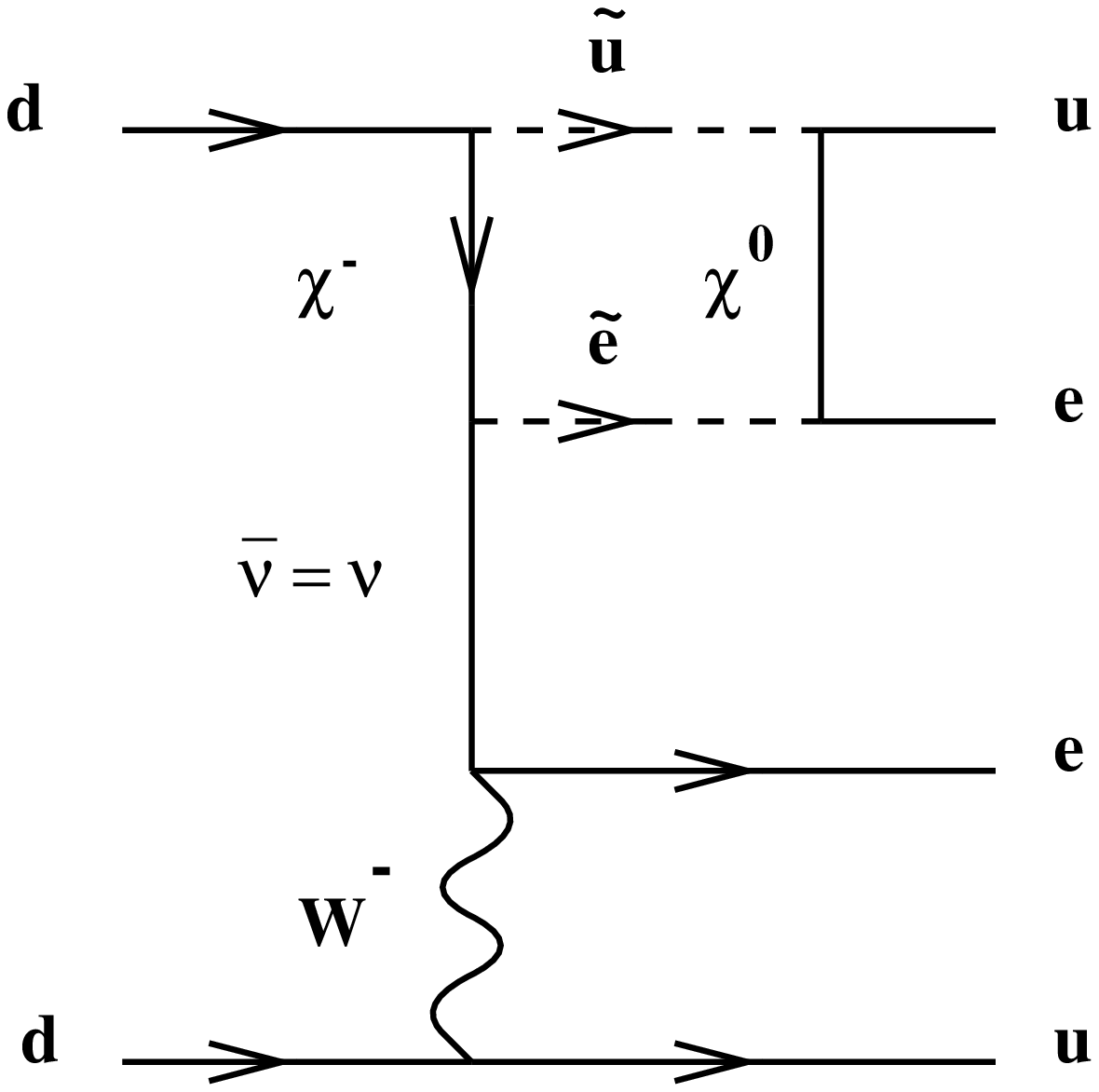}
\parbox{6cm}{
\vspace*{-45mm}
\hspace*{60mm}
\epsfxsize=50mm
\epsfbox{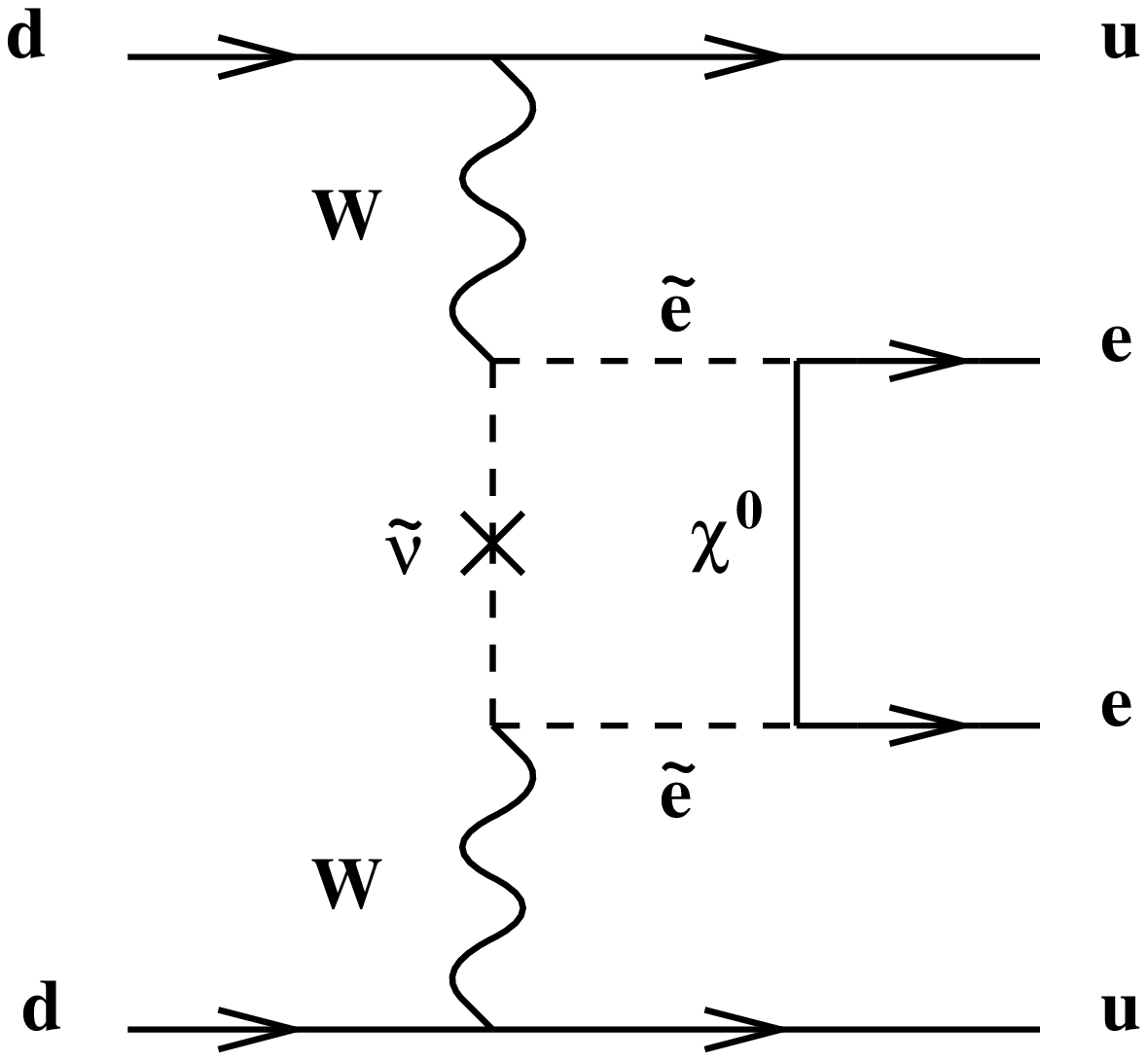}
\vspace*{8mm}
}
\caption{ Examples of $R_P$ conserving SUSY contributions
to $0\nu\beta\beta$ decay 
(from [Hir97a]).}}   
\label{fig7}
\end{figure}

It is also worthwile to notice that $0\nu\beta\beta$ decay is not only 
sensitive to $\lambda^{'}_{111}$. Taking into account the fact that the SUSY
partners of the left and right--handed quark states can mix with each other, 
one can derive limits on different combinations of $\lambda^{'}$ 
\cite{hir96,7,bab95,Paes99b} (see Fig. \ref{fig6}).
The dominant diagram 
of this type is the one where the exchanged scalar particles are the
$\tilde{b}-\tilde{b}^C$ pair. Under some assumptions (e.g. the MSSM mass 
parameters to be approximately equal to the ``effective'' SUSY breaking scale 
$\Lambda_{SUSY}$), one obtains \cite{hir96}
\be
\lambda_{11i}^{'}\cdot \lambda_{1i1}^{'}\leq \epsilon_i^{'} 
\Big( \frac{\Lambda_{SUSY}}{100 GeV} \Big)^3
\ee
and
\be
\Delta_n \lambda^{'}_{311} \lambda_{n13} \leq \epsilon \Big(\frac
{\Lambda_{SUSY}
}{100 GeV}\Big)^3
\ee

\begin{figure}
\epsfxsize=100mm
\vspace*{-3cm}
\epsfbox{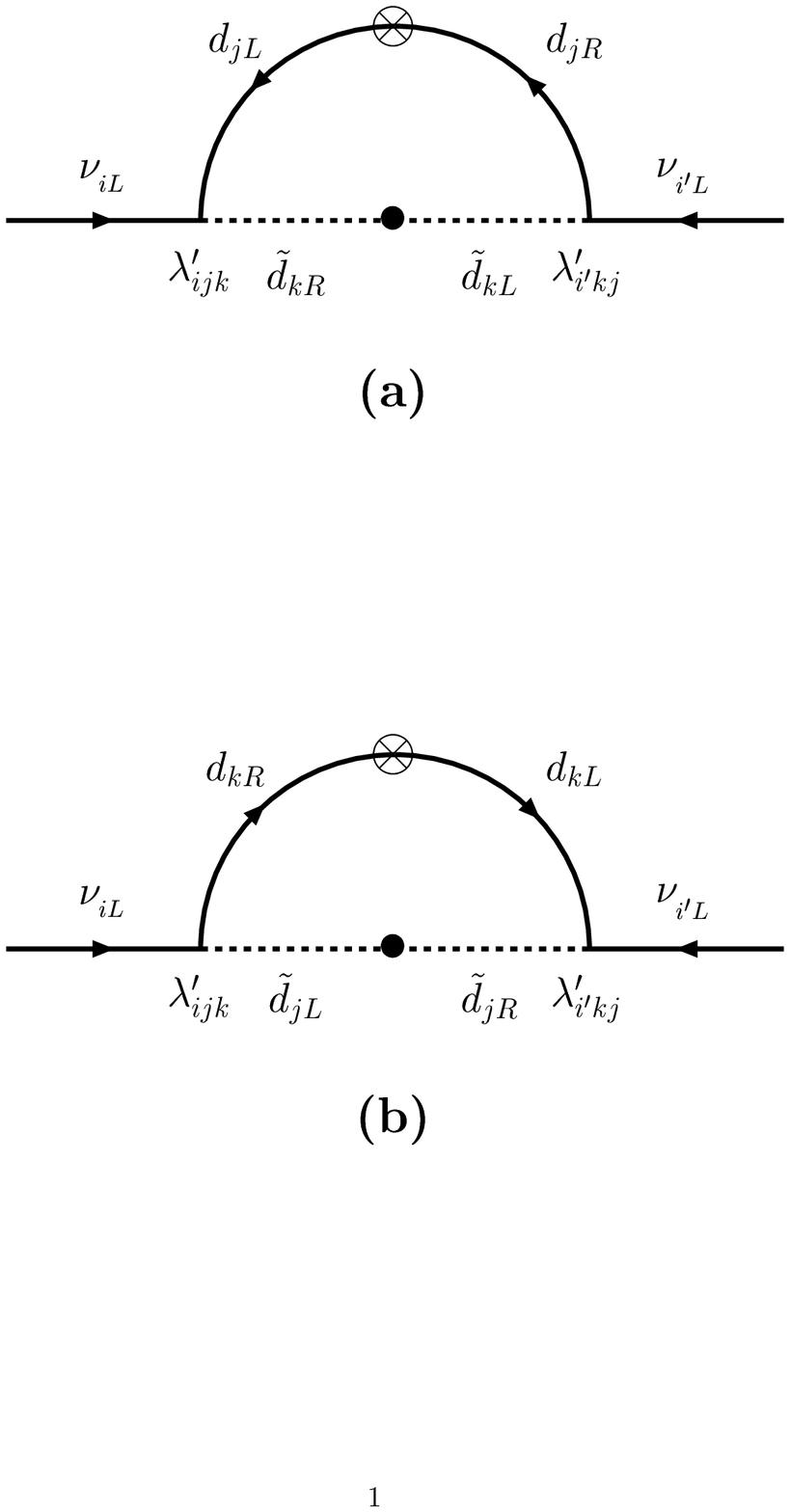}
\vspace*{-3cm}
\caption{
The $\lambda'$-induced one loop diagrams
contributing to Majorana masses for the neutrinos.}
\label{rpvfiggen}
\end{figure}

Further constraints on R parity violating Supersymmetry may be obtained
directly from the neutrino mass bound \cite{Bha99}.
Products of trilinear couplings $\lambda$ and/or
$\lambda'$ may generate a complete neutrino mass matrix through
one-loop self-energy graphs \cite{trilinear,recent}.
Let us first consider the effects of the $\lambda'$ interactions. The
relevant part of the Lagrangian can be written as 
\begin{equation}
 - {\cal L}_{\lambda'} = \lambda'_{ijk} \left[\bar{d}_k P_L \nu_i
  \tilde{d}_{jL} + \bar{\nu}^c_i P_L d_j \tilde{d}^*_{kR}\right]
+ ~{\rm h.c.}   
\label{lagrangian}
\end{equation}
Majorana mass terms for the left-handed neutrinos, given by 
\begin{equation}
{\cal
L}_M = -\frac{1}{2} m_{\nu_{ii'}} \bar{\nu}_{Li} \nu^c_{Ri'} +~{\rm
h.c.}, 
\end{equation}
are generated at one loop. Fig. \ref{rpvfiggen} shows the
corresponding diagrams. The induced masses are given by
\begin{equation} 
m_{\nu_{ii'}} \simeq {{N_c \lambda'_{ijk} \lambda'_{i'kj}}
\over{16\pi^2}} m_{d_j} m_{d_k}
\left[\frac{f(m^2_{d_j}/m^2_{\tilde{d}_k})} {m_{\tilde{d}_k}} +
\frac{f(m^2_{d_k}/m^2_{\tilde{d}_j})} {m_{\tilde{d}_j}}\right],
\label{mass}
\end{equation}     
where $f(x) = (x\ln x-x+1)/(x-1)^2$. Here, $m_{d_i}$ is the down quark
mass of the $i$th generation inside the loop, $m_{\tilde{d}_i}$ is an
average of $\tilde{d}_{Li}$ and $\tilde{d}_{Ri}$ squark masses, and
$N_c = 3$ is the colour factor. In deriving Eq.~(\ref{mass}), it was
assumed that the left-right squark mixing terms in the soft part of
the Lagrangian are diagonal in their physical basis and proportional
to the corresponding quark masses, {\em i.e.} $\Delta m^2_{\rm LR} (i)
= m_{d_i} m_{\tilde{d}_i}$. With $\lambda$-type interactions, one obtains 
exactly similar results
as above. The quarks
and squarks in these equations will be replaced by the leptons and
sleptons of the corresponding generations. The colour factor $N_c = 3$
and $Q_d$ would be replaced by $1$
and $Q_e$, respectively. 
Among the different entries of the flavour space mass
matrix, only the $ee$-term has a {\em direct experimental} bound
obtained from neutrinoless double-beta decay. 

For an overview on our knowledge
on $\lambda^{'}_{ijk}$
from other sources we refer to \cite{Kol97a,Bha97,Bha99}.

Also R--parity {\it conserving} softly broken supersymmetry can give 
contributions to $0\nu\beta\beta$ decay, via the $B-L$--violating sneutrino 
mass term, the latter being a generic ingredient of any weak--scale SUSY
model with a Majorana neutrino mass \cite{Hir97,Kolb1}. 
These contributions are 
realized  at the level of box diagrams \cite{Kolb1} (Fig. 13).
The $0\nu\beta\beta$ half-life for contributions from sneutrino exchange
is found to be \cite{Kolb1}
\be
[{T_{1/2}^{0\nu\beta\beta}}]^{-1}=G_{01}\frac{4 m_p^2}{G^4_F} 
\Big|\frac{\eta^{SUSY}}{m^5_{SUSY}} M^{SUSY}\Big|,
\ee
where the phase factor $G_{01}$ is tabulated in \cite{74}, $\eta^{SUSY}$
is the effective lepton number violating parameter, which contains the
$(B-L)$ violating sneutrino mass $\tilde{m}_M$ and $M^{SUSY}$ is the nuclear 
matrix element \cite{11}.

\subsubsection*{ 1.3.1.3 Left--Right symmetric theories --
Heavy neutrinos and right--handed W Boson}

Heavy {\it right--handed } neutrinos appear quite naturally in left--right
symmetric GUT models. Since in such models the symmetry breaking
scale for the right--handed sector is not fixed by the theory, the
mass of the right--handed $W_R$ boson and the mixing angle between the mass 
eigenstates $W_1$, $W_2$ are free parameters. $0\nu\beta\beta$ decay taking
into account contributions from both, left-- and right--handed neutrinos
have been studied theoretically by \cite{11,49}. The former gives a more
general expression for the decay rate than introduced earlier by \cite{50}. 

\begin{figure}
\vspace*{6mm}
\epsfxsize=50mm
\epsfbox{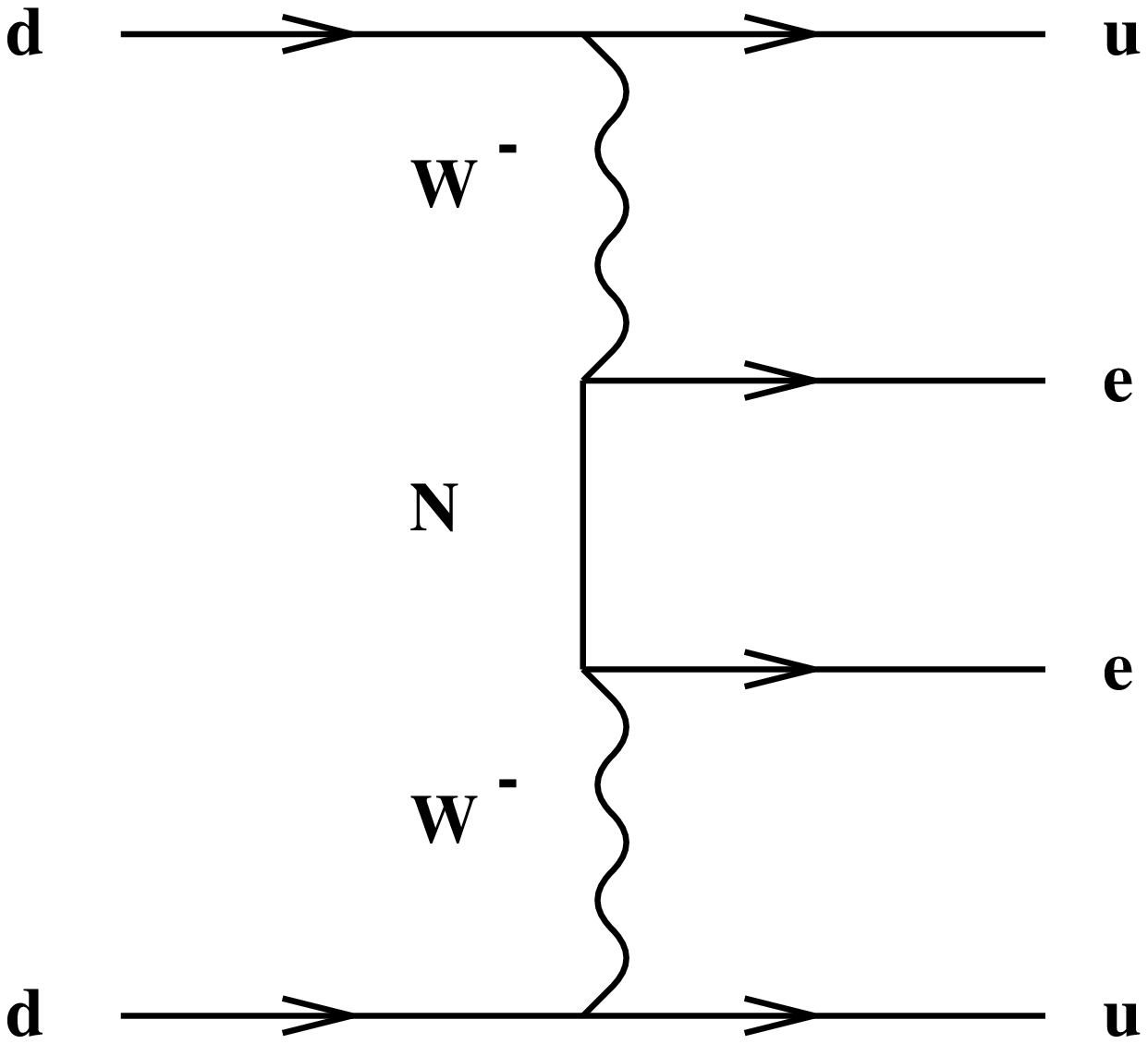}
\vspace*{-45mm}
\hspace*{60mm}
\epsfxsize=50mm
\epsfbox{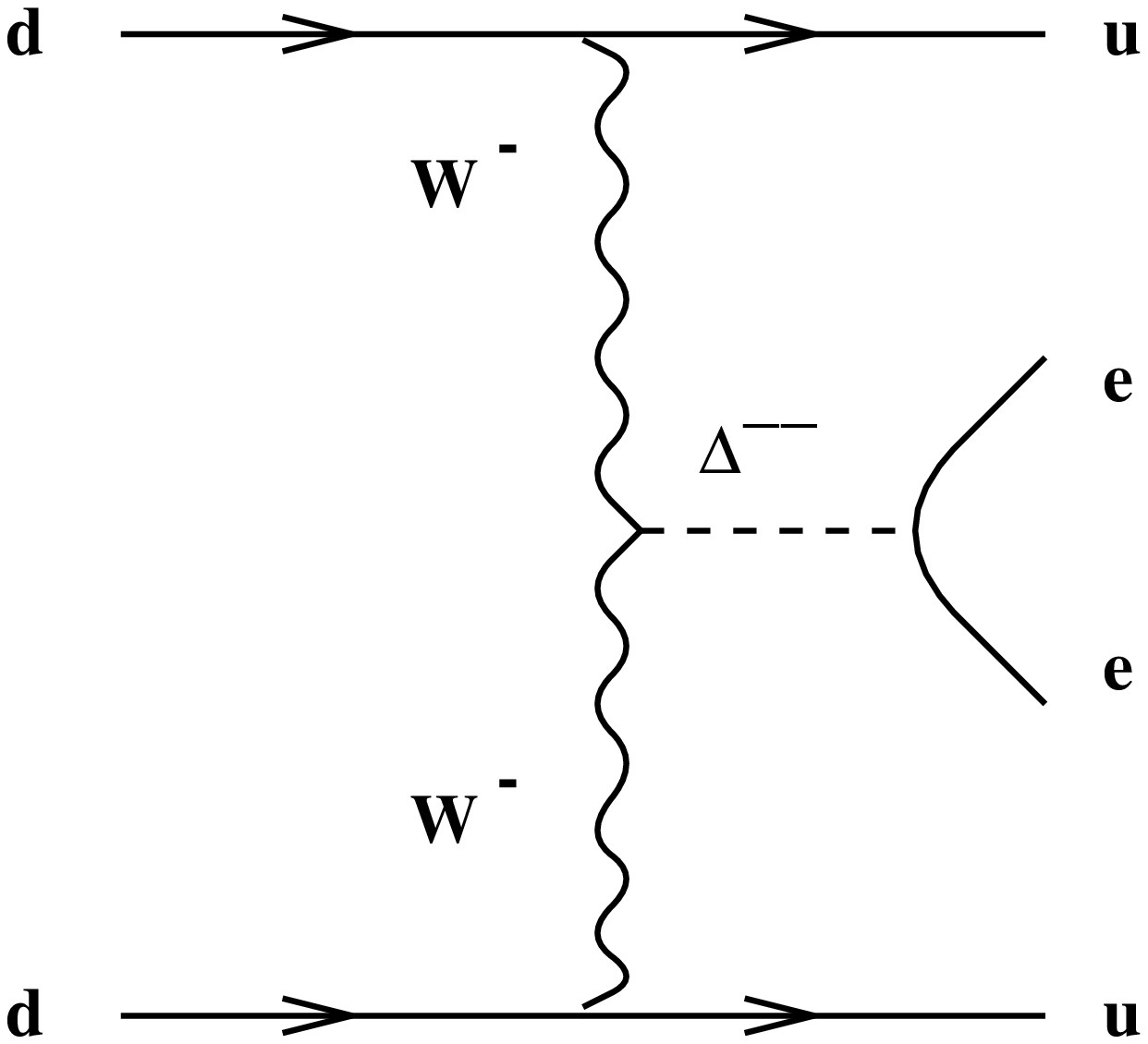}
\vspace*{8mm}
\caption{ Left: Heavy neutrino exchange contribution to neutrinoless
  double beta decay in left-right symmetric models, and Right: Feynman
  graph for the virtual exchange of a double-charged Higgs boson
(from [Hir96d]).}   
\label{fig8}
\end{figure}

The amplitude will be proportional to (see Fig. \ref{fig8})\cite{11}
\be
\Big( \frac{m_{W_{L}}}{m_{W_R}}^4 \Big) \Big(\frac{1}{m_N}+\frac{m_N}
{m^2_{\Delta^{--}_R}}\Big)
\label{ncs3}
\ee

Eq. \ref{ncs3} and the experimental lower limit of $0\nu\beta\beta$ decay leads 
to a constraint limit within the 3--dimensional parameter space
($m_{W_R}-m_N-m_{\Delta^{--}_R}$).

\subsubsection*{1.3.1.4 Compositeness}

\begin{figure}
\epsfxsize=80mm
\epsfbox{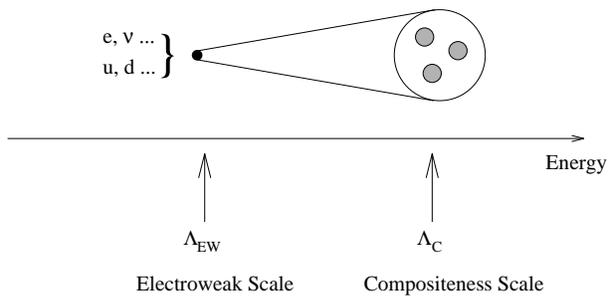}
\caption{The idea of compositeness. At a (still unknown) energy scale
  $\Lambda_C$ quarks and leptons might show an internal structure}   
\label{fig9}
\end{figure}

Although so far there are no experimental signals of a substructure of quarks 
and leptons, there are speculations that at some higher energy ranges beyond 1 
TeV or so there might exist an energy scale $\Lambda_C$ at which a
substructure of quarks and leptons (preons) might become visible 
\cite{8,45,51,Pan99} (Fig. \ref{fig9}). 

\begin{figure}
\epsfxsize=80mm
\hspace*{1cm}
\epsfbox{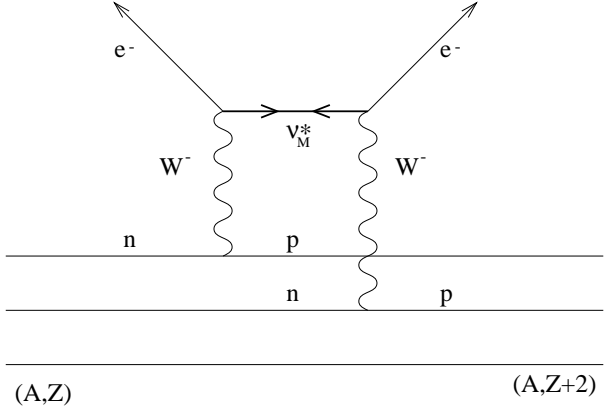}
\caption{Neutrinoless double beta decay ($\Delta$L = +2 process)
mediated by a composite heavy Majorana neutrino.}
\label{fig10}
\end{figure}

A possible low energy manifestation of compositeness could be neutrinoless
double beta decay, mediated by a composite heavy Majorana neutrino, 
which then should be a Majorana particle (Fig. \ref{fig10}).

Recent theoretical work shows (see \cite{8,9,Pan97,Tak97,Pan99}) 
that the mass bounds for such an excited neutrino 
which can be derived from double 
beta decay are at
 least of the same order of magnitude as those coming from the
direct search of excited states in high energy accelerators 
(see also subsection 1.3.2).

\subsubsection*{1.3.1.5 Majorons} 

The existence of new bosons, so--called Majorons, can play a significant
role in new physics beyond the standard model, in the history
of the early universe, in the evolution of stellar objects, in supernovae
astrophysics and the solar neutrino problem \cite{61,62,Kla92}.
In many theories of physics beyond the standard model neutrinoless 
double beta decay can occur with the emission of Majorons  
\be
2n\rightarrow2p+2e^{-}+\phi
\ee
\be
2n\rightarrow2p+2e^{-}+2\phi.
\ee 

To avoid an unnatural fine--tuning in recent years several 
new Majoron models were proposed \cite{68,69,70}, 
where the term
Majoron denotes in a more general sense light or massless bosons 
with couplings to neutrinos. 

The main novel features of these ``New Majorons'' are that they
can carry leptonic charge, that they need not be 
Goldstone bosons and that emission of two Majorons 
can occur. 
The latter can be scalar--mediated 
or fermion--mediated. For details we refer to
\cite{71,72}. 

The half--lifes are according to \cite{73,74} in some approximation given
by  
\be
[T_{1/2}]^{-1}=|<g_{\alpha}>|^{2}\cdot|M_{\alpha}|^{2}\cdot G_{BB_{\alpha}}
\ee
for $\beta\beta\phi$-decays, or
\be
[T_{1/2}]^{-1}=|<g_{\alpha}>|^{4}\cdot|M_{\alpha}|^{2}\cdot G_{BB_{\alpha}}
\ee
for $\beta\beta\phi\phi$--decays. The index ${\alpha}$ 
indicates that effective neutrino--Majoron coupling constants $g$, 
matrix elements $M$ and phase spaces $G$ differ for different models.

\subsubsection*{Nuclear matrix elements:}

There are five different nuclear matrix elements. Of
 these $M_{F}$ and $M_{GT}$ are the same which occur in $0\nu\beta\beta$ decay.
The other ones and the corresponding phase spaces have been calculated 
for the first time
by \cite{71,75}. The calculations of the matrix elements show 
that the new models predict, 
as consequence of the small matrix elements, 
 very large half--lives and that unlikely large
coupling constants would be needed to produce observable decay rates
(see \cite{71,75}).

\subsubsection*{1.3.1.6 Sterile neutrinos}

\begin{figure}[h!]
\epsfysize=8cm
\epsfbox{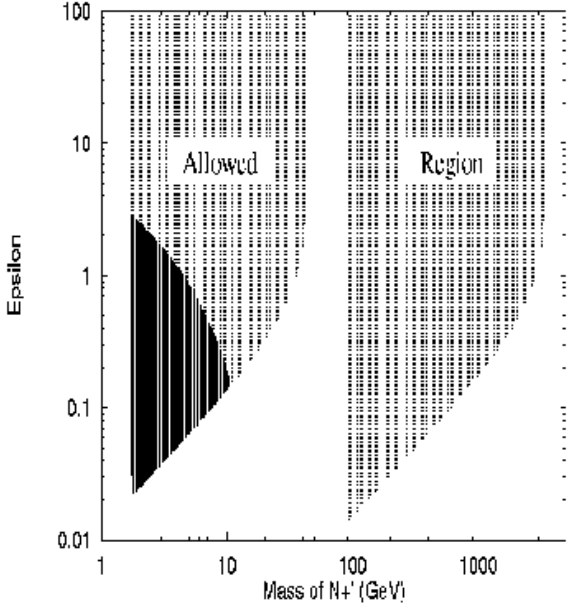}
\caption{Regions of the parameter space 
 ($\epsilon-M_{N'_{+}}$ )plane yielding
an observable signal (shaded areas) (from \cite{77}). Darker area: 'natural' 
region, lighter shaded: Finetuning needed, to keep $m_{\nu_e}$ below 1 eV.
$M_{N'_+}$: mass eigenstate, $\epsilon$: strength of lepton number violation in
mass matrix}
\label{fig11}
\end{figure}

Introduction of sterile neutrinos has been claimed to solve simultaneously the 
conflict between dark matter neutrinos, LSND and supernova nucleosynthesis
\cite{76} and light sterile neutrinos are part of popular 
neutrino mass scenarios
for understanding the various hints for neutrino
oscillations  and \cite{Moh96,Mohneu,Moh97a}. 
Neutrinoless double beta decay can also
investigate several effects
of {\it heavy} sterile neutrinos \cite{77} (Fig. \ref{fig11}).

If we assume having a light neutrino with a mass $\ll$ 1 eV, mixing with a much
 heavier (m $\ge$ 1 GeV) sterile neutrino can yield under certain conditions
a detectable signal in current $\beta\beta$ experiments.


\subsubsection*{1.3.1.7 Leptoquarks}

Interest on leptoquarks (LQ) has been renewed during the last few years 
since ongoing collider experiments have good prospects for searching 
these particles \cite{Lagr1}. LQs are vector or scalar particles 
carrying both lepton and baryon numbers and, therefore, have a 
well distinguished experimental signature. Direct searches of LQs in 
deep inelastic ep-scattering at HERA \cite{H196} placed lower limits 
on their mass $M_{LQ} \ge 225-275$ GeV, depending on the LQ type and 
couplings. 

\begin{figure}[h!]
\epsfxsize=50mm
\epsfysize=50mm
\hspace*{0.0cm}
\epsfbox{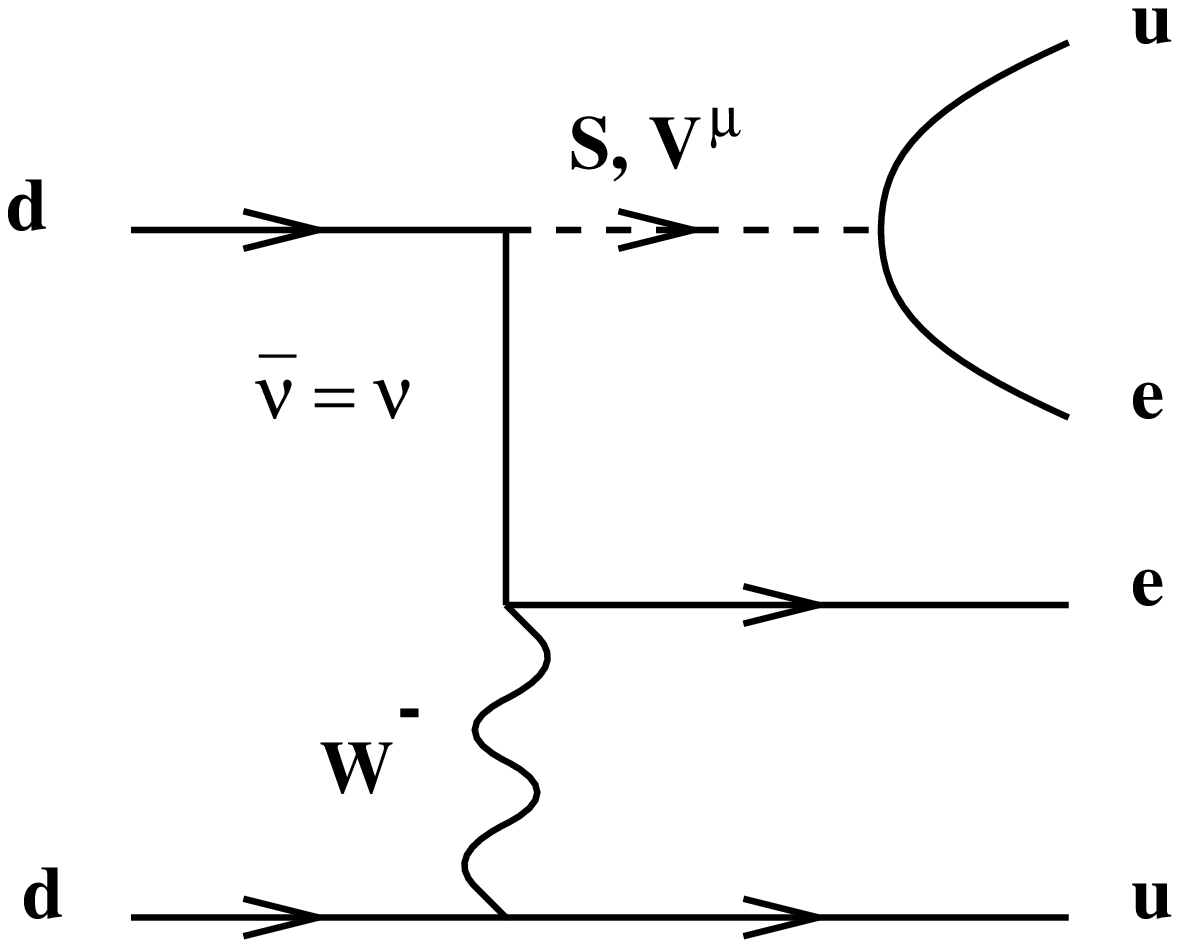}
\hspace*{1cm}
\epsfxsize=50mm
\epsfysize=50mm
\epsfbox{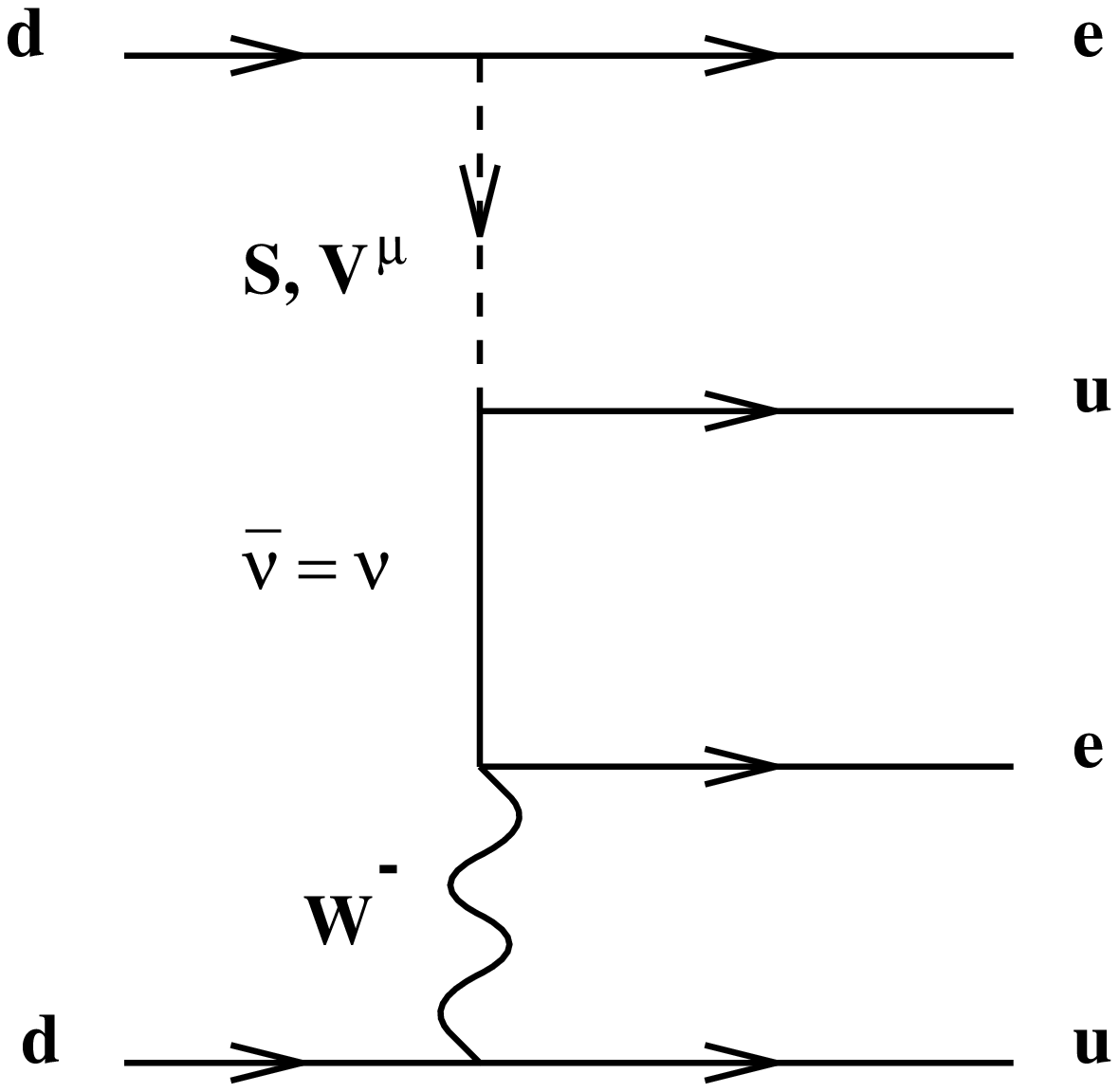}
\vskip5mm
\caption{ Examples of Feynman graphs for $0\nu\beta\beta$ decay 
within LQ models. $S$ and $V^{\mu}$ stand symbolically for scalar 
and vector LQs, respectively (from [Hir96a]).}
\label{feyn2}
\end{figure}

To consider LQ phenomenology in a model-independent fashion one 
usually follows some general principles in constructing the Lagrangian 
of the LQ interactions with the standard model fields. In order to 
obey the stringent constraints from (c1) helicity-suppressed 
$\pi \rightarrow e\nu$ decay, from (c2) FCNC processes and 
from (c3) proton stability, the following assumptions are commonly adopted: 
(a1) LQ couplings are chiral, (a2) LQ couplings are generation 
diagonal, and (a3) there are no diquark couplings. 

Recently, however, it has been pointed out \cite{hir96a} that possible 
LQ-Higgs interactions spoil assumption (a1): Even if one assumes 
LQs to be chiral at some high energy scale, LQ-Higgs interactions 
introduce after electro-weak symmetry breaking mixing between 
LQ states with different chirality. Since there is no fundamental 
reason to forbid such LQ-Higgs interactions, it seems difficult 
to get rid of the unwanted non-chiral interactions in LQ models. 

In such LQ models there appear contributions to $0\nu\beta\beta$ 
decay via the Feynman graphs of Fig.\ref{feyn2}. Here, $S$ and $V^{\mu}$ 
stand 
symbolically for scalar and vector LQs, respectively. 
The half--life for $0\nu\beta\beta$ decay arising from leptoquark 
exchange is given by \cite{hir96a}

\be
T_{1/2}^{0\nu}=|M_{GT}|^2 \frac{2}{G_F^2}[\tilde{C}_1a^2+C_4 b_R^2
+2 C_5 b_L^2].
\ee

with $a=\frac{\epsilon_S}{M_S^2}+\frac{\epsilon_{V}}{M_V^2}$, 
$b_{L,R}=\frac{\alpha_{S}^{(L,R)}}{M_S^2}+\frac{\alpha_V^{(L,R)}}{M_V^2}$,
$\tilde{C}_1=C_1 \Big(\frac{{\cal M}_1^{(\nu)}/(m_e R)}{M_{GT}-
\alpha_2 M_F}
\Big)^2$.

For the definition of the $C_n$ see \cite{74} and for
the calculation 
of the
matrix element ${\cal M}_{1}^{(\nu)}$ see \cite{hir96a}.
This allows to deduce information on leptoquark masses and leptoquark--Higgs
couplings (see subsection 1.3.2).

\subsubsection*{1.3.1.8 Special Relativity and Equivalence Principle}
Special relativity 
and the equivalence principle can be considered as the most 
basic foundations of the theory of gravity. 
Many experiments already have tested these principles to a very high 
level of
accuracy \cite{rel} for ordinary matter - generally for 
quarks and leptons of the first
generation. These precision tests of 
local Lorentz invariance -- violation of the equivalence 
principle should produce a similar effect \cite{will} -- probe for any 
dependence of the (non--gravitational) laws of physics on a laboratory's 
position, orientation or velocity relative to some preferred frame of
reference, such as the frame in which the cosmic microwave background is 
isotropic.  

A typical feature of the violation of local Lorentz invariance (VLI)
is that different species of matter have a characteristical 
maximum attainable speed.
This can be tested in various sectors of the standard model
through vacuum Cerenkov radiation \cite{gasp}, photon decay \cite{cole},
neutrino oscillations \cite{glash,nu1,nu2,hal,nu3} and $K-$physics
\cite{hambye,vepk}. These arguments can be extended
to derive new constraints from neutrinoless double
beta decay \cite{KPS}. 

The equivalence principle implies that spacetime is described by
unique operational geometry and hence universality of the gravitational 
coupling for all species of matter. In the recent years there
have been attempts to constrain a possible amount of 
violation of the equivalence principle (VEP) in the neutrino sector
from neutrino oscillation experiments \cite{nu1,nu2,hal,nu3}.
However, these bounds do not apply when the gravitational and the
weak eigenstates have small mixing. In a recent paper \cite{KPS} 
a generalized formalism of the neutrino sector has been given to test the VEP
and it has been shown that neutrinoless double beta decay also constrains the 
VEP. VEP implies different neutrino species to suffer from  
different gravitational potentials while propagating through the 
nucleus and hence the effect of different eigenvalues doesn't cancel
for the same effective momentum. 
The main result is that neutrinoless double beta decay can constrain
the amount of VEP even when the mixing angle is zero, {\it i.e.},
when only the weak equivalence principle is violated, for which 
there does not exist any bound at present.

\subsubsection*{1.3.2 Double Beta Decay Experiments: Present Status and Results}

\subsubsection*{1.3.2.1 Present Experiments and Proposals}

\begin{figure}
\parbox{10cm}{
\vspace*{-3cm}
\epsfxsize10cm
\epsffile{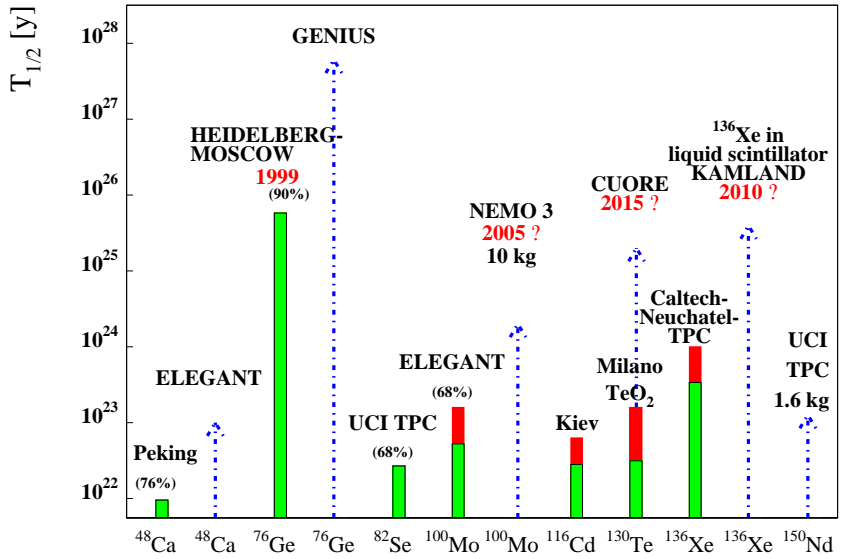}
\vspace*{-6cm}
}
\parbox{10cm}{
\hspace*{0.65cm}
\epsfxsize10cm
\epsffile{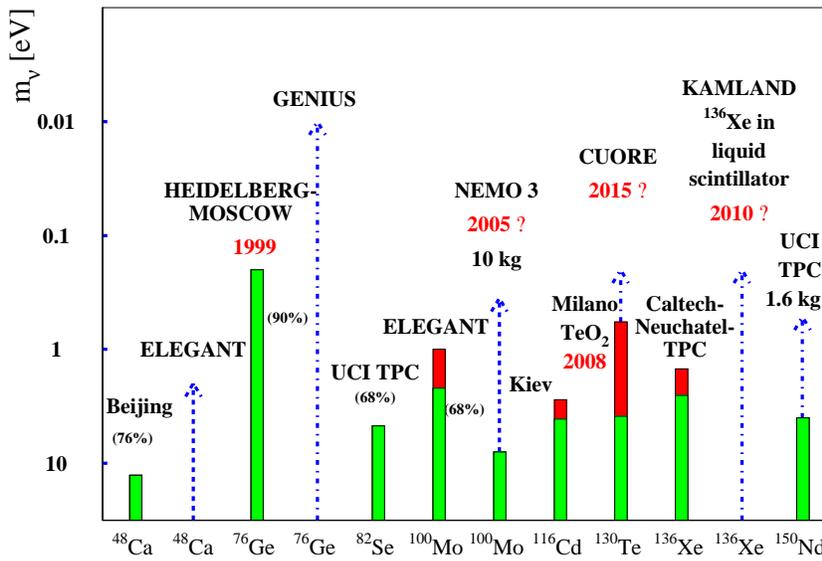}
}
\vspace*{-4cm}
\caption{ 
Present situation, 1999, and expectation for the near future
and beyond, of 
the most promising $\beta\beta$-experiments concerning accessible half life 
(upper) and neutrino mass limits (lower). The light-shaded parts of the bars 
correspond 
to the present status, the dark parts of the bars to 
expectations for running experiments, dashed lines to 
experiments under construction and dash-dotted lines to proposed
experiments.}
\label{mass_time}
\end{figure}

Fig. \ref{mass_time} shows an overview over measured 
$0\nu\beta\beta$ half--life limits and deduced mass limits. The largest 
sensitivity for $0\nu\beta\beta$ decay is obtained at present by active source 
experiments (source=detector), in particular $^{76}$Ge \cite{KK1,KK2,Bau99a}.

Only a few of the present most sensitive experiments may probe the 
neutrino mass 
in the next years into
the sub--eV region, the 
Heidelberg--Moscow experiment being the by far 
most advanced and most sensitive one, see Fig. \ref{mass_time}.


Figs. \ref{mass_time} show in addition to the present status 
the future perspectives of the main existing 
$\beta\beta$ decay experiments and includes some ideas for the future
which have been published.
The best presently existing limits besides the HEIDELBERG-MOSCOW 
experiment (light-shaded bars in Fig. \ref{mass_time}),
have been obtained with the isotopes: 
$^{48}$Ca \cite{87}, 
$^{82}$Se \cite{88}, 
$^{100}$Mo \cite{89}, 
$^{116}$Cd \cite{90},
$^{130}$Te \cite{91},
$^{136}$Xe \cite{92} and
$^{150}$Nd \cite{93}.
These and other double beta decay setups presently under construction or 
partly in operation 
such as NEMO \cite{94,Bar97}, 
the Gotthard $^{136}$Xe TPC experiment \cite{95}, 
the $^{130}$Te cryogenic experiment \cite{91},
a new ELEGANT $^{48}$Ca experiment using 30 g of $^{48}$Ca \cite{96},
a hypothetical experiment with an improved UCI TPC \cite{93} assumed to use 1.6 kg of $^{136}$Xe, 
 etc., will not reach or exceed the $^{76}$Ge limits.
The goal 0.3 eV aimed at for the year 2004 by the NEMO experiment 
(see \cite{98,Bar97}
and Fig. \ref{mass_time}) 
may even be very optimistic if claims about the effect of proton-neutron 
pairing on the $0\nu\beta\beta$ nuclear matrix elements by 
\cite{Pan96} will
turn out to be true, and also if the energy resolution will not be improved
considerably 
(see Fig. 1 in \cite{83}). 
Therefore, the conclusion given by \cite{Bed97c} concerning the
future SUSY potential of NEMO has no serious basis. 
As pointed out by Raghavan \cite{97}, even use of an 
amount of about 200 kg of 
enriched $^{136}$Xe or 2 tons of natural Xe added to the scintillator of the 
KAMIOKANDE detector 
or similar amounts added to BOREXINO (both primarily devoted to solar neutrino 
investigation) 
would hardly lead to a sensitivity larger 
than the present $^{76}$Ge experiment.
This idea is going to be realized at present by the KAMLAND
experiment \cite{Suz97}. 

It is obvious from Fig. \ref{mass_time} that {\it none}
of the present experimental approaches, or plans or even vague ideas has a
chance to surpass the border of 0.1 eV for the neutrino mass to lower values
(see also \cite{Nor97}).
At present there is only one way visible to reach the domain of lower 
neutrino masses,
suggested by \cite{KK1} and meanwhile investigated 
in some
detail concerning its experimental realization and physics potential in
\cite{Kla97d,Hel97,KK2,KK3,Bau99a,Kla99b}.

\subsubsection*{1.3.2.2
Present limits on beyond standard model parameters from double beta decay}

\begin{figure}[h!]  
\hspace*{15mm}
\epsfxsize=90mm
\epsfbox{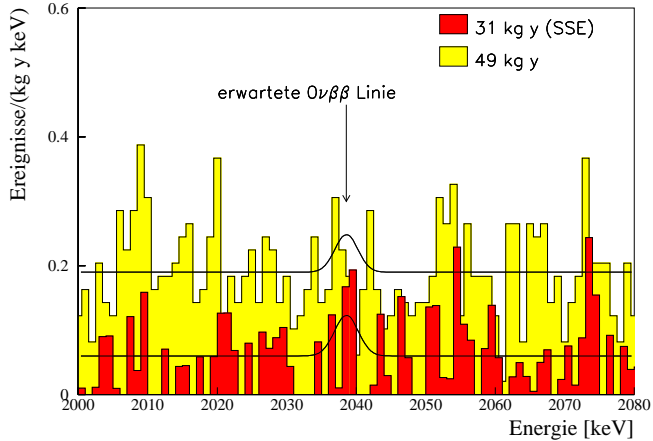}
\caption{Integral spectrum in the region of interest after 
subtraction of the first 200 days of measurement of each detector, 
leaving 49 kg yr and 31 kg y of measuring time without and with 
 pulse shape analysis respectively.
The solid
 curves correspond to the signal excluded
with $90 \% C.L.$ (with ${\rm T}_{1/2}^{0\nu} 
\geq 1.1 \times 10^{25} {\rm~ yr}$, 90\% C.L. and 
${\rm T}_{1/2}^{0\nu} \geq 1.8 \times 10^{25} {\rm~ yr}$, 90\% C.L.)}
\label{pfa}
\end{figure}

The sharpest limits from $0\nu\beta\beta$ decay are presently coming from
the Heidelberg--Moscow experiment \cite{84,KK2,Kla99a,Bau99a}. 
They will be given in the following.
With five 
enriched (86\% of $^{76}$Ge) detectors of a total mass of 11.5 kg 
taking data in the Gran Sasso underground laboratory, and with a background
of at present 0.06 counts/kg year keV in the region of the Q--value, 
the experiment has reached its final 
setup and is now 
exploring the sub--eV range for the mass of the electron neutrino.
Fig. \ref{pfa} shows the spectrum taken in a measuring time of 24  kgy with pulse 
shape analysis.

\subsubsection*{Half-life of neutrinoless double beta decay}
The deduced half-life limit for $0\nu\beta\beta$ decay is using the method 
proposed by \cite{PDG98} 

\be
T^{0\nu}_{1/2} > 1.1 \cdot 10^{25} y \hspace{2mm}(90\% C.L.)
\ee
\be
\hskip8mm     > 1.6 \cdot 10^{25} y \hspace{2mm}(68 \% C.L.).
\ee

\noindent
from the full data set with 49 kg yr and:\\

\be
T^{0\nu}_{1/2} > 1.8 \cdot 10^{25} y \hspace{2mm}(90\% C.L.)
\ee
\be
\hskip8mm     > 3.0 \cdot 10^{25} y \hspace{2mm}(68 \% C.L.).
\ee
 
\noindent
from the data with pulse shape analysis with a total exposure of 
31 kg yr.

{\sl {Neutrino mass}}\\


{\it {Light neutrinos:}} The upper limit of an (effective) electron
neutrino Majorana mass, deduced from the data with pulse shape
analysis, is, with the matrix element from \cite{29}

\be
\langle m_{\nu} \rangle < 0.36 eV \hspace{2mm}(90\% C.L.)
\ee
\be
\hskip10mm < 0.28 eV \hspace{2mm}(68 \% C.L.)
\ee

This is the sharpest limit for a Majorana mass of the electron neutrino so
far. With these values the Heidelberg--Moscow experiment starts to take 
striking influence on presently discussed neutrino mass scenarios, which arose 
in connection with the recent Superkamiokande results on solar and atmospheric
neutrinos. We mention a few examples:

The new $0\nu\beta\beta$ result excludes already now simultaneous 3$\nu$
solutions for hot dark matter, the atmospheric neutrino problem and the small
mixing angle MSW solution \cite{Adh98}. This means that Majorana neutrinos 
are ruled out, if the small mixing angle solution of the solar neutrino
problem is borne out -- if we insist on neutrinos as hot dark matter
candidates. According to \cite{Min97} degenerate neutrino mass schemes for hot 
dark matter, solar and atmospheric anomalies and CHOOZ are already now 
excluded (with 68 \% C.L.) for the small {\it and} 
large mixing angle MSW solutions
(without unnatural finetuning). If starting from recent dark matter models 
\cite{Pri98} including in addition to cold and hot dark matter also a 
cosmological constant $\Lambda \neq 0$, these conclusions remain also valid, 
except for the large angle solution which would not yet be excluded by
$0\nu\beta\beta$ decay (see \cite{Kla99b}).

According to \cite{Bar98} simultaneous 3$\nu$ solutions of solar and 
atmospheric neutrinos, LSND and CHOOZ (no hot dark matter!) predict
$\langle m_{\nu} \rangle\simeq 1.5 eV$ for the degenerate case 
($m_i \simeq 1 eV$) and
$\langle m \rangle \simeq 0.14 eV$ for the hierarchical case.    
This means that the first case is being tested already by the present 
Heidelberg--Moscow result. A model producing the neutrino masses based on a
heavy scalar triplet instead of the seesaw mechanism derives from the solar
small angle MSW allowed range of mixing, and accomodating the atmospheric 
neutrino problem, $\langle m_{\nu} \rangle$ =0.17-0.31 eV \cite{Ma99}. Also
this model is already close to be disfavored. Looking into 4-neutrino 
scenarios, according to \cite{Giu99} there are only two schemes with
four neutrino mixing that can accomodate the results of {\it all} 
neutrino oscillation experiments (including LSND). In the first of the schemes,
where $m_1 < m_2 \ll m_3 < m_4$, with 
solar (atmospheric) neutrinos oscillating between $m_3$ and $m_4$ ($m_1$ and 
$m_2$), and 
$\Delta m^2_{LSND}= \Delta m_{41}^2$,
the HEIDELBERG--MOSCOW $0\nu\beta\beta$ bound excludes \cite{Giu99} the
small mixing angle MSW solution of the solar neutrino problem, for both 
$\nu_e \rightarrow \nu_{\tau}$, and $\nu_e \rightarrow \nu_s$ transitions. 
Including recent astrophysical data yielding $N_{\nu}^{BBN}\leq 3.2$ 
(95 \% C.L.) \cite{Bur99}, the oscillations of solar neutrinos occur mainly
in the $\nu_e \rightarrow \nu_s$ channel, and {\it only} the small angle 
solutions is allowed by the fit of the solar neutrino data \cite{Bah98,Fuk99}.
This means that $0\nu\beta\beta$ excludes the whole first scheme.

In the second scheme $m_1 < m_2 \ll m_3 < m_4$, with solar (atmospheric) 
neutrinos oscillating between $m_1$ and $m_2$ ($m_3$ and $m_4$), 
the present neutrino 
oscillation experiments indicate an effective Majorana mass of 
$7 \cdot 10^{-4} eV \leq |\langle m \rangle| \leq 2 \cdot 10^{-2} eV$. This
could eventually be measured by GENIUS (see below). For a similar recent 
analysis see \cite{Bil99}. For further detailed analyses of neutrino mass
scenarios 
in the light of present and future
neutrino experiments including double beta 
decay we refer to \cite{Kla99b}. 

\subsubsection*{Superheavy neutrinos:}     

For a superheavy {\it left}--handed neutrino we deduce 
\cite{79,14,Bel98} exploiting the 
mass dependence of the matrix 
element (for the latter 
see \cite{28}) a lower limit (see also Fig. 36)
\be
\langle m_{H} \rangle \ge 9 \cdot 10^7 GeV.
\ee
Assuming the bound on the mixing matrix, $U^2_{ei}<5 \cdot 10^{-3}$
\cite{Bel98}, and
assuming no cancellation between the involved states, this limit implies a 
bound on the mass eigenstate
\be
M_i > 4.5 \cdot 10^5 GeV.
\ee

\subsubsection*{Right--handed W boson}

\begin{figure}[h!]
\hspace*{5mm}
\epsfxsize=80mm
\hspace*{4mm}
\epsfbox{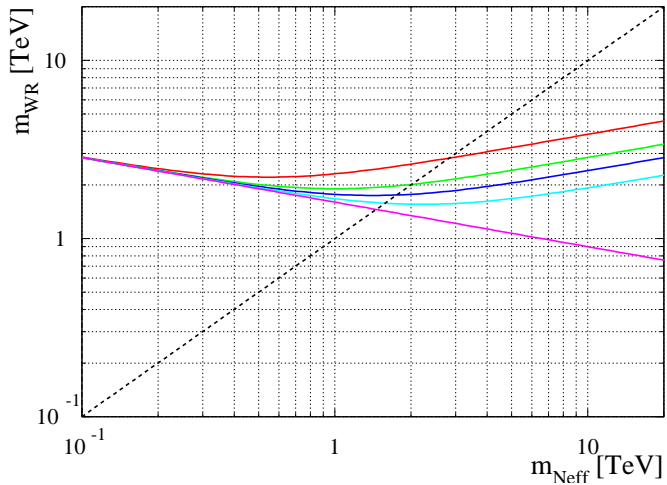}
\vspace*{-2cm}
\caption{ Limits on the mass of the right-handed W-boson from 
neutrinoless double beta decay (full lines) and vacuum stability 
 (dashed line). The five full lines correspond to the following 
masses of the doubly charged higgs, $m_{\Delta^{--}}$: 0.3, 
1.0, 2.0, 5.0 and $\infty$ [TeV] downward (from \cite{11}).} 

\label{fig15}
\end{figure}

For the right--handed W boson we obtain (see Fig. \ref{fig15}) 
a lower limit of 
\be
m_{W_R} \ge 1.4 TeV
\ee
(see \cite{11,KKP}).

\subsubsection*{SUSY parameters -- R--parity breaking and sneutrino mass}

\begin{figure}[h!]
\vspace*{-4.0cm}
\epsfxsize=120mm
\epsfbox{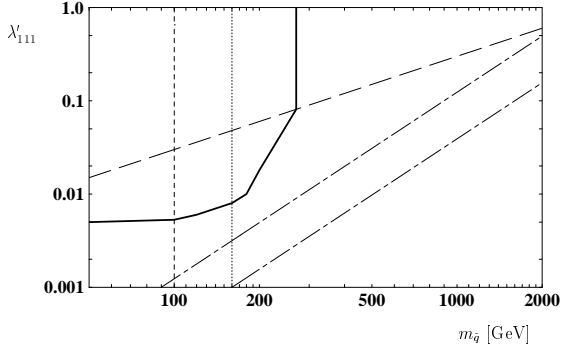}
\vspace*{-7cm}
\caption{Comparison of limits on the R--parity violating MSSM parameters 
from different experiments in the $\lambda'_{111}$--$m_{\tilde{q}}$
plane. The ashed line is the limit from charged current universality
according to \cite{113}. The vertical line is the limit from the data
of Tevatron 
\cite{114}. The thick full line is the region which might be explored by HERA
\cite{115}. The two dash--dotted lines to the right are the limits obtained
from the half--life limit for $0\nu\beta\beta$ decay of $^{76}$Ge, for
gluino masses of (from left to right) $m_{{\tilde{g}}}=$1TeV and 100 GeV,
respectively. The regions to the upper left of the lines are forbidden.
 (from [Hir95])}
\label{fig16}
\end{figure}

The constraints on the parameters of the minimal supersymmetric standard model
 with explicit R--parity violation deduced \cite{6,hir96c,hir96} 
from the $0\nu\beta\beta$
half--life limit are more stringent than those from other
low--energy processes and from the largest high energy
accelerators (Fig. \ref{fig16}). The limits are 
\be
\lambda^{'}_{111} \leq 4 \cdot 10^{-4} \Big(\frac {m_{\tilde{q}}}{100 GeV} 
\Big)^2 \Big(\frac {m_{\tilde{g}}}{100 GeV} \Big)^{\frac{1}{2}}
\ee
with  $m_{\tilde{q}}$ and  $m_{\tilde{g}}$ denoting squark and gluino masses,
respectively, and with the assumption $m_{\tilde{d_R}} \simeq m_{\tilde{u}_L}$.
This result is important for the discussion of new physics in the connection
with the high--$Q^2$ events seen at HERA. It excludes the possibility of 
squarks of first generation (of R--parity violating SUSY) being produced in the
high--$Q^2$ events \cite{Cho97,Alt97,Hir97b}.

We find further \cite{Paes99b} 
\be
\lambda_{113}^{'}\lambda_{131}^{'}\leq 3 \cdot 10^{-8}
\ee
\be
\lambda_{112}^{'}\lambda_{121}^{'}\leq 1 \cdot 10^{-6}.
\ee 

The constraints on coupling products derived from the double beta decay 
neutrino mass limit
\cite{Bha99} are presented in 
tab. \ref{tabrpv}. As is obvious from the table, the double beta decay 
neutrino mass limits improve previous bounds on products of R--parity 
violating couplings by 1-5 orders of magnitude.

\begin{table}
\begin{center}
\begin{tabular}{ccc}
\hline
\hline
$\lambda^{(')}_{ijk}\lambda^{(')}_{i^{'}kj}$
&                     Our &  Previous  \\  
& Bounds & Bounds \\
\hline
\hline
$m_{ee}<0.36$ eV && \\ \hline
$\lambda^{'}_{133}\lambda^{'}_{133}$ & $5.0 \cdot 10^{-8}$ & $4.9
\cdot 10^{-7}$ \\
$\lambda^{'}_{132}\lambda^{'}_{123}$ & $1.0 \cdot 10^{-6}$ & $1.6
\cdot 10^{-2}$ \\
$\lambda^{'}_{122}\lambda^{'}_{122}$ & $3.0 \cdot 10^{-5}$ & $4.0
\cdot 10^{-4}$ \\
$\lambda_{133}\lambda_{133}$ & $9.0 \cdot 10^{-7}$ & $9.0 \cdot
10^{-6}$ \\
$\lambda_{132}\lambda_{123}$ & $2.0 \cdot 10^{-5}$ & $2.0 \cdot
10^{-3} $ \\
$\lambda_{122}\lambda_{122}$ & $2.0 \cdot 10^{-4}$ & $1.6 \cdot
10^{-3}$\\ \hline
\end{tabular}
\caption{Correlation among neutrino mass bounds from neutrinoless double
beta decay and upper limits
on RPV couplings. We have used $m_d$=9
MeV, $m_s$= 170 MeV, $m_b$=4.4 GeV \protect{\cite{PDG98}}. 
For
$\lambda$-products, $m_{\tilde{d}}$ should be read as
$m_{\tilde{e}}$. The relevant scalars are always assumed to have a
common mass of 100 GeV.
\label{tabrpv}}
\end{center}
\end{table}

For the $(B-L)$ violating sneutrino mass $\tilde{m}_{M}$ the following limits 
are obtained \cite{Hir97a}
\ba{rconv2}
\tilde{m}_M &\leq& 1.3 \Big(\frac{m_{SUSY}}{100 GeV}\Big)^{\frac{3}{2}}GeV,
\hskip5mm \chi \simeq \tilde{B}\\
\tilde{m}_M &\leq& 7 \Big(\frac{m_{SUSY}}{100 GeV}\Big)^{\frac{7}{2}}GeV,
\hskip5mm \chi \simeq \tilde{H}
\ea
for the limiting cases that the lightest neutralino is a pure Bino $\tilde{B}$,
as suggested by the SUSY solution of the dark matter problem \cite{jkg96},
or a pure Higgsino. Actual values for $\tilde{m}_M$ for other choices of the
neutralino composition should lie in between these two values.
 
Another way to deduce a limit on the `Majorana' sneutrino mass $\tilde{m}_M$
is to start from the experimental neutrino mass limit, since the sneutrino 
contributes to the Majorana neutrino mass $m_M^{\nu}$ at the 1--loop level
proportional to $\tilde{m}^2_M$
\cite{Hir97a}.
Starting from the mass limit determined for the electron neutrino  by 
$0\nu\beta\beta$ decay this leads to 
\be
\tilde{m}_{M_{(e)}} \leq 14 MeV    
\ee
This result is somewhat dependent on neutralino masses and mixings. 
A non--vanishing `Majorana' sneutrino mass would result in new processes 
at future colliders, like sneutrino--antisneutrino oscillations.
Reactions at the Next Linear Collider (NLC) like the SUSY analog to inverse
neutrinoless double beta decay $e^-e^-\rightarrow \chi^-\chi^-$ (where $\chi^-$
denote charginos) or single sneutrino production, e.g. by 
$e^-\gamma \rightarrow \tilde{\nu}_e \chi^-$ could give information on the 
Majorana sneutrino mass, also. This is discussed by \cite{Hir97,Hir97a,Kolb1}.
A conclusion is that future
accelerators can give information on second and third generation sneutrino
Majorana masses, but for first generation sneutrinos cannot compete with
$0\nu\beta\beta$--decay.

\subsubsection*{Compositeness}

\begin{figure}[h!]
\epsfxsize=80mm
\hspace*{1.5cm}
\epsfbox{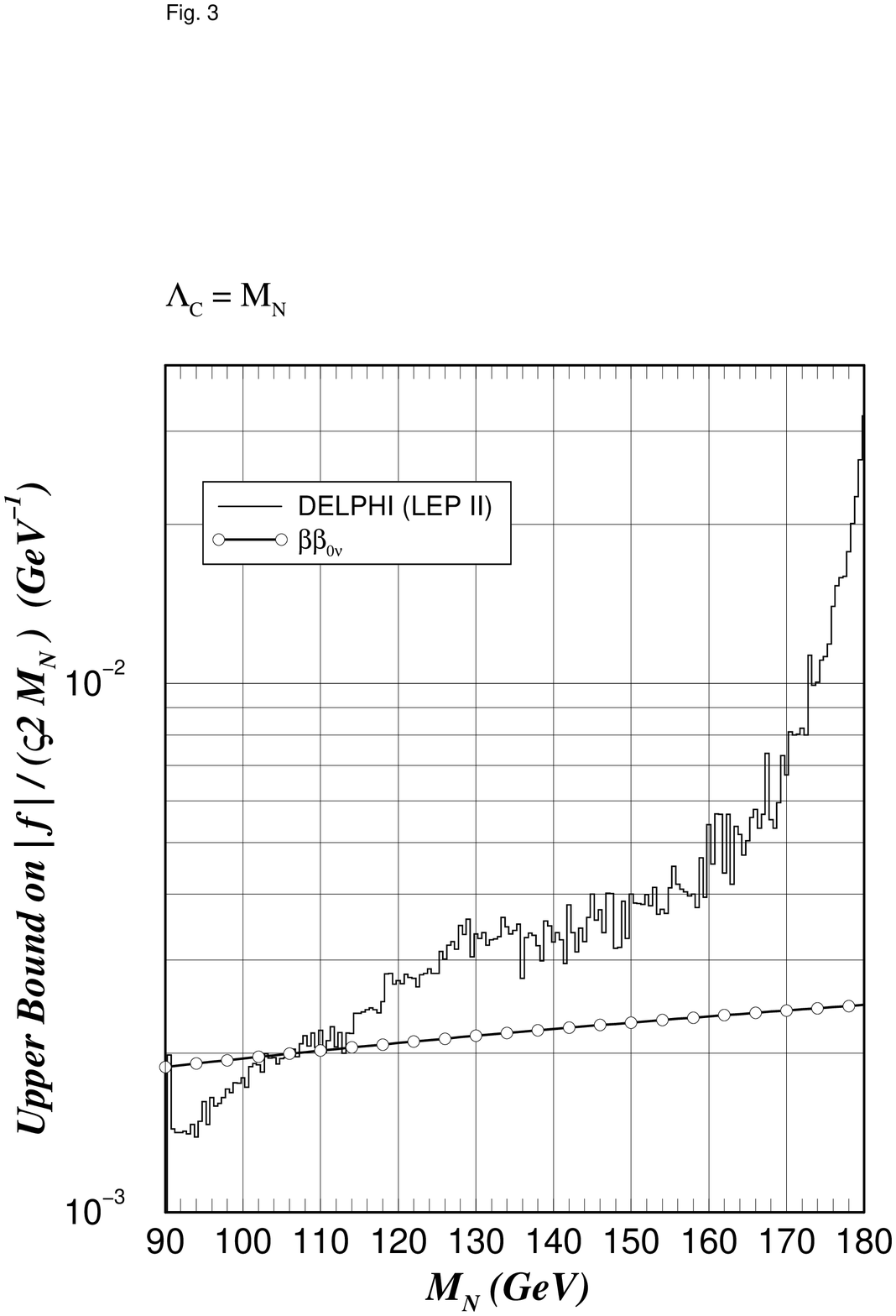}
\caption{Comparison between the 0$\nu\beta\beta$ 
  (Heidelberg-Moscow experiment) and the LEPII upper
  bound on the quantity $|$f$|$($\sqrt{2}M_N$) as a function of the heavy 
  neutrino mass M$_N$, with the choice $\Lambda_C$ = M$_N$. Regions
  above the curves are excluded (from [Pan99]).}
\label{fig17}
\end{figure}

Evaluation of the $0\nu\beta\beta$ half--life limit assuming 
exchange of excited
Majorana neutrinos $\nu^*$ yields for the mass of the
 excited neutrino a lower bound of \cite{Pan97,Tak97,Pan99}. 
\be
m_{N} \geq 3.4 m_W
\ee
for a coupling of order {\cal O}(1) and $\Lambda_c \simeq m_N$. Here,
$m_W$ is the W--boson mass. Fig. \ref{fig17} shows that this result is 
more stringent than the result obtained by LEPII..

\subsubsection*{Leptoquarks}

Assuming that either scalar or vector leptoquarks contribute
to $0\nu\beta\beta$ decay, the following constraints on the 
effective LQ parameters  (see subsection 1.3.1) can be derived \cite{hir96a}:
\ba{dbd_constraint}
\epsilon_I \leq 1.0 \times 10^{-9}
\left(\frac{M_I}{100\mbox{GeV}}\right)^2, \\
\alpha_I^{(L)} \leq 1.3 \times 10^{-10}
\left(\frac{M_I}{100\mbox{GeV}}\right)^2, \\
\alpha_I^{(R)} \leq 2.8  \times 10^{-8}
\left(\frac{M_I}{100\mbox{GeV}}\right)^2.
\ea

Since the LQ mass matrices appearing in $0\nu\beta\beta$ 
decay are ($4\times4$) 
matrices \cite{hir96a}, it is difficult to solve their diagonalization 
in full generality algebraically. However, if one assumes that only 
one LQ-Higgs coupling is present at a time, the (mathematical) problem is 
simplified greatly and one can deduce from, for example, 
eq. 1.41 that either 
the LQ-Higgs coupling must be smaller than $\sim 10^{-(4-5)}$ or there can not 
be any LQ with e.g. couplings of electromagnetic strength with masses below
$\sim 250 GeV$. These bounds from $\beta\beta$ decay are of interest in 
connection with recently discussed evidence for new physics from HERA
\cite{Hew97,Bab97,Kal97,Cho97}. Assuming that actually leptoquarks have
been produced at HERA, double beta decay (the Heidelberg--Moscow experiment)
would allow to fix the leptoquark--Higgs coupling to a few $10^{-6}$
\cite{Hir97b}. It may be noted, that after the first 
consideration of leptoquark--Higgs coupling in \cite{hir96a} recently
Babu et al. \cite{Bab97b} noted that taking into account 
leptoquark--Higgs coupling reduces the leptoquark mass lower bound deduced
by TEVATRON -- making it more consistent with the value of 200 GeV 
required by 
HERA. 
\vspace*{3mm}

{\sl {Special Relativity and Equivalence Principle}}\\

\begin{figure}[h!]
\epsfysize=80mm
\hspace*{15mm}
\epsfbox{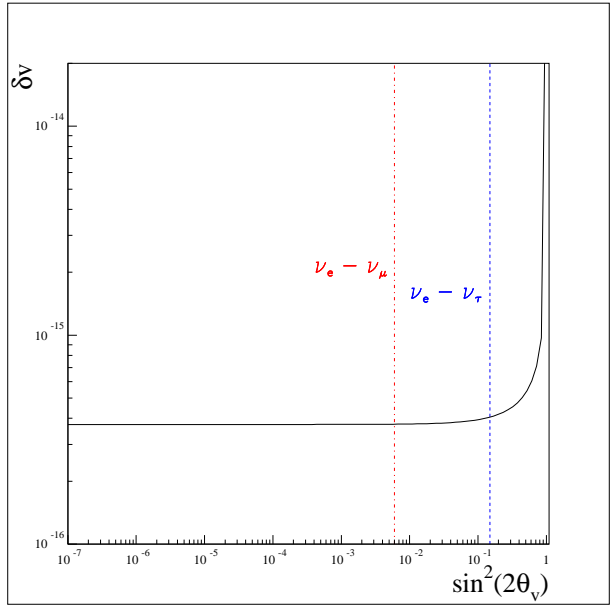}
\vspace*{5mm}
\caption{ Double beta decay bound (solid line)
on violation of Lorentz invariance 
in the neutrino sector, excluding the region to the upper left. 
Shown is a double logarithmic plot 
in the $\delta v$--$\sin^2(2 \theta)$ parameter space. 
The bound becomes most stringent for the
small mixing region, which has not been constrained from any
other experiments. For comparison the bounds obtained from neutrino oscillation
experiments (from \protect{\cite{hal}})
in the $\nu_{e} - \nu_{\tau}$ (dashed lines) and in the
$\nu_e - \nu_\mu$ (dashed-dotted lines) channel, excluding the region to the 
right, are shown (from \protect{\cite{KPS}).}}
\label{fig6a}
\end{figure}

{\it Violation of Lorentz invariance (VLI):} The bound obtained from the 
Heidelberg--Moscow experiment is
\be
\delta v < 2 \times 10^{-16}~~~~ {\rm for}~~~ \theta_v=\theta_m =0
\ee
where $\delta v=v_1-v_2$ is the measure of VLI in the neutrino sector.
$\theta_v$ and $\theta_m$ denote the velocity mixing angle and the weak 
mixing angle, respectively.
In Fig. \ref{fig6a} (from \cite{KPS}) the bound implied by double beta decay is 
presented for the entire
range of $sin^2(2 \theta_v)$, and compared with bounds obtained from
neutrino oscillation experiments (see \cite{hal}).

\nopagebreak

{\it Violation of equivalence principle (VEP):}
Assuming only violation of the weak equivalence principle, there does not 
exist any bound on the amount of VEP. It is this region of the parameter space
which is most restrictively bounded by neutrinoless double beta decay.
In a linearized theory the gravitational part of the Lagrangian to first order
in a weak gravitational field $g_{\mu\nu}=\eta_{\mu\nu}+    h_{\mu\nu}$
($h_{\mu\nu}= 2\frac{\phi}{c^2}  {\mbox diag}(1,1,1,1)$)
can be written as ${\cal L} = -\frac{1}{2}(1+g_i)h_{\mu\nu}T^{\mu\nu}$,
where  $T^{\mu\nu}$  is the  stress-energy  in the  gravitational
eigenbasis. In the presence of VEP the $g_i$ may differ.
We obtain \cite{KPS} the following bound from the Heidelberg--Moscow 
experiment, for $\theta_v=\theta_m=0$:
  \ba{99}
\phi \delta g &<& 2 \times 10^{-16} ~ ({\rm for~} \bar{m}<13 
{\rm eV})\nn \\
\phi \delta g &<& 1 \times 10^{-18} ~ ({\rm for~} \bar{m}<0.08 
{\rm eV}).
\ea
Here $\bar{g}=\frac{g_1+g_2}{2}$ can be considered as the standard 
gravitational coupling, for which the equivalence principle applies.
$\delta g=g_1 - g_2$.
The bound on the VEP thus, unlike the one for VLI, will depend on the choice
for the Newtonian potential $\phi$.

\subsubsection*{Half--life of $2\nu\beta\beta$ decay}

The Heidelberg--Moscow experiment 
produced for the first time a high statistics $2\nu\beta\beta$
spectrum ($\gg$ 20000 counts, to be compared with the 40 counts on which the 
first detector observation of $2\nu\beta\beta$ decay by \cite{Ell87} 
(for the decay of $^{82}$Se) had to rely). 
The deduced half--life is \cite{HM2000}

\be
T^{2\nu}_{1/2} = (1.55 \pm 0.01(stat.)^{+0.03}_{-0.02}(norm.)^{+0.16}_{-0.13}(syst.))\cdot 10^{21} y
\ee

This result brings $\beta\beta$ research for the first time into the region 
of `normal' nuclear spectroscopy and allows for the first time statistically
reliable investigation of Majoron--accompanied decay modes.

\subsubsection*{Majoron--accompanied decay}

From simultaneous fits of
the $2\nu$ spectrum and one selected Majoron mode, experimental limits 
for the half--lives of the decay modes of
the newly introduced Majoron models \cite{72} are given
for the first time \cite{71,HM96}.

The small matrix elements and phase spaces for these modes 
\cite{71,75} already determined that these 
modes by far cannot be seen
in experiments of the present sensivity if we assume typical values for the
neutrino--Majoron coupling constants around $\langle g \rangle = 10^{-4}$.

\subsubsection*{1.3.3 The GENIUS Potential for Double Beta Decay} 

\subsubsection*{Neutrino mass matrix and neutrino oscillations}

\begin{figure}[h!]

\epsfxsize=10cm
\epsfbox{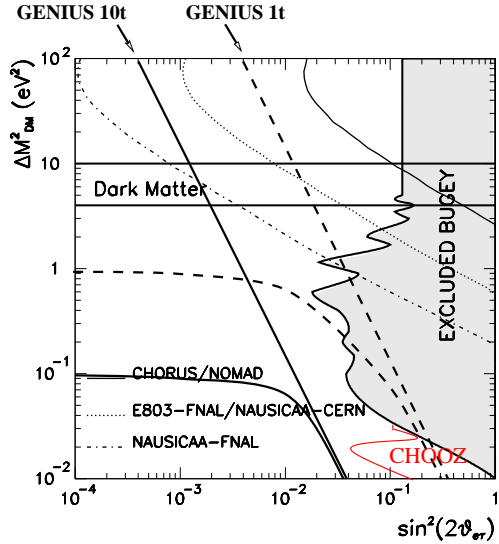}

\vspace*{-5cm}
\caption{ Current limits and future experimental sensitivity 
on $\nu_e - \nu_{\tau}$ oscillations. The shaded area is currently 
excluded from reactor experiments. The thin line is the estimated 
sensitivity of the CHORUS/NOMAD experiments. The dotted and dash-dotted 
thin lines are sensitivity limits of proposed accelerator experiments, 
NAUSICAA and E803-FNAL [Gon95]. 
The thick lines show the sensitivity of GENIUS (broken line: 
1 t, full line: 10 t), for two examples of mass ratios. The straight lines are 
for the strongly hierarchical case (R=0), while the lines bending to the left 
assume R=0.01.  
(from [Kla97c])}
\label{fig20}
\end{figure}

GENIUS will allow a large step in sensitivity for probing the neutrino mass. 
It will allow to probe the effective neutrino Majorana mass down to 
10$^{-(2-3)}$ eV, and thus 
surpass  the existing experiments being sensitive on the mass eigenstate
by a factor of 50-500.
GENIUS will test the structure of the neutrino mass matrix and thereby also 
neutrino oscillation parameters 
\footnote{The double beta observable, the effective neutrino mass 
(eq. 10), can be expressed 
in terms of the usual neutrino oscillation parameters, once an assumption
on the ratio of $m_1/m_2$ is made. E.g., in the simplest two--generation case
\be
\langle m_{\nu} \rangle=|c_{12}^2 m_1 + s_{12}^2 m_2 e^{2 i \beta}|,
\ee
assuming CP conservation, i.e. $e^{2 i \beta}=\eta=\pm 1$, and 
$c_{12}^2 m_1 << \eta s_{12}^2 m_2$,
\be
\Delta m^2_{12}\simeq m_2^2=\frac{4 \langle m_{\nu} \rangle^2}{(1-\sqrt{1-
sin^2 2 \theta})^2}
\ee
A little bit more general, keeping corrections of the order $(m_1/m_2)$ 
one obtains
\be
m_2=\frac{ \langle m_{\nu} \rangle}{|(\frac{m_1}{m_2})+\frac{1}{2}
(1-\sqrt{1-sin^2 2 \theta})(\pm 1 - (\frac{m_1}{m_2}))|}.
\ee
For the general case see \cite{Kla97d}.}
superior in sensitivity to many present
dedicated terrestrial neutrino oscillation experiments and will provide
complementary informations to recent proposals for the future in this field.
Already in the 
first stage GENIUS will test degenerate or inverted neutrino 
mass scenarios, discussed in the literature as possible solutions of current 
hints to finite neutrino masses (see \cite{Kla99b,Giu99,Cza99,Vis99}).
If the $10^{-3}$ eV
level is reached, GENIUS will allow to test the large angle and for degenerate
models even the small angle MSW 
solution of the solar neutrino problem. It will also allow to test the 
hypothesis of a shadow world underlying introduction of a sterile neutrino \cite{Moh97a}.
The Figures 26, 27, 28, 29 show some examples of this potential (for more
details see \cite{Kla97d,KK1,KK2,KK3,Kla99a}. Fig. \ref{fig20} compares the
potential of GENIUS with the sensitivity of CHORUS/NOMAD and with the
proposed future experiments NAUSIKAA-CERN and NAUSIKAA-FNAL 
-- now renamed to TOSCA and COSMOS,
looking for 
$\nu_e \leftrightarrow \nu_{\tau}$ oscillations, for different assumptions on
$m_1/m_2$. 

\begin{figure}[h!]
\vskip0mm 
\hskip5mm
\epsfxsize=90mm
\epsfbox{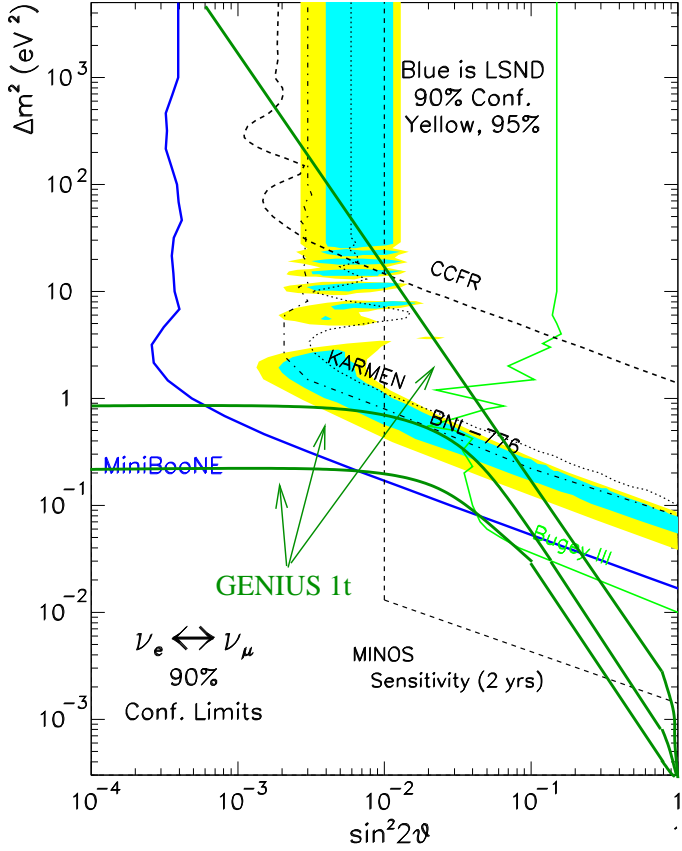}
\caption{LSND compared to the sensitivity of GENIUS 1t 
for $\eta^{CP} = +1$ and three ratios $R_{12}$, from top to bottom 
$R_{12}= 0, 0.01, 0.02$ (from [Kla97c])}
\label{fig21}
\end{figure}

Already in the worst case for double beta decay of $m_1/m_2=0$
GENIUS 1 ton is more sensitive than CHORUS and NOMAD.
For quasi--degenerate models, for example $R=0.01$ already, GENIUS 1 ton would 
be more sensitive than the planned future experiments TOSCA and COSMOS.

Fig. \ref{fig21} shows the potential of GENIUS for checking the LSND indication for
neutrino oscillations (original figure from \cite{Lou98}). 
Under the assumption 
$m_1/m_2 \geq 0.02$ and $\eta=1$, GENIUS 1 ton will be sufficient to find 
$0\nu\beta\beta$ decay if the LSND result is to be explained in terms of $\nu_e
\leftrightarrow \nu_{\mu}$ oscillations. This sensitivity is comparable to --
and for small and large mixing angles better than -- the 
one of the dedicated project MINIBOONE and
might be of particular interest 
also since the upgraded KARMEN will not completely cover \cite{Dre97} the full
allowed LSND range.
Fig. 28 shows the situation for $\nu_e$ - $\nu_{\mu}$
oscillations in reactor and accelerator experiments 
(assuming sin$^2 \theta_{13}$ = 0). 
The original figure
is taken from [Gel95]. The GENIUS 10 ton sensitivity is
superior to the one obtained by CHOOZ
and could compete with the long baseline project MINOS,
even in the worst case of $m_{\nu_e}<<m_{\nu_{\mu}}$.
In
the quasi-degenerate models GENIUS would be much more sensitive -
see Fig. \ref{fig21}.

\begin{figure}[h!]
\epsfxsize=80mm
{\epsfbox{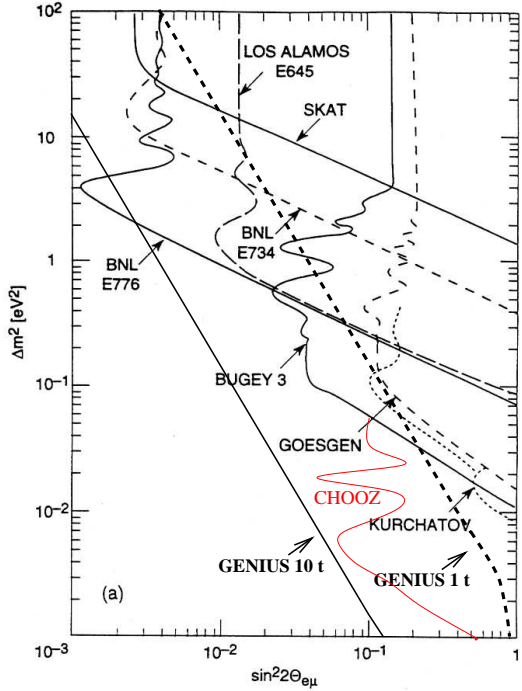}
\caption{Current limits on $\nu_e - \nu_{\mu}$ oscillations. 
Various existing experimental limits from reactor and accelerator 
experiments are indicated, as summarized in ref. [Gel95]. In addition, 
the figure shows the expected sensitivities for GENIUS with 1 ton
(thick broken line) and GENIUS with 10 tons (thick, full line) 
(from [Kla97c])}}
\label{fig22}
\end{figure}

Fig. \ref{fig24} shows a summary of currently known constraints
on neutrino oscillation parameters (original taken from \cite{Hat94}), but 
including the $0\nu\beta\beta$ decay sensitivities of GENIUS 1 ton and GENIUS 
10 tons, for different assumptions on $m_1/m_2$ (for $\eta^{CP}=+1$,
for $\eta^{CP}=-1$ see \cite{Kla97d}).
It is seen that already GENIUS 1 ton tests all degenerate or quasi--degenerate
($m_1/m_2 \geq \sim 0.01$) 
neutrino mass models in any range where neutrinos are 
interesting for cosmology, and also the atmospheric neutrino problem, if it is 
due to $\nu_e \leftrightarrow \nu_{\mu}$ oscillations. GENIUS in its 10 ton
version would directly test the large angle solution of the solar neutrino 
problem and in case of almost degenerate neutrino masses, also the
small angle solution.

After this overview we discuss the potential of GENIUS for the various 
neutrino mass scenarios in some detail, putting some emphasis on the relations 
to the solar and atmospheric neutrino oscillation experiments and on the 
complementarity  of recent and future projects in these fields including the
investigation of cosmological parameters like 
by the future satellite experiments MAP and PLANCK.

In a three neutrino framework the 
atmospheric neutrino data are assumed to be described by  
$\nu_{\mu} - \nu_{\tau}$ oscillations \cite{Smi99}
with: 
\begin{equation}
\Delta m^2_{atm} = (1 - 10)~10^{-3} {\rm eV}^2~,~~~
\sin^2 2\theta_{atm} = 0.8 - 1,     
\end{equation}
as the leading mode. Also small contributions of 
other modes are not excluded. 

For solar neutrinos different 
possibilities are considered \cite{Smi99}
which in general lead to different expectations 
for the double beta decay: 

1. Small mixing MSW solution with 
\begin{equation}
\Delta m^2_{\odot} = (0.4 - 1) \cdot 10^{-5} {\rm eV}^2~,~~~
\sin^2 2\theta_{\odot} = (0.3 - 1.2) \cdot 10^{-2}
\label{small}
\end{equation}  

2. Large mixing MSW solution with  
\begin{equation}
\Delta m^2_{\odot} = (0.1 - 3)\cdot 10^{- 4} {\rm eV}^2~,~~~
\sin^2 2\theta_{\odot} = (0.7 - 1)
\label{large}
\end{equation}

3. Vacuum oscillation solutions
\begin{equation}
\Delta m^2_{\odot} = (0.6 - 6)\cdot 10^{- 10} {\rm eV}^2~,~~~
\sin^2 2\theta_{\odot} = (0.6 - 1)
\label{VO}   
\end{equation}

The so-called MSW low solution gives a worse fit to the data and will not be 
considered in the following (see however subsection 1.4).

Expressing eq. \ref{obs} in terms of oscillation parameters we get
\begin{equation}
\langle m \rangle = |U_{e1}|^2 m_0 +
e^{i\phi_{21}}|U_{e2}|^2 \sqrt{\Delta m^2_{21} + m_0^2}
+
e^{i\phi_{31}}|U_{e3}|^2 \sqrt{\Delta m^2_{31} + m_0^2}~, 
\label{mee}
\end{equation}
where $\phi_{ij}$  are the relative phases of 
masses $m_i$ and $m_j$. 
Assuming $m_1$ to be the lightest state we
have absorbed $m_1^2$ 
in the definition of  $m_0^2$, $m_0^2 \rightarrow m_0^2 + m_1^2$
so that now $m_0 \geq 0$.

The crucial assumption in order to link neutrino oscillations and the
double beta observable eq. \ref{mee}
concerns the grade of 
degeneracy in the
\epsfysize=180mm
\epsfbox{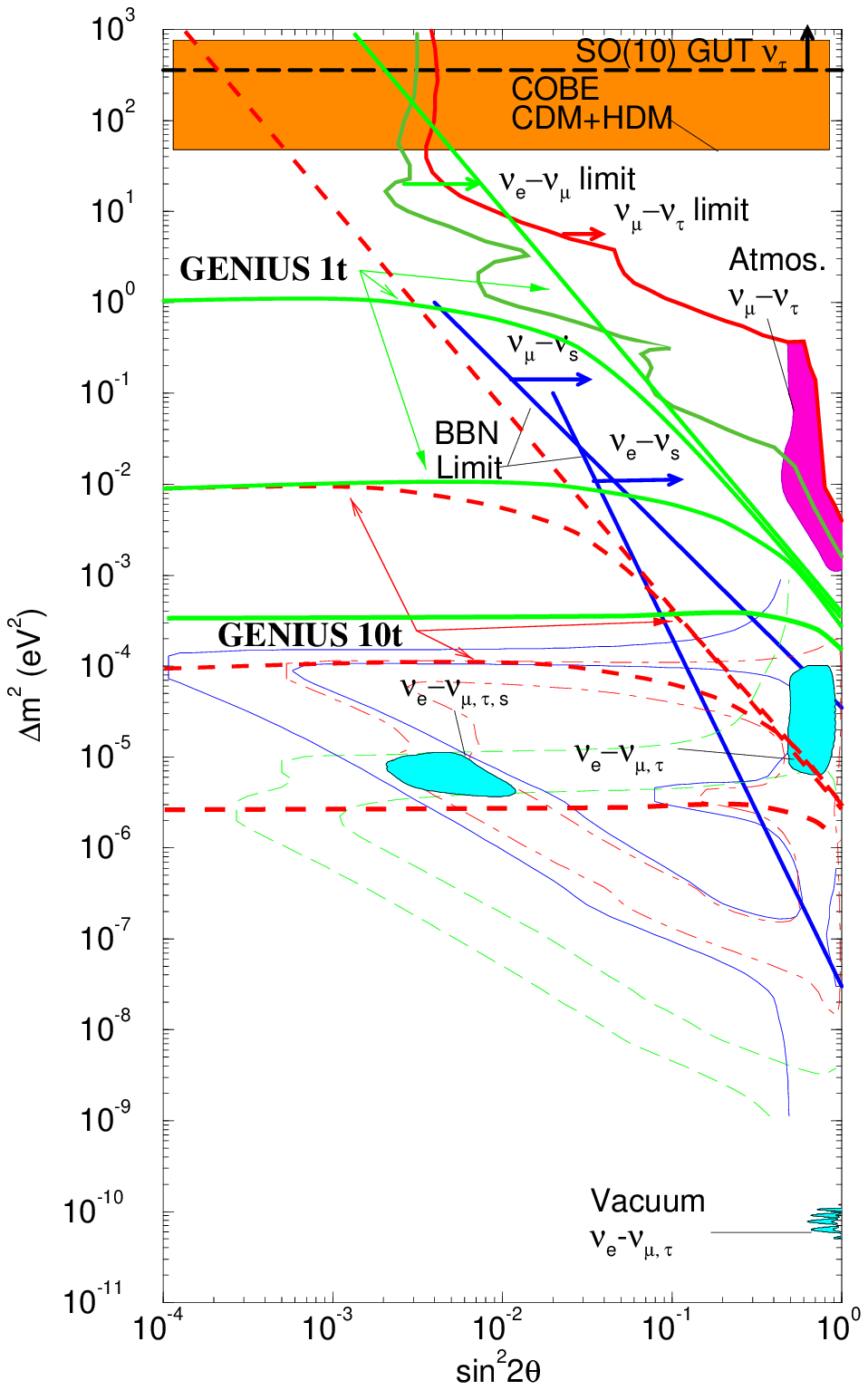}
\pagebreak
\begin{figure}
\caption{~Summary of currently known constraints on neutrino 
oscillation parameters. The (background) figure without the \znbb{} 
decay constraints can be obtained from
http://dept.physics.upenn.edu/\~\-www/neutrino/solar.html. Shown are 
the vacuum and MSW solutions (for two generations of neutrinos) 
for the solar neutrino problem, 
the parameter range which would solve the atmospheric neutrino problem 
and various reactor and accelerator limits on neutrino oscillations. 
In addition, the mass range in which neutrinos are good hot dark matter 
candidates is indicated, 
as well as limits on neutrino oscillations into sterile states from 
considerations of big bang nucleosynthesis. Finally the 
thick lines indicate the sensitivity of GENIUS (full lines 1 ton, 
broken lines 10 ton) to neutrino oscillation parameters for three values 
of neutrino mass ratios $R = 0, 0.01$ and $0.1$ (from top to bottom).
For GENIUS 10 ton also the contour line for $R=0.5$ is shown.  
The region beyond the lines would be excluded.
While already the 1 ton GENIUS would be sufficient to constrain degenerate 
and quasi-degenerate neutrino mass models, the 10 ton version of 
GENIUS could cover a significant new part of the parameter space, 
including the large angle MSW solution to the solar neutrino problem, 
even in the worst case of $R=0$. For $R\geq 0.5$ it would even probe the 
small angle MSW solution (see \cite{klapneut,KKP}).}  
\label{fig24}
\end{figure}
neutrino mass spectrum, which may be
described by the  value 
of $m_0$. Three possibilities are determined  by  
the relative values of $m_0^2$, $\Delta m^2_{21}$ and 
$\Delta m^2_{31}$:  

\begin{itemize}

\item
neutrino schemes with strong hierarchy: 
$m_0^2 \ll \Delta m^2_{21} \ll  \Delta m^2_{31}$, 

\item 

with partial degeneracy: 
$\Delta m^2_{21} \ll m_0^2  \ll  \Delta m^2_{31}$, 

\item 

and with complete degeneracy: 
$\Delta m^2_{21} \ll  \Delta m^2_{31}  \ll m_0^2$. 
\end{itemize}

\subsubsection*{1.3.4 Schemes with  mass hierarchy \label{hs}}

\begin{figure}[htb]
\epsfxsize=10cm
\hspace*{0.8cm}
\epsfbox{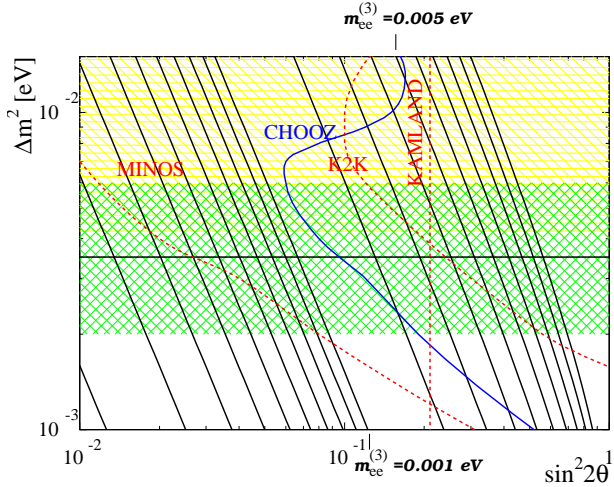}
\caption{
Iso-mass ($\langle m \rangle$) lines 
in the single maximal mixing
scheme with hierarchical mass pattern. 
From the upper right downward $\langle m \rangle$ = 
0.01, 0.009, 0.008, 0.007, 0.006, 
0.005, 0.004, 0.003, 0.002, 0.001, 0.0009, 0.0008, 0.0007, 0.0006, 0.0005,
0.0004, 0.0003 eV. Also shown are  the regions 
favored by 
the atmospheric neutrino data of Super--Kamiokande 
with current bestfit and Kamiokande (lower and upper shaded areas respectively
according \protect{\cite{Kaj99}})
and the borders of regions excluded by  CHOOZ and
BUGEY (solid lines) as well as the expected final sensitivity of CHOOZ
(according to \protect{\cite{Dec99}})
and KAMLAND (dashed) as well as of MINOS and K2K (dash-dotted)
(according to \protect{\cite{Zub98}}). For less hierarchical scenarios 
additional contributions from the first state arise (see fig. \ref{smix3}).
(from \protect{\cite{Kla99b}})}
\label{smix1}
\end{figure}

\begin{figure}[htb]
\epsfxsize=75mm
\hspace*{1.5cm}
\epsfbox{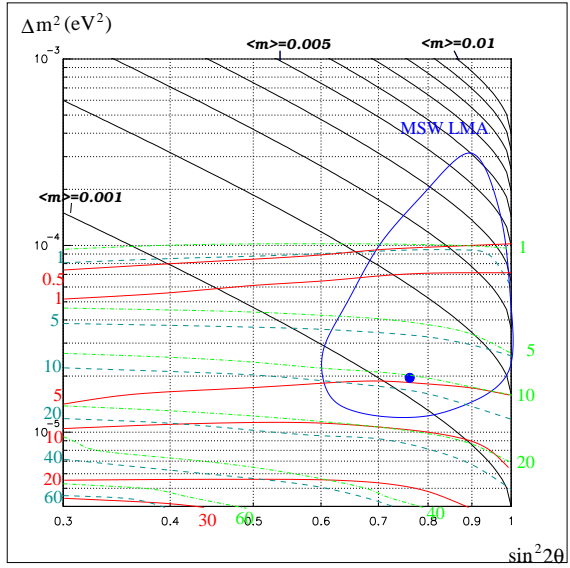} 
\caption{
The iso-mass $\langle m \rangle^{(2)}$ lines, determining the 
contribution of the second 
state in the $\Delta m_{12}^2 - \sin^2 2 \theta_{12}$ plane for the
hierarchical scheme with the LMA MSW solution.
From the upper right downward: 
$\langle m \rangle^{(2)}$ =
0.01,
0.009, 0.008, 0.007, 0.006, 0.005, 0.004, 0.003, 0.002, 0.001 eV.
Also shown is the MSW LMA 90 \% C.L. allowed region from the combined rate
analysis of Homestake, Gallex, Sage and Super-K with the BP98 SSM and the
Super--Kamiokande Day-Night variation \protect{\cite{Fuk99}} 
with the point showing the bestfit (rates only) according to 
\protect{\cite{Bah98}}. 
The solid, dashed and dash-dotted lines correspond to contours of constant 
day-night assymmetry $A_{n-d}= Q_n-Q_d/Q_n+Q_d$ of average rates $Q$
in Super-Kamiokande, SNO and ICARUS respectively,
according to 
\protect{\cite{bahkra}}. 
KAMLAND should observe a 
disappearance signal and the 10 ton version of GENIUS should see double beta 
decay in this model.
(from \protect{\cite{Kla99b}})
}
\label{smix2}
\end{figure}

\begin{figure}[!tt]
\epsfxsize=8cm
\hspace*{1.5cm}
\epsfbox{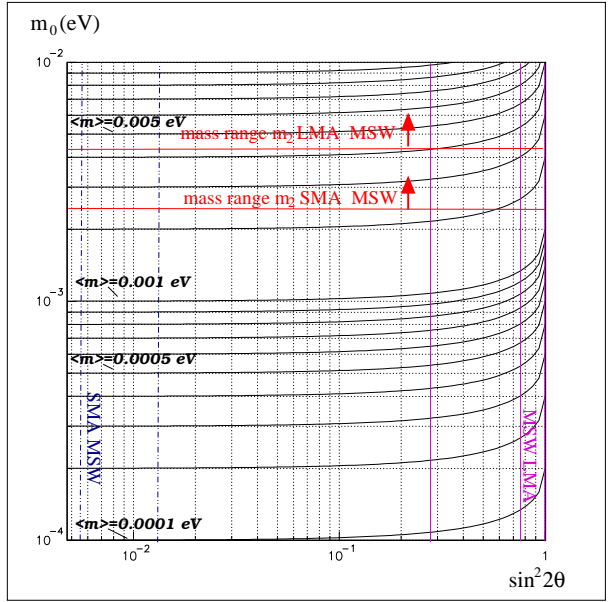} 
\caption{
Contribution of the first state in hierarchical 
models with $m_1>0$. 
Shown is $\langle m \rangle$ in the $m_0-sin^2(2 \theta_{12})$ plane, together
with the favored regions for LMA MSW (solid)
and SMA MSW (dash-dot-dot), the SMA extends further to smaller mixing, where 
$\langle m \rangle=m_0=const.$ 
The horizontal solid lines indicate the region above which the assumption 
$m_1^2 \ll \Delta m^2 = m_2^2$ 
is not valid anymore (Super-K bestfit for MSW LMA and SMA).
Vacuum oscillations are not included here, since 
for this case the model will be partially degenerate before any significant 
contribution to $\langle m \rangle$ arises. It is easy to see that sizable
contributions from the first state could lead to observable double beta rates 
even in hierarchical models.
(from \protect{\cite{Kla99b}})
}
\label{smix3}
\end{figure}

In the hierarchical case, 
\begin{equation}
m_0^2 \ll \Delta m^2_{21} \ll  \Delta m^2_{31}~, 
\end{equation}
the absolute values of two heavy neutrinos are completely 
determined by the mass squared differences (that is, 
by the oscillation parameters):  
\begin{equation}
m_3^2 = \Delta m^2_{31} =  \Delta m^2_{atm},~~    
m_2^2 = \Delta m^2_{21} = \Delta m^2_{\odot},~~ m_1^2 = m_0^2,  
\end{equation}
and the only freedom is the choice of the value of $m_1$.
In this scheme
there is no explanation of  the LSND result,  and
the contribution to the Hot Dark Matter component of the universe is
small:  $\Omega_{\nu} < 0.01$.
Different solutions of the solar neutrino problem
lead to different implications for the effective neutrino mass. 
It is useful to discuss the contributions of the mass eigenstates separately.
They are shown in figs. \ref{smix1}-\ref{smix3}. 

In the {\it single maximal (large) mixing scheme} $\nu_{\mu}$ and $\nu_{\tau}$ 
are mixed strongly in $\nu_{2}$ and $\nu_{3}$. 
The electron flavor is weakly mixed:  
it is mainly
in $\nu_{1}$ with small admixtures in the heavy states. 
The solar neutrino data are explained by  
$\nu_e \rightarrow \nu_{2}$ resonance conversion inside the Sun.

For double beta decay searches this scheme is a kind of worst--case 
scenario.  
Due to the hierarchy of masses and the small admixture of $m_{\mu}$
in $m_1$ 
$\langle m \rangle$ 
is dominated by $m_3 \sim \Delta m_{13}$:
\be
\langle m \rangle ^{(3)} \simeq U_{e3}^2 m_3,
\label{meffsmh}
\ee
which is severely constrained by the CHOOZ experiment (see fig. \ref{smix1}).
In terms of oscillation parameters the effective neutrino mass  can be
written as
\be
\langle m \rangle  \simeq \frac{1}{2}\sqrt{\Delta m_{atm}^2} \cdot
\left(1- \sqrt {1- \sin^2 2 \theta}\right).
\label{third}
\ee 
The contribution from the second mass (first term) 
is $ < 10^{-5}$ eV. 
As follows from fig. \ref{smix1} in the range of 
$\Delta m^2$ relevant for  the solution 
of
the atmospheric 
neutrino problem $\langle m \rangle$ can reach 
$\langle m \rangle \approx \langle m \rangle^{(3)} =  (3 - 4)\cdot 10^{-3}$ eV
and in the best fit range: 
$\langle m \rangle \approx    2\cdot 10^{-3}$ eV.
Thus the 10 ton 
GENIUS experiment could access 
the unexcluded region of $\langle m \rangle$, while the observation of
neutrinoless double beta decay induced by the neutrino 
mass mechanism with
$\langle m \rangle > 6 \cdot 10^{-3}$eV, the final sensitivity of the 1 ton 
version, would rule out the single maximal 
scenario with maximal mass hierarchy.

{\it Bi-large mixing:} The previous scheme can be modified in such a way that 
solar neutrino data  are explained by large angle MSW conversion. 
The contribution from the third state is the same as in eq.
\ref{third}. 

However now the contribution from the second level can be 
significant: both mixing parameter and the mass are  now larger.  
The 
contribution
from the second  state equals
\be
\langle m \rangle^{(2)} = m_2  |U_{e2}^2| = \sqrt{\Delta m_{\odot}^2} \sin^2
\theta_{\odot}  \sim (0.8 - 6) \cdot 10^{-3}  {\rm eV}. 
\label{second}
\ee
Providing a sensitivity of $\langle m \rangle =0.001$ eV 
GENIUS could cover the main part of the
large
mixing angle MSW solution of the solar neutrino deficit and could 
be complementary to the search of day-night effects (see fig. \ref{smix2}).

{\it Contributions of the first state:}
For a non-vanishing $m_0$ a further contribution for both schemes is implied
by an offset over the oscillation pattern. This contribution arising from
$m_1$ is shown in fig. \ref{smix3}. The total effective neutrino mass can 
easily be determined from figs. \ref{smix1} - \ref{smix3} by adding the
single contributions.

This is true as 
long as $m_1^2 \ll \Delta m_{12}^2 \simeq m_2^2$. 
As can be seen from fig. \ref{smix3} this additional contribution 
($< \langle m \rangle^{(1)} +2 m_0$)   easily may 
reach $10^{-2}$ eV without leaving the hierarchical pattern, shifting the
effective neutrino mass to observable values for the 1 ton version of GENIUS. 
However, also cancellation of 
the contributions may appear. In any case neutrino oscillations are not 
sensitive on this quantity, pushing GENIUS into some key position for testing 
the mass of the first state.

\subsubsection*{Schemes with partial degeneracy\label{pd}}

In the partially degenerate case,  
\be
\Delta m^2_{21} \ll m_0^2  \ll  \Delta m^2_{31}~, 
\ee
the two light neutrinos have close masses determined by $m_0$
and the heaviest mass is determined by the oscillation 
parameter: 
\be             
m_1^2 \approx m_2^2 \approx m_0^2~, ~~~ m_3^2 \approx  
\Delta m^2_{31} =  \sqrt{\Delta m_{atm}^2}~. 
\ee
The expression for the effective neutrino mass can be 
written as 
\be
\langle m \rangle = m_0 (\sin^2 \theta_{\odot} + 
e^{i \phi_{21}} \cos^2 \theta_{\odot}) + 
e^{i \phi_{31}}\sqrt{\Delta m_{atm}^2} \sin^2 \theta_{atm},
\ee
where $\theta_{atm}$ determines the admixture of the $\nu_e$ 
in $\nu_3$. 
For the small mixing MSW solution of the solar neutrino problem we get 
\be
\langle m \rangle \approx m_0  +
e^{i \phi_{31}}\sqrt{\Delta m_{atm}^2} \sin^2 \theta_{atm}.  
\ee
For the {\it large mixing angle MSW solution} 
cancellation of contributions from the
lightest states may occur, so that even for relatively large 
$m_0$ the contribution from the third neutrino state 
gives 
the
main contribution, which is severely constrained by CHOOZ (see fig. 
\ref{smix1}).

\begin{figure}[!h] 
\epsfxsize=8cm
\hspace*{1.5cm}
\epsfbox{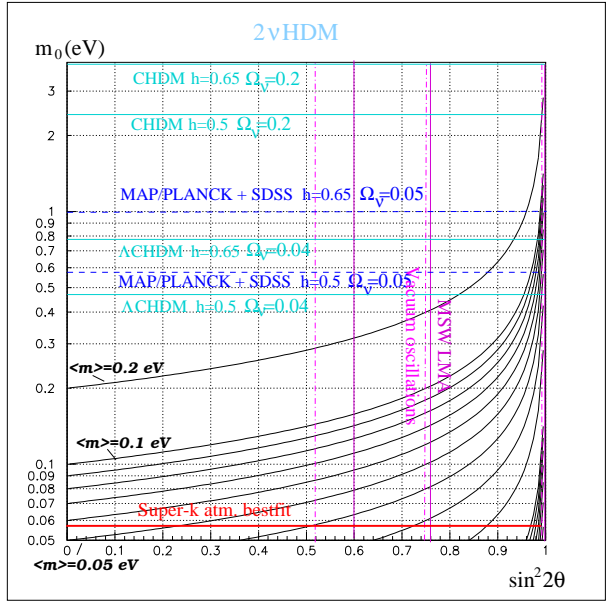} 
\caption{
Plotted are iso-mass lines in the $m_0-sin^2(2 \theta)$ plane for the case of 
cancellation between degenerate states $m_1$ and $m_2$ in partially
degenerate scenarios
with two neutrinos contributing to the hot dark matter. 
Mass splitting is neglected, since 
$m_1-m_2 \ll m_0$ and $m_3-m_1$ changes the cosmological considerations 
less than 
10 \%. 
Shown are the bestfits for CHDM (according to 
\protect{\cite{eric98}}),
and $\Lambda$CHDM (according to 
\protect{\cite{Pri98}})
for different values of the Hubble constant. Also shown is the sensitivity of
MAP/Planck combined with SDSS according to \protect{\cite{Eis98}}.
The regions of the MSW LMA and vac. osc. have been taken from
\protect{\cite{Bah98}}.
Also shown is the bestfit for atmospheric neutrinos, which gives a 
lower limit for $m_0$ in inverse hierarchical models.
Combined with the neutrino oscillation results and the
precision determinations of cosmological parameters GENIUS will allow to
give precise informations about mixing and the absolute mass scale in 
partially degenerate scenarios. Assuming, e.g. a worst case $m_0=0.06$ eV
just above the atmospheric neutrino bestfit, the MSW LMSA or vacuum solution 
would imply $\langle m \rangle$ = 0.03 eV
(from \protect{\cite{Kla99b}}). }
\label{smix5}
\end{figure}

The partially
degenerate spectrum with 
\be
\Delta m^2_{21} \ll  \Delta m^2_{31}  = m_0^2~.
\ee
leads to a scheme with inverse mass hierarchy: 
\be
m_1^2 \approx m_2^2 \approx m_0^2~ = \Delta m_{atm}^2, ~~~ 
m_3^2 \ll  m_0^2. 
\ee
The effective Majorana mass can be written as 
\be
\langle m \rangle \approx   \sqrt{\Delta m_{atm}^2}(\sin^2 \theta_{\odot} +
e^{i \phi_{21}} \cos^2 \theta_{\odot})~,  
\label{mee3}
\ee
where we have neglected the small contribution from the third state: 
$m_3 U_{e3}^2$ ($U_{e3}^2 < 5 \cdot 10^{-2}$).
The two heavier eigenstates contribute to the hot dark matter (HDM)
\be
\Omega_{\nu}=\frac{2 m_1}{91.5 eV}h^{-2}
\ee
Assuming the {\it vacuum oscillation 
solution} 
(for inverse hierarchy no level--crossing in the sun and thus no MSW effect 
appears), both addition yielding $\langle m \rangle \simeq 0.03-0.1$ eV
as well as compensation of the contributions from the two 
heavy states can  occur. In the case
of the
bi-maximal scheme the compensation is  complete. Again, additional 
contributions from the lightest state, here $m_3$ may be possible.
In Fig. \ref{smix5} the sensitivity in the $m_0 - \sin^2 2 \theta$
plane is shown together with the favored regions of the
``Just-so'' vacuum solution.  
Allowing for CP violation here just implies the cancellation to be less 
effective. All values for $\phi_{12},\phi_{13}$ imply a behavior of the mass 
eigenstates settled between the extreme values of 0 and $\pi$. 
As can be seen, GENIUS may provide sensitive informations about the mixing
and the grade of cancellation among the states, being complementary to
determinations of the  sum of the neutrino mass eigenstates due to studies
of the power spectra of the cosmic microwave background or galaxies by
MAP and Planck or the SDSS (Sloan Digital Sky Survey) \cite{Eis98}.

\begin{figure}[!h]
\epsfxsize=8cm
\hspace*{1.5cm}
\epsfbox{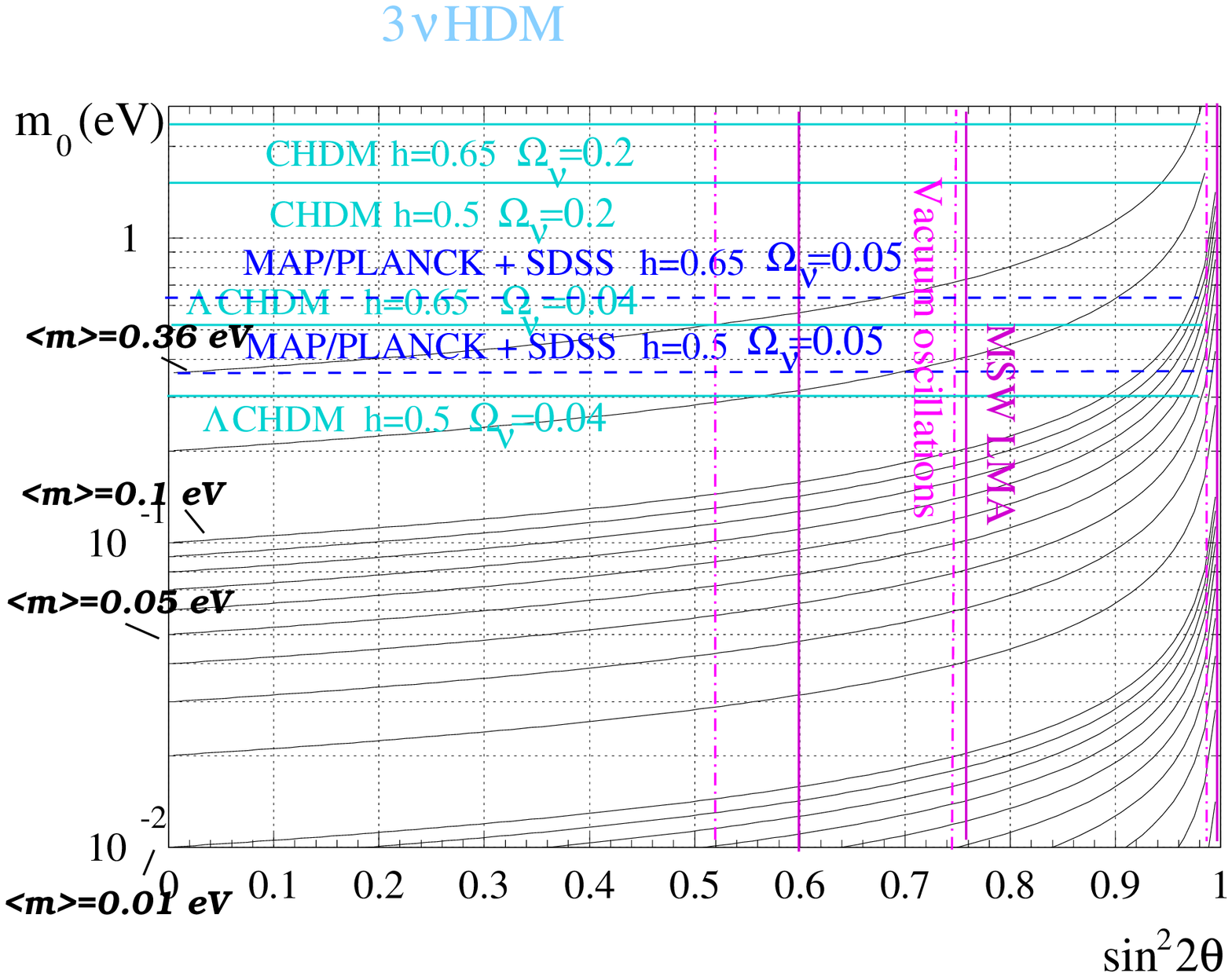} 
\caption{
As figure \protect{\ref{smix5}} for totally degenerate scenarios, that means
three neutrinos are contributing to the hot dark matter.
Combined with the neutrino oscillation results and the
precision determinations of cosmological parameters GENIUS will allow to
give precise informations about mixing and the absolute mass scale in 
degenerate scenarios. E.g. assuming an overall scale of 0.3 eV corresponding
to the $\Lambda$CHDM models with $\Omega=0.04$ and $h=0.5$, 
the bestfit of either the 
MSW or the vacuum solution would imply $\langle m \rangle$ = 0.15 eV 
(from \protect{\cite{Kla99b}}).}
\label{smix4}
\end{figure}

\subsubsection*{Schemes with  complete degeneracy}

In degenerate schemes the
common mass is much larger than the mass splittings: 
\be 
\Delta m^2_{21} \ll  \Delta m^2_{31}  \ll m_0^2. 
\ee 
In this case the effective neutrino mass is
\be
\langle m \rangle = (|U_{e1}|^2 
+ |U_{e2}|^2 e^{i\phi_{21}}
+ |U_{e3}|^2 e^{i\phi_{31}}) m_0,
\ee 
which
is determined by mixing angles and relative phases of the mass
terms. 

In the case of the small mixing MSW solution ($U_{e1}^2 \gg U_{e2}^2,
U_{e3}^2$) no substantial cancellation appears and 
$\langle m \rangle \approx m_0$. 
The same expression can be obtained for any solution of the solar neutrino
problem if the CP violating phases are zero: $\phi_{12}=\phi_{13}=0$ or
$\phi_{12}=0$,$U^2_{e3}\simeq 0$. 
Double beta decay and neutrino oscillations decouple. 
The effective neutrino  
mass  can be restricted by cosmological observations. 
The  contribution of  neutrinos to the
HDM in the universe is 
\be
\Omega_{\nu}=\frac{3 m_0}{91.5 eV}h^{-2}
\ee
(see Fig. \ref{smix5}). In bimaximal schemes $\langle m \rangle$ 
is exactly vanishing. 
However, comparing with the quark sector, this case seems to be rather 
unnatural. For $U^2_{e3}\simeq 0$ and $\phi_{12}=\pi$ in eq. \ref{mee}
the double beta observable becomes 
\be 
\langle m \rangle \simeq m_0 \sqrt{1 - \sin^2 2\theta}.
\ee
As in fig. \ref{smix5} 
in fig. \ref{smix4} the sensitivity in the $m_0 - \sin^2 2 \theta$
plane is shown together with the favored regions of the solar
MSW large mixing angle solution as well as the ``Just-so'' vacuum solution.  
Again GENIUS provides a tool of unique sensitivity for determining mixings 
and the grade of cancellation among the states.

\subsubsection*{LSND and four neutrino scenarios}
\begin{figure}[htb]
\vspace*{2cm} 
\hspace*{1cm}
\epsfxsize=9cm
{\epsfbox{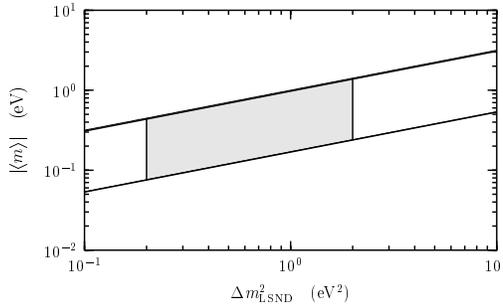}
\vspace*{-9cm} 
\caption{
Four neutrinos in the scheme with direct mass hierarchy: The shaded area
shows the possible value of the effective Majorana mass $\langle m \rangle$
in the range of $\Delta m^2_{LSND}$ for the case of the MSW LMA solution of 
the solar neutrino problem.
This case can be easily checked by the 1 ton 
version of GENIUS (from \protect{\cite{Bil99}}).}
\label{bilfig99}}
\end{figure}

Additional neutrino states being singlets under the usual SU(2) have been 
discussed \cite{Smi99} in order to account for the LSND anomaly with 
\be
\Delta m^2_{LSND}\simeq1 eV^2.
\ee
The viable schemes contain two light states responsible for the solution of
the solar neutrino problem and two heavy states in the range relevant 
for structure formation in the universe and for oscillations of atmospheric 
neutrinos (see also the discussion in \cite{Giu99}).
 $\nu_{\mu}$ and $\nu_{\tau}$ are strongly 
mixed in two heavy mass eigenstates $\nu_2$ and $\nu_3$ with 
\be
\sqrt{m_3^2 - m_2^2} \equiv \sqrt{\Delta m_{ATM}^2} \ll  
m_3  \approx m_{HDM}, 
\ee
so that 
$\nu_{\mu} \leftrightarrow \nu_{\tau}$ oscillations 
solve the atmospheric neutrino problem.  
$\nu_e$ and $\nu_s$ are weakly mixed in the two lightest 
mass states. Resonance conversion   $\nu_e \rightarrow \nu_s$
solves the solar neutrino problem.
As has been pointed out in \cite{Giu99} the inverse scheme requires strong 
cancellation in the heavy states to fit the present bound from the 
Heidelberg--Moscow experiment. This requires large mixing of $\nu_s$ and 
$\nu_e$, which is excluded by BBN bounds on the number of neutrino species.
Since this issue is still rather controversial (see \cite{Giu99} and 
references therein) it may be interesting to study the situation
of a weakened BBN bound. In this case the inverse hierarchical scheme is still 
unexcluded in combination with the MSW LMA or vacuum solution and \cite{Bil99}
\be
7 \cdot 10^{-2} eV < \langle m \rangle < 1.4 eV.  
\ee
In fig. \ref{bilfig99} $\langle m \rangle$ is plotted as a function of 
$\Delta m^2$ for the case of the LMA MSW solution.

Turning to the hierarchical scheme
the contribution to $m_{ee}$ from the pair of heavy mass 
states can be written as 
\be 
m_{ee}^{(23)} = U_{e2}^2 m_2 +  U_{e3}^2 m_3 \approx 
(|U_{e2}|^2 +  |U_{e3}|^2 e^{i\phi}) m_{3}~. 
\ee
The masses  $m_{3} \approx m_{2}$  can be relevant for cosmology, 
their value determines the
splitting between pairs of the heavy and
the light states and can induce the oscillations observed by LSND: 
\be 
m_{3} =  \sqrt{\Delta m_{LSND}^2} = \frac{1}{2} m_{HDM}~.  
\ee
Therefore, 
\be
\langle m \rangle^{(23)} = 
(|U_{e2}|^2 +  |U_{e3}|^2 e^{i \phi}) \sqrt{\Delta m_{LSND}^2}. 
\ee
Taking the bound from Bugey into account
this is leading to
\be
7 \cdot 10^{-4} <\langle m \rangle^{(23)}< 2 \cdot 10^{-2}, 
\ee
which may yield a positive signal in GENIUS \cite{Giu99}.
The contribution of the light states corresponds to the situation
in hierarchical schemes discussed above. It may become 
significant in the case of strong cancellation in the heavy states.

\subsubsection*{Summary}
In summary, GENIUS can play an important role in reconstructing the neutrino
mass spectrum. In strongly hierarchical schemes the magnitude of the double beta 
observable depends crucially on the assumed solution for the solar neutrino 
problem. While with assuming the MSW SMA or vacuum oscillations the observation
of $0\nu\beta\beta$ with $\langle m \rangle > 6 \cdot 10^{-3}$ would rule out
the scheme, in scenarios with MSW LMA GENIUS (10 tons) should observe a 
positive signal for the main part of the MSW LMA solution, being complementary
to the search for day-night effects in present and future solar neutrino 
experiments such as Superkamiokande, SNO or ICARUS. 
In any case GENIUS may provide a unique possibility to determine
the mass of the lightest state. Even more stringent restrictions may be 
obtained in partially or completely degenerate schemes, motivated by giving
sizable contributions to the hot dark matter in the universe. 
In such scenarios already
the present half life limit of the Heidelberg--Moscow experiment requires 
strong cancellation between the mass eigenstates. GENIUS could help to
determine the mixing in such schemes with extreme accuracy, providing 
informations being complementary to precision tests of cosmological 
parameters by MAP and Planck. In four neutrino scenarios GENIUS has good
perspectives for testing the LSND signal.
For further recent discussions of the potential of GENIUS for probing neutrino 
masses we refer, e.g., to \cite{Kla99b,Giu99,Cza99,Vis99,Bil99}. 

\subsubsection*{1.3.3.2 GENIUS and super--heavy left--handed neutrinos:}

\begin{figure}[h!]
\epsfysize=85mm
\epsfbox{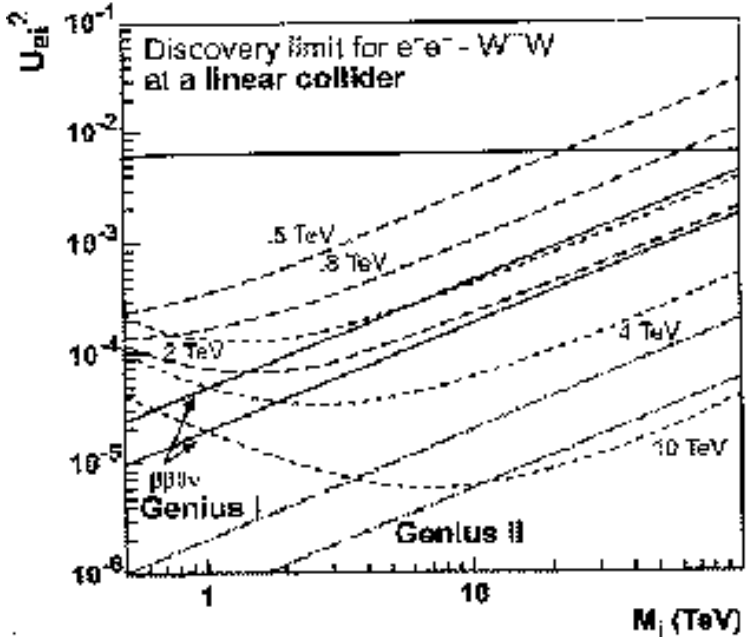}
\caption{Discovery limit for $e^- e^- \rightarrow W^- W^-$ at a linear collider 
as function of the mass $M_i$ of a heavy left--handed neutrino, and of
$U_{ei}^2$ for $\sqrt{s}$ between 500 GeV and 10 TeV. In all cases the
parameter space above the line corresponds to observable events. 
Also shown are the limits set by the Heidelberg--Moscow $0\nu\beta\beta$
experiment as well as the prospective limits from GENIUS. The areas {\rm above}
the $0\nu\beta\beta$ contour lines are {\rm excluded}. The horizontal
line denotes the limit on neutrino mixing, $U_{ei}^2$, from LEP.
Here the parameter space above the line is excluded. (from \cite{Bel98}).
}
\label{fig25}
\end{figure}

Fig. \ref{fig25} (from \cite{Bel98}) compares the sensitivity of GENIUS for heavy 
left-handed neutrinos (as function of $U_{ei}^2$, for which the present
LEP limit is $U_{ei}^2 \leq 5 \cdot 10^{-3}$ \cite{Nar95}) with the discovery 
limit for $e^- e^- \rightarrow W^- W^-$  at Next Linear Colliders. The 
observable in $0\nu\beta\beta$ decay is 
\be
\langle m^{-1}_{\nu} \rangle_H = \sum_i ~^{''}U^2_{ei} \frac{1}{M_i}. 
\ee
Also shown are the present limits from the Heidelberg--Moscow experiment
(denoted by $0\nu\beta\beta$) assuming different matrix elements. It is
obvious that $0\nu\beta\beta$ is more sensitive than any reasonable future
Linear Collider.  

\subsubsection*{1.3.3.3 GENIUS and left--right symmetry:}

If GENIUS is able to reach down to $\emass \le 0.01$ eV, it would at 
the same time be sensitive to right-handed $W$-boson masses up to 
$m_{W_R} \ge 8$ TeV (for a heavy right-handed neutrino mass of 
$1$ TeV) or $m_{W_R} \ge 5.3$ TeV (at $\langle m_N \rangle = m_{W_R}$) 
\cite{Kla97d}. 
Such a limit would be comparable to the one expected for LHC, 
see for example \cite{Riz96}, which quotes a final sensitivity 
of something like $5-6$ TeV. Note, however that in order to 
obtain such a limit the experiments at LHC need to accumulate 
about $100 fb^{-1}$ of statistics. A 10 ton version of
 GENIUS 
could even reach a sensitivity of $m_{W_R} \ge 18$ TeV (for a heavy 
right-handed neutrino mass of
$1$ TeV) or 
$m_{W_R} \ge 10.1$ TeV (at $\langle m_N \rangle = m_{W_R}$).

This means that already GENIUS 1 ton could be sufficient to definitely
test recent supersymmetric left--right symmetric models having the 
nice features of solving the strong CP problem without the need for an axion 
and having automatic R--parity conservation \cite{Kuc95,Moh96}.

\subsubsection*{1.3.3.4 GENIUS and $R_p$--violating SUSY:}

\begin{figure}[h!]
\vskip-25mm 
\hskip10mm
\epsfxsize=100mm
\epsfysize=120mm
\epsfbox{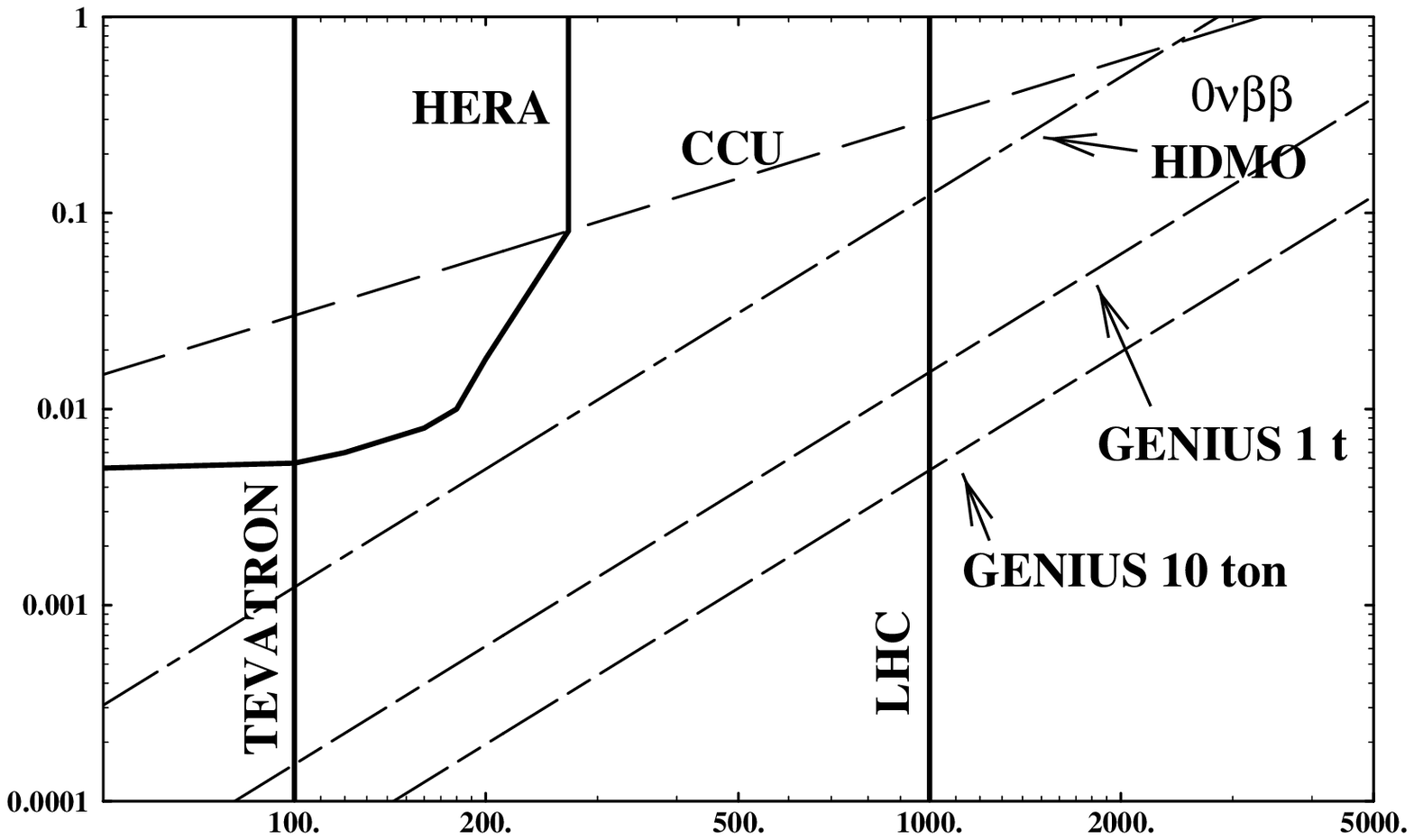}
\vskip-80mm
\noindent
$\lambda'_{111}$ 
\vskip45mm 
\hskip90mm $m_{\tilde q}$ [GeV] 
\bigskip
\caption{ Comparison of sensitivities of existing and future 
experiments on \rp SUSY models in the plane $\lambda'_{111}-m_{\tilde q}$. 
Note the double logarithmic scale! Shown are the areas currently excluded 
by the experiments at the TEVATRON, the limit from charged-current 
universality, denoted by CCU, and the limit from absence of \znbb{} 
decay from the Heidelberg-Moscow collaboration (\znbb{} HDMO). 
In addition, the estimated sensitivity of HERA and the LHC is compared to the 
one expected for GENIUS in the 1 ton and the 10 ton version (from [Kla97c]).}
\label{fig26}
\end{figure}

The improvement on the R--parity breaking Yukawa coupling $\lambda^{'}_{111}$
(see subsection 1.3.1) is shown in Fig. \ref{fig26}.
The full line to the right is the expected sensitivity of the 
LHC -- in the 
limit of large statistics. The three dashed--dotted lines denote (from top
to bottom) the current constraint from the Heidelberg--Moscow experiment
and the sensitivity of GENIUS 1 ton and GENIUS 10 tons, all
 for the 
conservative case of a gluino mass of 1 TeV. If squarks would be heavier than 
1 TeV, LHC could not compete with GENIUS. However, for typical squark masses  
below 1 TeV, LHC could probe smaller couplings.
However, one should keep in 
mind, that LHC can probe squark masses up to 1 TeV only with several years of 
data taking.

The potential of GENIUS on R-parity breaking coupling products 
derived from
the neutrino mass bounds
is shown 
in tab. \ref{tabrpv2} \cite{Bha99}.
GENIUS in the 1(10) ton version would provide
a further improvement by 1(2) orders of magnitude compared to the 
Heidelberg--Moscow experiment. 

\begin{table}[!h]
\begin{center}
\begin{tabular}{ccc}
\hline
\hline
$\lambda^{(')}_{ijk}\lambda^{(')}_{i^{'}kj}$
&                     Our &  Previous  \\  
& Bounds & Bounds \\
\hline
\hline
$m_{ee}<0.01 (0.001)$ eV & [GENIUS 1(10)t] & \\ \hline
$\lambda^{'}_{133}\lambda^{'}_{133}$ & $1.5 \cdot 10^{-9(-10)}$ & $4.9
\cdot 10^{-7}$ \\
$\lambda^{'}_{132}\lambda^{'}_{123}$ & $3.7 \cdot 10^{-8(-9)}$ & $1.6
\cdot 10^{-2}$ \\
$\lambda^{'}_{122}\lambda^{'}_{122}$ & $9.2 \cdot 10^{-7(-8)}$ & $4.0
\cdot 10^{-4}$\\
$\lambda_{133}\lambda_{133}$ & $2.6 \cdot 10^{-8(-9)}$ & $9.0 \cdot
10^{-6}$ \\
$\lambda_{132}\lambda_{123}$ & $4.3 \cdot 10^{-7(-8)}$ & $2.0 \cdot
10^{-3} $ \\
$\lambda_{122}\lambda_{122}$ & $7.1 \cdot 10^{-6(-7)}$ & $1.6 \cdot
10^{-3}$\\ \hline
\end{tabular}
\caption{Correlation among neutrino mass bounds from GENIUS and upper limits
on RPV couplings. We have used $m_d$=9
MeV, $m_s$= 170 MeV, $m_b$=4.4 GeV \protect{\cite{PDG98}}. 
For
$\lambda$-products, $m_{\tilde{d}}$ should be read as
$m_{\tilde{e}}$. The relevant scalars are always assumed to have a
common mass of 100 GeV.
\label{tabrpv2}
}
\end{center}
\end{table}

\subsubsection*{1.3.3.5 GENIUS and $R_p$--conserving SUSY:}
Since the limits on a `Majorana--like' sneutrino mass $\tilde{m}_M$ scale
with $(T_{1/2})^{1/4}$, GENIUS 1 ton (or 10 tons)
would test `Majorana' sneutrino masses lower
by factors of about 7(20), compared with present constraints 
\cite{Hir97,Hir97a,Hir97b}. 

\subsubsection*{1.3.3.6 GENIUS and Leptoquarks:}Limits on the lepton--number violating parameters as defined previously 
improve as $\sqrt{T_{1/2}}$. This means that for leptoquarks in the range
of 200 GeV LQ--Higgs couplings down to (a few) $10^{-8}$ could be explored. 
In other words, if leptoquarks interact with the standard model Higgs boson
with a coupling of the order ${\cal O}(1)$, either $0\nu\beta\beta$ must be 
found, or LQs must be heavier than (several) 10 TeV.

\subsubsection*{1.3.3.7 GENIUS and composite neutrinos}
GENIUS in the 1(10) ton version would improve the limit on the excited
Majorana neutrino mass deduced from the Heidelberg--Moscow experiment
(eq. 32) to
\be
m_N\geq  1.1 (2.3) \hskip3mm TeV
\ee

A recent detailed study \cite{Pan99} shows that while the HEIDELBERG--MOSCOW
experiment already exceeds the sensitivity of LEPII in probing compositeness,
GENIUS will reach the sensitivity of LHC. With the $0\nu\beta\beta$ half life
against decay by exchange of a composite Majorana 
neutrino given by \cite{Pan99}
\be
T_{1/2}^{-1}=\Big(\frac{f}{\Lambda_c}\Big)^4 \frac{m_A^8}{M_N^2} 
|{\cal M}_{FI}|^2 \frac{G_{01}}{m_e^2}
\ee
where $M_N$ is the composite neutrino Majorana mass, and $f$ denotes the 
coupling with the electron. Fig. 38 shows the situations of GENIUS and LHC. 

\begin{figure}[h!]
\epsfysize=10cm
\hspace*{1.5cm}
{\epsfbox{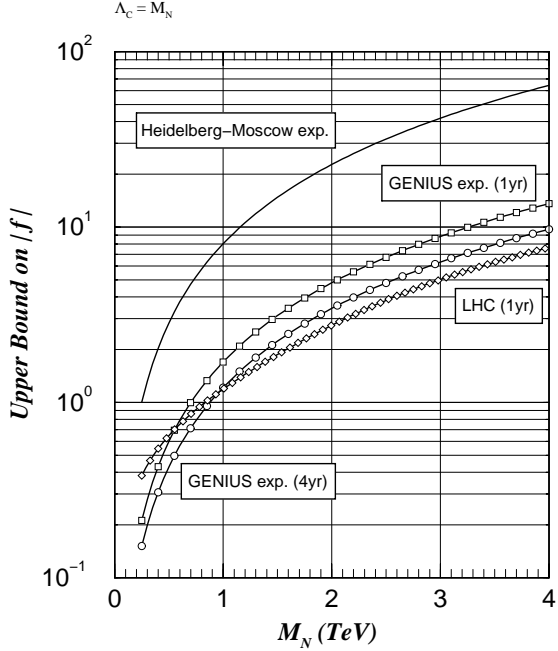}
\caption{Sensitivity of LHC and GENIUS to compositeness parameters (assuming 
$\Lambda_C=M_N$). Regions above the curves are excluded. The LHC bound is weaker
than the GENIUS bound for $M_N<550 (1000)$ GeV. 
(from \cite{Pan99})
}}
\label{fig27}
\end{figure}

\subsubsection*{1.3.3.8 GENIUS, special relativity and equivalence principle
in the neutrino sector}

The already now strongest limits given by the Heidelberg--Moscow experiment
discussed in subsection 1.3.2 would be improved by 1--2 orders of magnitude.
It should be stressed again, that while neutrino oscillation bounds 
constrain the region of large mixing of the weak and gravitational 
eigenstates, these bounds from double beta decay apply even in the case
of no mixing and thus probe a totally unconstrained region in the parameter 
space.

\subsection{The Solar Neutrino Potential of GENIUS} 
\subsubsection{Introduction}

The study of neutrinos coming from the Sun is a very active area of research.
Results from five solar neutrino experiments are now available.
These experiments measure the solar neutrino flux with different energy
thresholds and using very different detection techniques.
All of them, the Chlorine experiment at Homestake \cite{chlor},
the radiochemical Gallium experiments, GALLEX \cite{gallex} and SAGE 
\cite{sage}, the water Cerenkov detectors Kamiokande \cite{kamiok} and
Super-Kamiokande \cite{SuperK},
measure a deficit of the neutrino flux 
compared to the predictions of the standard solar model (SSM) \cite{SSM}.
Recently it has been stated out that it is imposible to construct
a solar model which would reconcile all the data \cite{hiroshi}. 
Moreover, a global analysis of the data of all the experiments do not leave
any room for the $^7$Be neutrinos \cite{Bah98b}.
On the other hand the  predictions of the SSM have 
been confirmed by helioseismology \cite{basu} to a high precision. 
An explanation of the results of solar neutrino experiments 
seems to require new physics beyond the standard model of electroweak 
interaction.

If neutrinos have non-zero masses and if they mix in analogy to the quark 
sector, then conversions between different neutrino flavours become
possible. Flavour conversions can occur in different physical scenarios,
depending on certain parameters on neutrino masses and mixing angles.
One oscillation scenario makes use of the 
MSW-mechanism \cite{msw85}, where the solar $\nu_e$ transform into other
neutrino flavours or into sterile neutrinos  as they pass through a thin  
resonance region near the solar core. 
The other scenario assumes that the neutrinos
oscillate in the vacuum between the Sun and the Earth \cite{gla87}, which means
that the oscillation length `just so` matches the Earth-Sun distance.

\subsubsection{The solar neutrino spectrum}

\begin{figure}[h!]
\epsfxsize=12cm
\epsfbox{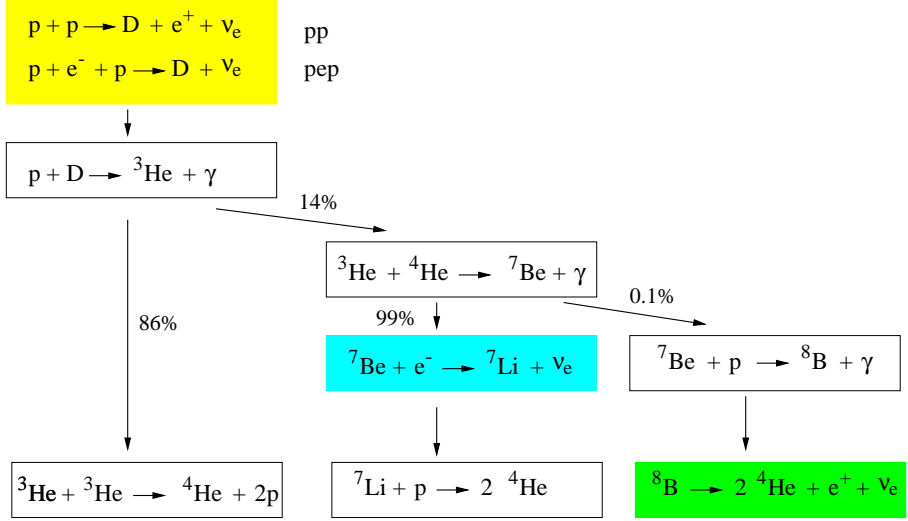}
\caption{Nuclear reactions in the pp-chain in the Sun.}
\label{ppchain}
\end{figure}

The Sun acquires its energy by nuclear reactions taking place in the core,
mainly via the so-called pp-chain (see Fig. \ref{ppchain}).

\begin{figure}[h!]
\epsfxsize=12cm
\epsfbox{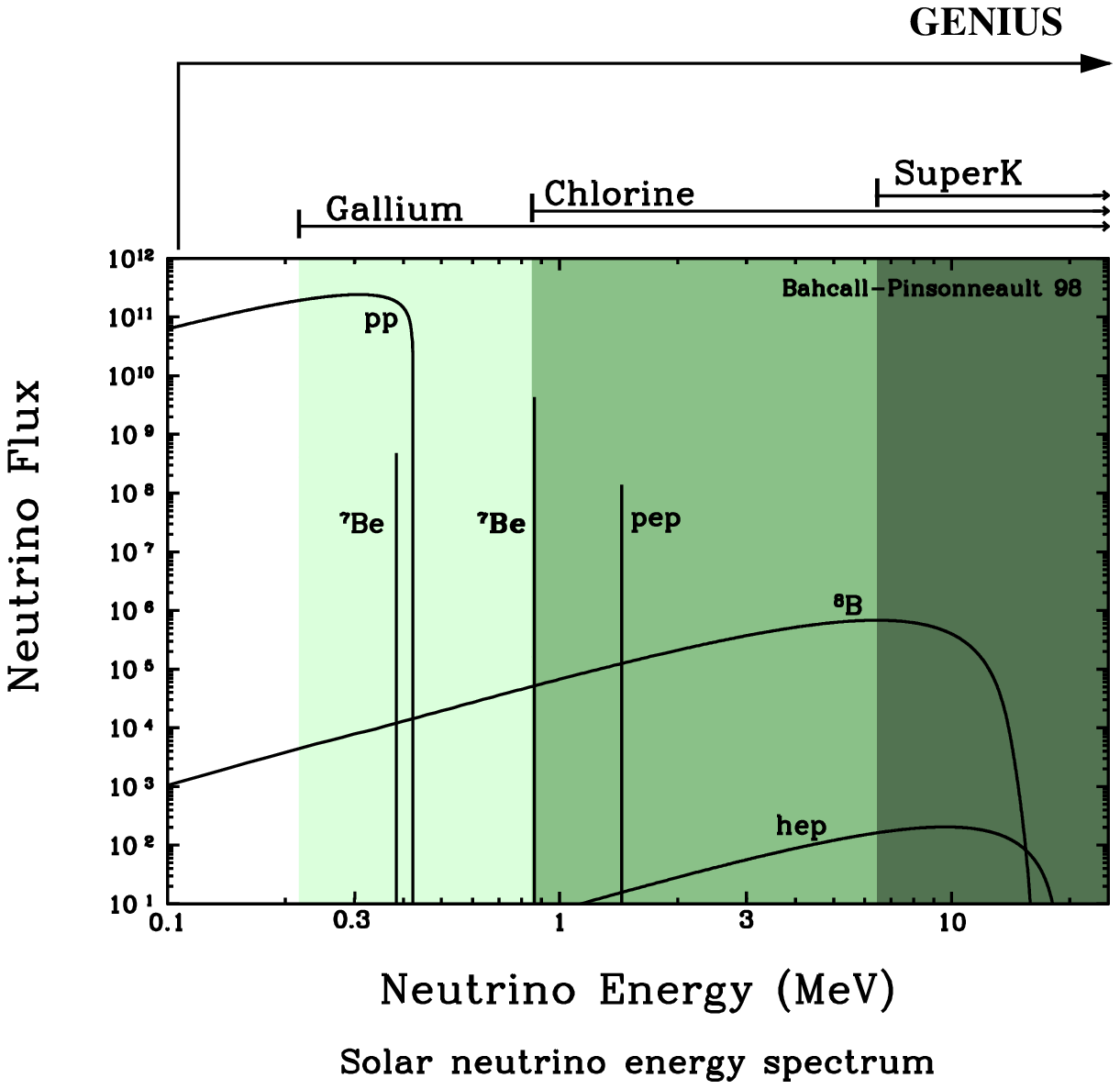}
\caption{Predicted solar neutrino spectrum in the SSM (from \cite{bah91}).}
\label{nuspec}
\end{figure}

The neutrino spectrum predicted by the SSM for the pp-chain
is shown in Fig. \ref{nuspec}. The dominant part of the flux is
emitted at energies below 1 MeV.

The pp neutrinos, emitted in the reaction p+p $\rightarrow$ D+e$^+$+$\nu_e$,
have a continuous energy spectrum with the endpoint at 420 keV.
Their flux is most accurately predicted in the SSM, since it is strongly
restricted by the solar luminosity and by helioseismological measurements.
The other main features of the solar neutrino spectrum are a strong
monoenergetic line at 861 keV, from the reaction  $^7$Be+ e$^-$$\rightarrow$
$^7$Li+$\gamma$+$\nu_e$, the $^7$Be neutrinos, and a continuous
spectrum of neutrinos extending up to 15 MeV, due to the reaction
$^8$B$\rightarrow$2$\alpha$+e$^+$+$\nu_e$, the $^8$B neutrinos.
Table \ref{nufluxes} gives the solar neutrino fluxes in the SSM with
their respective uncertainties (from \cite{Bah98c}).

\subsubsection{Present status of the solar neutrino experiments}

\begin{figure}[h!]
\epsfysize=10cm
\epsfbox{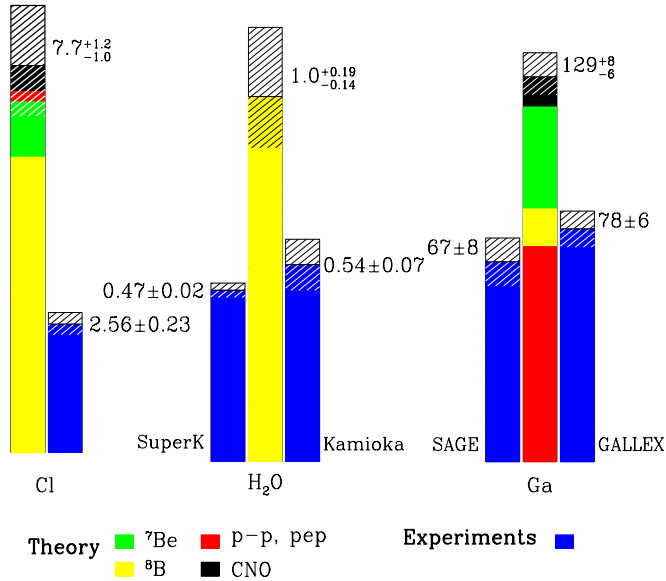}
\caption{Comparison of the total rates predicted in the SSM and the
observed rates in the present solar neutrino experiments, 
from \cite{Bah96}.}
\label{theoexp}
\end{figure}

The solar neutrino problem has been known for two decades,
since the Homestake experiment reported its first result.
At that time, however, it was not clear if the difference between
the chlorine measurement and the standard solar model prediction
was due to experimental systematics or the uncertainties in the 
SSM or if it was a sign of new physics.
Meanwhile, the observed discrepancy was confirmed by other
four solar neutrino experiments (see Fig. \ref{theoexp}, from \cite{Bah96}).
Model independent analysis performed by many authors (see \cite{hiroshi}
and references therein) 
suggest that the solar neutrino problem can only be solved if
some additional assumptions are made in the standard electroweak
theory. The most generic assumption is to give neutrinos a mass,
which leads to neutrino oscillations in vacuum or matter.

Oscillations between two neutrino species are characterized by two
parameters: $\Delta$m$^2$, the difference of the squared mass eigenstates,
and $\theta$, the mixing angle between the mass eigenstates.
 
\begin{figure}[h!]
\epsfysize=10cm
\epsfbox{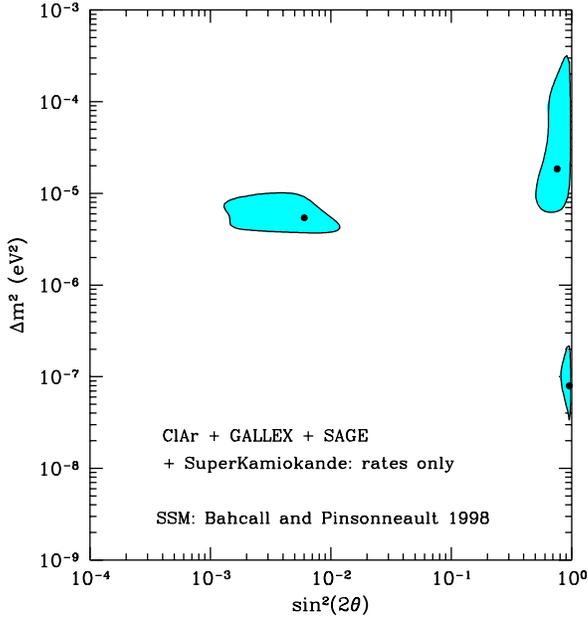}
\caption{The allowed  regions (99\% C.L.) in $\Delta m^2$ ---
$\sin^22\theta$ parameter space for the MSW solution, from \cite{Bah98}. }
\label{solutions}
\end{figure}

The Ga experiments, sensitive to the low-energy pp and $^7$Be neutrinos,
combined with the Homestake and Super-Kamiokande experiments, which
are sensitive to the high-energy $^8$B neutrinos, strongly restrict
the allowed range of $\Delta$m$^2$ and $\theta$ for all oscillation 
scenarios.
There exist four parameter areas compatible with the results of
all existing solar neutrino experiments: the  
large mixing angle solution (LMA), the small mixing angle solution (SMA), 
the low mass solution (LOW) and the vacuum oscillation solution
with strong mixing (see Fig. \ref{solutions} for the MSW-solutions).
Up to date, there is no clear evidence for one of the above solutions.
To clarify the situation, there is great demand for additional solar
neutrino experiments, especially at energies below 1 MeV.

Borexino \cite{borexprop} is now being built up especially to measure
the flux of $^7$Be neutrinos in real time. 
It will use 300 tons of organic scintillator
(100 tons of fiducial volume) to detect recoil electrons from
elastic neutrino-electron scattering. Since the scintillator has
no directional information and the signal is characterized only by the
scintillation light produced by the recoil electron, very stringent
constraints on the radiopurity of the scintilator and on the activity
of all detector materials are imposed.

\begin{table}
\hspace*{2.4cm}
\begin{tabular}{lc}
Source & Flux (10$^{10}$ cm$^{-2}$s$^{-1}$)\\
\hline
pp & 5.94 $\pm$ 0.01 \\
pep & 1.39$\times$10$^{-2}$$\pm$ 0.01 \\
$^7$Be & 4.80$\times$10$^{-1}$$\pm$ 0.09 \\
$^8$B & 5.15$\times$10$^{-4}$$\pm$ 0.19 \\
\end{tabular}
\caption{Solar Standard Model predictions of the neutrino fluxes, from
\cite{Bah98c}}
\label{nufluxes}
\end{table}

So far, there exist three proposals to measure the pp-flux in real time, 
HERON \cite{heron}, HELLAZ \cite{hellaz} and LENS \cite{lens}.

The HERON project will use $^4$He in its superfluid state (at 20 mK) as the 
target medium. The detection reaction is elastic neutrino-electron
scattering, the electron recoil energy is converted into low-energy
elementary excitations of the helium, rotons, which can be detected.
For a fiducial volume of seven tons, the total SSM predicted event rate
is 14 per day (8 events per day from the pp neutrinos). HERON would
measure only the energy distribution of recoiling electrons, without
a direct determination of the neutrino energy.

In the HELLAZ project a large TPC (2000 m$^3$) filled with gaseous helium
at high pressure (5 atm.) and low temperature (77 K) will serve as a target.
It is planned to measure both the kinetic energy and the scattering angle
of recoil electrons from elastic neutrino-electron scattering and thus
to determine the solar neutrino energy.
The kinetic energy of recoil electrons is measured by counting the individual
electrons in a ionisation cloud generated by the energy loss of the recoil
electron due to ionisation in the helium gas.
The expected event rate for 2$\times$10$^{30}$ target electrons is 7 per day 
and 4 per day for pp neutrinos and $^7$Be neutrinos, respectively.

LENSE would be a complementary approach to the above detectors using flavour
independent elastic scattering from electrons.
The method of neutrino detection is neutrino capture in $^{82}$Se, 
$^{160}$Gd or $^{176}$Yb.
The neutrino captures occur to excited states of the final nuclides,
providing a strong signature against radioactive background.
The thresholds for neutrino capture are 173 keV for$^{82}$Se, 244 keV
for $^{160}$Gd and 301 keV for  $^{176}$Yb. Three different techniques
for implementation as a solar neutrino detector are explored at present
\cite{lenseloi}:
liquid scintillator loaded with Yb or Gd, scintillating crystals of silicates
of Gd (GSO) and time projection chambers with a gaseous compound of isotopic 
$^{82}$Se.

All of these projects are still in a stage of research and development,
they have not yet shown full feasibility for implementation as a solar
neutrino detector.

\subsubsection{ Time signatures of solar neutrinos}

Due to the eccentricity of the Earth orbit, seasonal variations
in the flux of solar neutrinos are expected.
The number of neutrinos of all flavours reaching the Earth is larger
when the Earth is closer to the Sun than when it is farther away and
should vary with 1/R$^2$, where R is the Eart-Sun distance,
R=R$_{0}$(1-$\epsilon$cos(2$\pi$t/year)). R$_{0}$= 1AU and $\epsilon$=0.017.
The neutrino flux thus shows a seasonal variation of about 7\%  from
maximum to minimum.
This variation can in principle be used by a real time solar neutrino
experiment to extract the neutrino signal independently of background
(if the background is stable in time) and is limited only by 
statistics.

Beyond the so-called `normal' seasonal variation, an anomalous 
seasonal variation is predicted for the $^7$Be neutrino flux
in case of the vacuum oscillation solution,
since their oscillation length in this case is comparable to the seasonal
variation of the Earth-Sun distance due to the eccentricity of the 
Earth orbit.
The flux variations in this case are much larger than for the normal 
seasonal variation, 
they could serve as a unique
signature of vacuum oscillations \cite{gla87}.

\begin{figure}[h!]
\epsfysize=10cm
\epsfbox{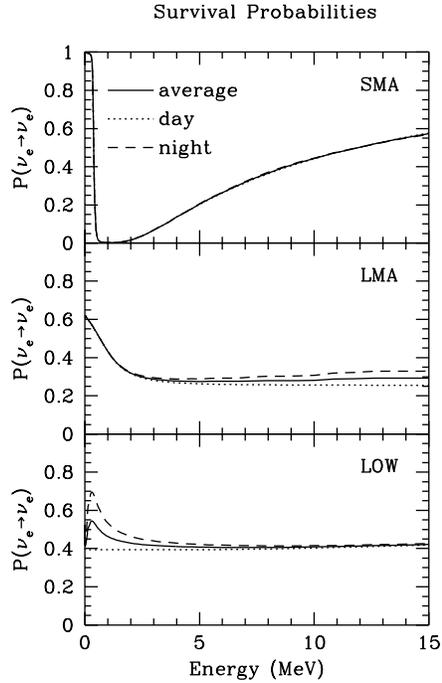}
\caption{Survival probabilities for an electron neutrino created in the Sun 
for the three MSW solutions, from \cite{bahkra}. SMA, LMA, LOW stand
for the small mixing angle, the large mixing angle and the low
$\Delta$m$^2$ MSW-solutions. }
\label{daynight}
\end{figure}

If neutrinos oscillate via the MSW-effect, then a regeneration of 
electron-neutrinos while passings through the Earth is predicted 
\cite{bahc89}. 
The so-called day/night-effect is neutrino energy dependent, its detection
would be a strong evidence for the MSW-effect. In Fig. \ref{daynight}
(from \cite{bahkra}) the $\nu_e$ survival probabilities for the MSW solutions
computed for the day-time and night-time are shown. At
low energies only  the LOW solution shows significant differences between the
day- and night-time survival probability.
Therefore this solution could be tested by a real-time detector
of low energy solar neutrinos, in particular by measuring the pp and 
$^7$Be neutrino flux.

\subsubsection{GENIUS as a solar neutrino detector}

The goal of the GENIUS project as a dark matter detector is to
achieve the background level of 10$^{-3}$ events/kg y keV in the energy
region below 100 keV.  Such a low background in combination with
a target mass of at least  1 ton of natural (or enriched) Ge opens
the possibility to measure the solar pp- and $^7$Be-neutrino flux 
in real time with a very low energy threshold.

\subsubsection{Signal Detection}

The detection reaction is  the elastic scattering process 
$\nu$ +  e$^- \rightarrow$ $\nu$ +  e$^-$.
The maximum electron recoil energy is 261 keV for the pp-neutrinos and 665 
keV for the  $^7$Be-neutrinos \cite{bahc89}. 
The energy of the recoiling electrons is detected through ionisation
in  high purity Ge detectors.
GENIUS in its 1 ton version would consist of an array of about 400 HPGe 
detectors, 2.5  kg each. Thus, the sensitive volume would be naturally divided 
into 400 cells which helps in background discrimination, since a 
neutrino interaction is taking place in a single cell.

\subsubsection{Signal Rates}

The dominant part of the signal in GENIUS is produced by 
pp-neutrinos (66 \%) and the  $^7$Be-neutrinos (33\%).

A target mass of 1 ton (10 tons) of natural or enriched Ge corresponds
to about 3$\times$10$^{29}$ (3$\times$10$^{30}$) electrons.

With the cross section for elastic neutrino-electron scattering 
\cite{bahc89}:\\

\noindent
$\sigma_{\nu_{e}}$ = 11.6 $\times$ 10$^{-46}$cm$^2$ \hspace*{3mm} pp\\
$\sigma_{\nu_{e}}$ = 59.3 $\times$ 10$^{-46}$cm$^2$ \hspace*{3mm} $^7$Be\\

\noindent
and the neutrino fluxes \cite{Bah98c}:\\

\noindent
$\phi_{pp}$ = 5.94 $\times$10$^{10}$ cm$^{-2}$s$^{-1}$\\
$\phi_{^{7}Be}$ = 0.48 $\times$10$^{10}$ cm$^{-2}$s$^{-1}$\\

\noindent
the expected number of events calculated  
in the standard solar model (BP98 \cite{SSM}) can be estimated:\\

\noindent
R$_{pp}$ = 69 SNU = 1.8 events/day (18 events/day for 10 tons)\\
R$_{^7Be}$ = 28.5 SNU = 0.6 events/day (6 events/day for 10 tons),\\

\noindent
The event rates for full $\nu_e \rightarrow \nu_{\mu}$ conversion 
are 0.48 events/day for pp-neutrinos and 0.14 events/day for
$^7$Be-neutrinos for 1 ton of Ge and ten times higher for 10 tons 
(see also Table \ref{rates})

\begin{table}
\begin{tabular}{lcc}
Case & Events/day & Events/day  \\
& 11-665 keV& 11-665 keV\\
&  (1 ton) & (10 tons)\\
\hline
SSM & 2.4  & 24 \\
Full $\nu_e \rightarrow \nu_{\mu}$ conversion & 0.62 & 6.2\\
\end{tabular}
\caption{Neutrino signal rates in GENIUS for 1 ton (10 tons) of 
Germanium.}
\label{rates}
\end{table}

\subsubsection{Background requirements}

GENIUS is conceived such that the external background from the natural 
radioactivity of the environment and from muon interactions is reduced
to a minimum, the main background contributions coming from the
liquid nitrogen shielding and the Ge detectors themselves.
To measure the low-energy solar neutrino flux, a nitrogen shielding
of 13 m in diameter is required.
Regarding the radiopurity of li\-quid nitrogen, the values reached at
present by the Borexino collaboration for their liquid scintillator
would be sufficient. Much attention has to be paid to the cosmogenic
activation of the Ge crystals at the Earth surface. In case of one day 
exposure, five years of deactivation below ground are required.
The optimal solution would be to produce the detectors in an underground 
facillity.

Table \ref{sol_backgr} shows the expected background events in the
energy region 11-260 keV and 11-665 keV.
 
\begin{figure}[!ht]
\epsfxsize=100mm
\epsfbox{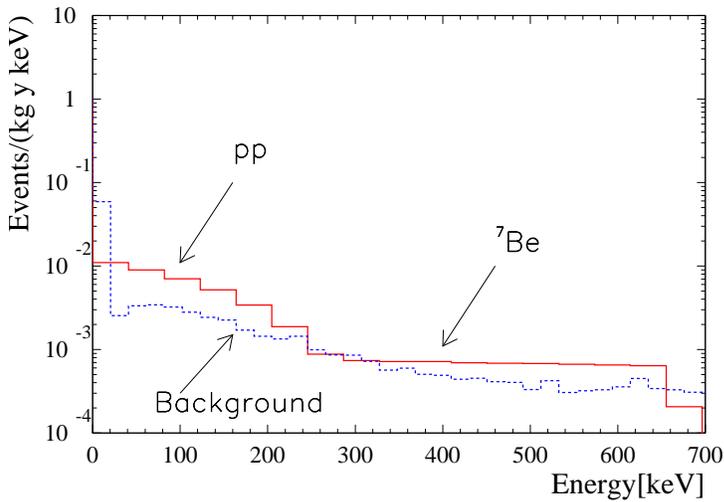}
\caption{Simulated spectra of the low energy neutrino signal (in the
  SSM) and the total background in GENIUS (1 ton of natural germanium).}
\label{simspektrum}
\end{figure}

Fig. \ref{simspektrum} shows the simulated spectrum of the low-energy neutrino
signal in GENIUS, together with the total expected background.

If the signal to background ratio S/B will be greater than 1, than the 
pp- and $^7$Be-neutrino flux can be measured by spectroscopic techniques
alone. 
If S/B $<$ 1,  one can make use of a solar signature in order to derive
the flux.

\begin{table}
\begin{tabular}{lc}
Energy region & Events/day \\
\hline
11 - 260 keV & 1.4 \\
11 - 665 keV & 1.8 \\
\end{tabular}
\caption{Expected background events in the GENIUS experiment (1 ton of Germanium).}
\label{sol_backgr}
\end{table}

The eccentricity of the Earth's orbit induces a seasonal variation
of about 7\% from maximum to minimum. Even if the number of background
events is not known, the background event rate and the signal event rate
can be extracted independently by fitting the event rate to the seasonal
variation. The only assumption is that the background is stable in time
and that enough statistics is available. 

In case of a day/night - variation of the solar neutrino flux,
GENIUS would be sensitive to the LOW MSW solution of the
solar neutrino problem (compare Fig. \ref{daynight}).

GENIUS could be the {\sl{first detector to detect the solar pp neutrinos
in real time.}}
 Although this imposes
very strong purity restrictions for all the detector components, with a
liquid nitrogen shielding of 13 m in diameter and production of the Germanium 
detectors below ground, it should be feasible to achieve such a low
background level.
The advantages are the well understood detection technique
(ionization in a HPGe detector), the excellent energy resolution (1 keV 
at 300 keV), low energy threshold (about 11 keV) and the measurement
of the recoiling electrons in real time.

The good energy resolution for detecting the recoiling electrons
would allow for the first time to measure the 1.3 keV predicted 
shift of the average energy of the beryllium neutrino line.
This shift is a direct measure of the central temperature of the Sun
\cite{bah93}.

\section{The GENIUS experiment}

\subsection{Design, detection technique, threshold}

GENIUS will operate an array of 40 or 300 'naked' Ge crystals (natural 
Ge for WIMP-detection and measurement of the pp-flux, enriched
$^{76}$Ge for double beta decay searches) in a cylindrical vessel
filled with liquid nitrogen.
The basic idea of the GENIUS setup relies on the fact that most of the 
contributions of measured spectra in conventional low level detectors
result from the cryostat system and the shielding material. If these
can be eliminated reasonably, the sensitivity of the experiment
increases accordingly (linearly in the case of Dark Matter and solar
Neutrino search). It is therefore essential to keep away radioactive
materials and sources as far as possible from the detector itself.
In case of the GENIUS project this will be accomplished by the use of
liquid nitrogen as a cooling medium and as the shielding material against
natural radioactivity of the environment at
once. Liquid nitrogen has the advantage that it can be processed to a
very high purity through fractional distillation.
In this way practically all radioactive impurities in the material
near the detectors, which are known to produce the main part of the
radioactive background, are eliminated.

\subsubsection{Detector Size}
Due to its rather low density (0.8 g/cm$^3$), the nitrogen
shielding has to be several meters in diameter.
The dimensions of the vessel depend on the gamma- and
n-flux in the Gran Sasso Laboratory and on the intrinsic 
radiopurity of the steel vessel.
The required background conditions imply a tank size of 12m
diameter and height (13 m for solar neutrino detection),
if no other shielding is used in addition to the nitrogen (see Chapter 
3.1). The tank size could be reduced to some limited extent against the $\gamma$-radiation
and neutron flux from outside the tank, by replacing part of the outer 
nitrogen by other shielding material, e.g.  
lead (2m of nitrogen could be replaced by a layer of 10.8 cm Pb).
The minimal diameter of the nitrogen tank
would physically be determined by the contamination of the vessel 
and the shielding material and the distance between the lead and the
tank wall from the crystals. It has been calculated that a nitrogen
tank with $\sim$ 8m in diameter would be the minimum for this purpose.
This gives some flexibility to adapt the setup to the different sizes
of the halls in the Gran Sasso.
Of course the cost of the project would be increased in a
non-negligible  way by such a lead layer
(10.8~cm correspond to $\sim$1000 tons).
We have also considered
other alternative setups (see, e.g. \cite{Kla98i}).
We see, however, no other reasonable way
to accomplish the goal of
reducing the background to the required level than to use a tank of
the above dimensions.
These considerations show also that an intermediate size test setup
(as discussed in chapter 3.1) as a first step of the full setup seems
unreasonable. 

Figure \ref{confA} shows the design of the experiment,
which could be located, e.g., in the Gran Sasso Underground
Laboratory,  or in the WIPP laboratory.

\subsubsection{Detection Technique}
The proposed detection technique for GENIUS is ionization in a
Germanium detector. 
The detectors would be coaxial HPGe crystals of p-type, weighting
about 2.5 kg each. For p-type crystals, the outer contact is n$^+$ and 
the surface dead layer has a thickness of se\-veral hundred
micrometers. This prevents the detection of $\beta$-particles and
gamma rays of low energy from outside the crystals. 
The optimal working temperature is 77 K.
Besides the energy signal, the pulse shape of the interactions can be
recorded in view of background discrimination. 

The energy resolution of GENIUS would be about 0.3\%,  the energy 
threshold about 11 keV.

\subsection{Signals and signatures}

\subsubsection{Dark  Matter}

The  signal for a dark matter WIMP with mass between 20 GeV and  1 TeV is 
expected  in the energy region below 100 keV. The event rates for the 
neutralino as the lightest supersymmetric particle 
range in most SUSY models from   10$^{-2}$ to  10$^{2}$ events/kg y keV.
The low-energy spectrum in GENIUS is dominated by the 2$\nu \beta\beta$ signal 
from the decay of $^{76}$Ge.  For natural Germanium (7.8\% $^{76}$Ge) 
an event rate of 3$\times$10$^{-2}$ events/kg y keV from 2$\nu \beta\beta$ decay is expected.
Therefore, the 2$\nu \beta\beta$- signal has to be subtracted.
Another possibility is to make use of the predicted seasonal 
modulation of the WIMP flux.  Due to the motion of the Sun in the 
galactic halo and the Earth motion around the Sun, a flux variation of 
7\%  between two extremes is expected \cite{freese} . 

\subsubsection{Neutrinoless double beta decay}

The expected signature for the 0$\nu \beta\beta$ decay of $^{76}$Ge  
is a peak at the energy of 2038.56$\pm$0.32 keV \cite{hyka91}. The event rate for 1 ton 
of enriched $^{76}$Ge and an effective Majorana neutrino mass of 0.01 
eV is 0.3 events/yr.  Due to the good energy resolution of Ge 
detectors (typically better than 0.3 \% ), the 0$\nu \beta\beta$ signal 
is not affected by the 2$\nu \beta\beta$ spectrum.

\subsubsection{Solar neutrinos}

The reaction used to detect solar neutrinos is the elastic neutrino 
electron scattering: $\nu$ +  e$^- \rightarrow$ $\nu$ +  e$^-$.
The maximum electron recoil energy is 261 keV for the pp-neutrinos and 665 
keV for the  $^7$Be-neutrinos \cite{bahc89}. 
The detection rates for the pp and $^7$Be-fluxes, calculated for the 
SSM \cite{SSM}, are R$_{pp}\simeq$ 70 SNU and R$_{^7Be}\simeq$ 26 SNU (1 SNU = 
10$^{-36}$/(s target atom)). For one ton of natural (or enriched) 
 Ge (corresponding 
to 3$\times$ 10$^{29}$ electrons), the total rates are R$_{pp}\simeq$ 
1.8 events/day  and R$_{^7Be}\simeq$ 0.65 events/day, assuming the
detection of all electrons. This is about ten times higher than the rates 
in present radiochemical Gallium (GALLE and SAGE) experiments.
The event rates for full $\nu_e \rightarrow \nu_{\mu}$ conversion 
are 0.48 events/day for pp-neutrinos and 0.14 events/day for
$^7$Be-neutrinos.
GENIUS can measure only the energy distribution of the recoiling 
electrons, whereas the energy of the incoming neutrinos is not directly 
determined. However, due to the excellent energy resolution of the 
detectors and the 
difference in the elastic scattering cross section of electron and 
muon neutrinos, a comparison of the energy spectrum of recoiling 
electrons with the theoretical prediction of the SSM can be made.
Due to its relatively high counting rate, GENIUS would be able to test the 
LOW MSW
flavour conversion solution \cite{bahkra} via the day-night modulation of the
neutrino flux  and the vacuum-oscillation solution via the seasonal
flux variation.

\subsection{Technical study of detector operation}

To demonstrate the feasibility of operating Ge detectors in liquid nitrogen,
instead of in a vacuum--tight cryostat system \cite{kno89}, 
a first experiment has been successfully performed in the low level
laboratory in Heidelberg with one naked p--type Ge crystal immersed in a 50
l dewar \cite{KK3}. 
Already in this attempt we could not see any deterioration in the
detector performance relative to our conventionally operated detectors.

\begin{figure}[!t]
\hspace*{1.1cm}
\epsfxsize9cm
\epsfbox{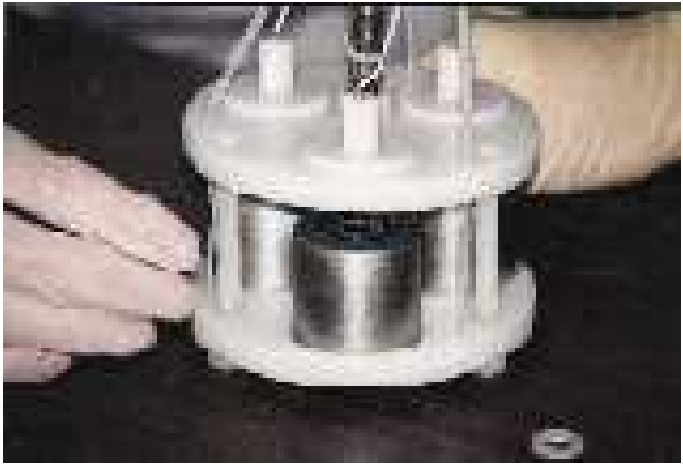}
\caption{The three--crystal holder-system with germanium crystals mounted 
shortly before cooling. Some crystal-to-FET cables can be seen.}
\label{canberra2}
\end{figure}

In a second phase the goal was to look for possible interferences between 
two or more naked Ge crystals, to test different cable lengths between
FETs and crystals and to design and test a preliminary holder system
of high molecular polyethylene.
We performed a technical study operating three germanium detectors on a
common plastic holder system inside liquid nitrogen \cite{Bau98}. 
All crystals were of p--type and weighted about 300 g each.

A picture of the three--crystal holder--system can be seen in 
figure ~\ref{canberra2}. 
Two thin polyethylene plates (1 cm thick) are used to fix the contacts 
to the crystals. The FETs are placed close to the liquid nitrogen 
surface but kept inside. Cables having three different lengths (2, 4 and 
6 m) connect the three crystals to their FETs.

The main purpose of the experiment was to test the behaviour of the
crystals in the low energy region: energy resolution, energy
threshold, crosstalk between the detectors and possible signs of microphonic events caused by nitrogen boiling. 
The general performance 
of the crystals is as stable as already seen with a single detector inside 
liquid nitrogen. We couldn't observe any cross talk using only p--type
detectors (same polarity for the HV-bias), since cross talk signals have the wrong 
polarity and are filtered by the amplifier.

\begin{figure}[!tt]
\epsfxsize12cm
\epsfbox{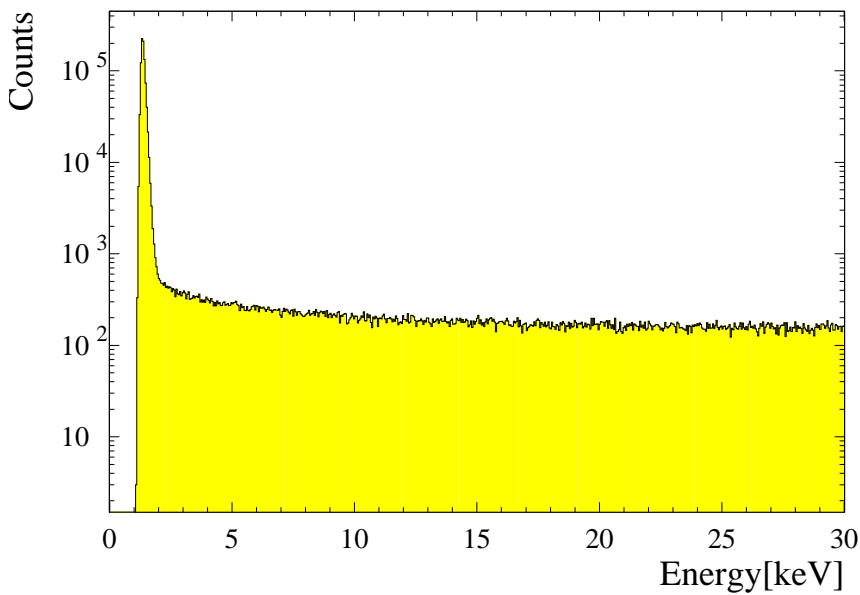}
\caption{Background spectrum of a naked, unshielded Ge crystal in liquid nitrogen. 
Note the low energy threshold of 2 keV of the detector.}
\label{topfgs_back}
\end{figure}

\begin{figure}[h]
\epsfxsize12cm
\epsfbox{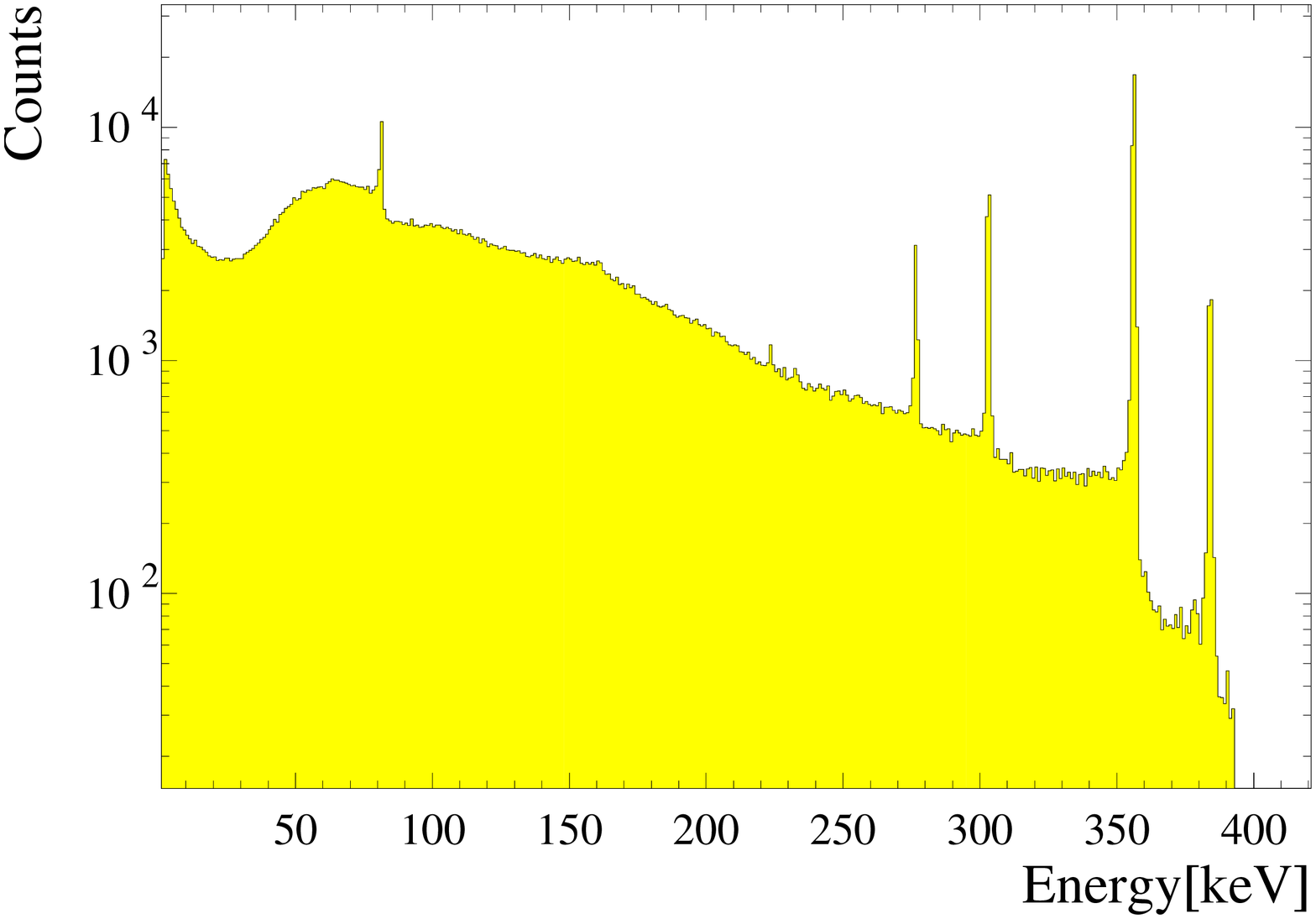}
\caption{Calibration $^{133}$Ba spectrum of a naked Ge crystal in
  liquid nitrogen.  
The energy resolution is 1 keV at 300 keV.}
\label{baspec}
\end{figure}

Figure \ref{topfgs_back} shows a background spectrum and
figure \ref{baspec} a $^{133}$Ba calibration spectrum of one of the naked Ge
detectors in liquid nitrogen. The
cable length between detector and FET was 6 m (winded up in loops). 
We achieved an energy  resolution of 1.0 keV at 300 keV and a threshold of 2 keV. 
No microphonic events due to nitrogen boiling beyond 2 keV could be detected. 
We conclude that the performance of the Ge detectors is as good (or even
better) as for conventionally operated crystals, even with 6 m cable
lengths between crystal and FET. 

\begin{figure}[h]
\epsfysize10cm
\hspace*{1.2cm}
\epsfbox{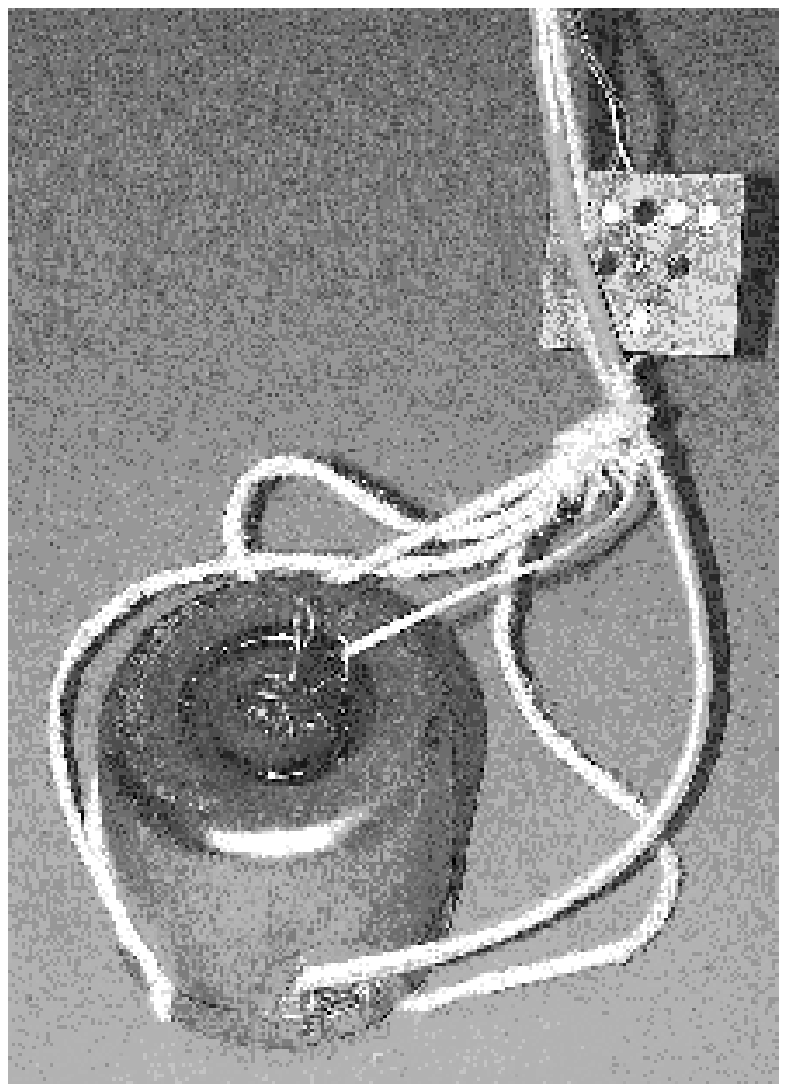}
\caption{A naked Ge crystal suspended on kevlar wires. Only 3 g of
  material in total (kevlar and electrical contacts) were used.}
\label{kevlar}
\end{figure}

\begin{figure}[h]
\epsfxsize12cm
\epsfbox{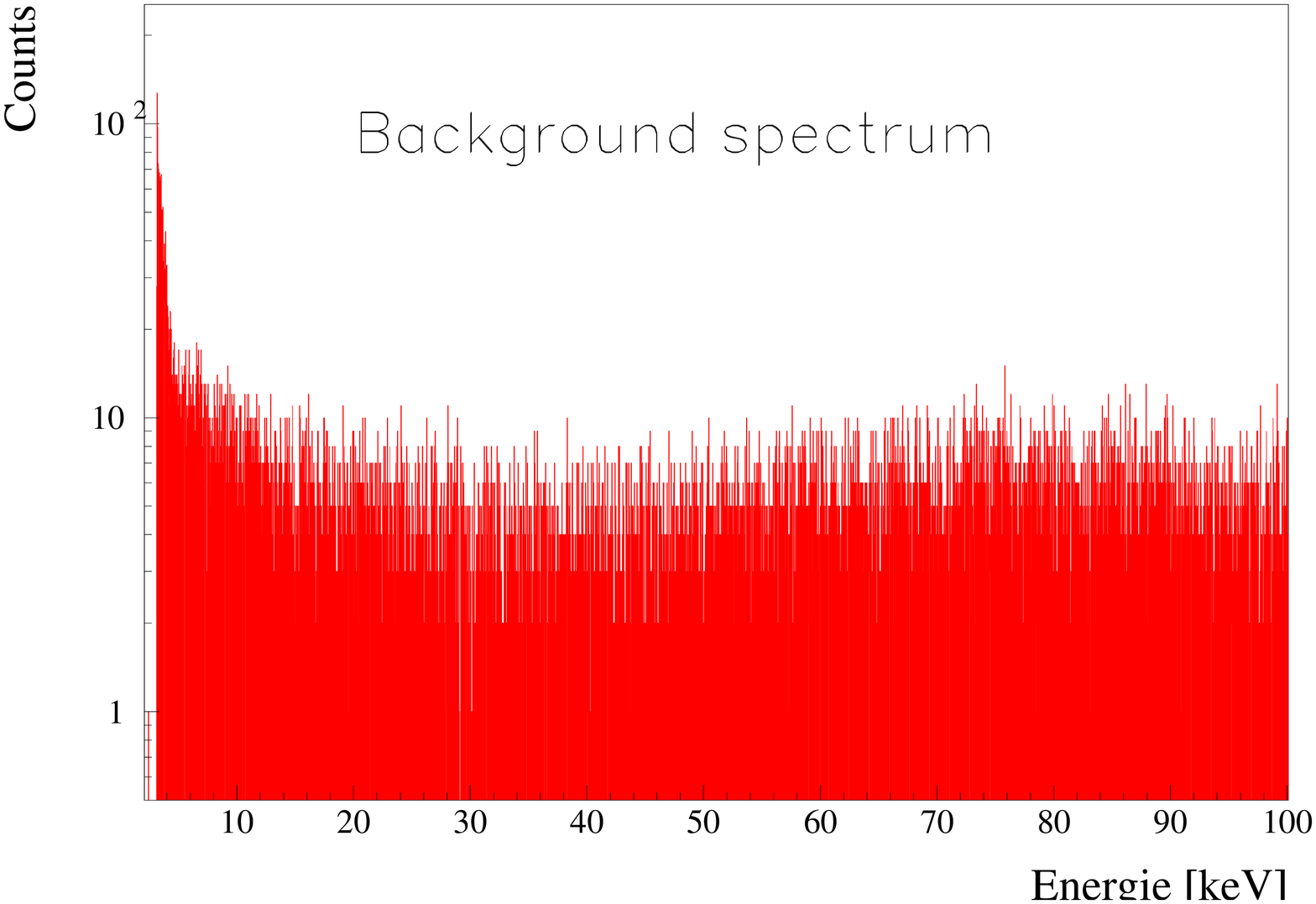}
\caption{Background spectrum of a naked, unshielded 400g Ge crystal in liquid nitrogen. 
The energy threshold is 2.5 keV.}
\label{back}
\end{figure}

\begin{figure}[h]
\epsfxsize12cm
\epsfbox{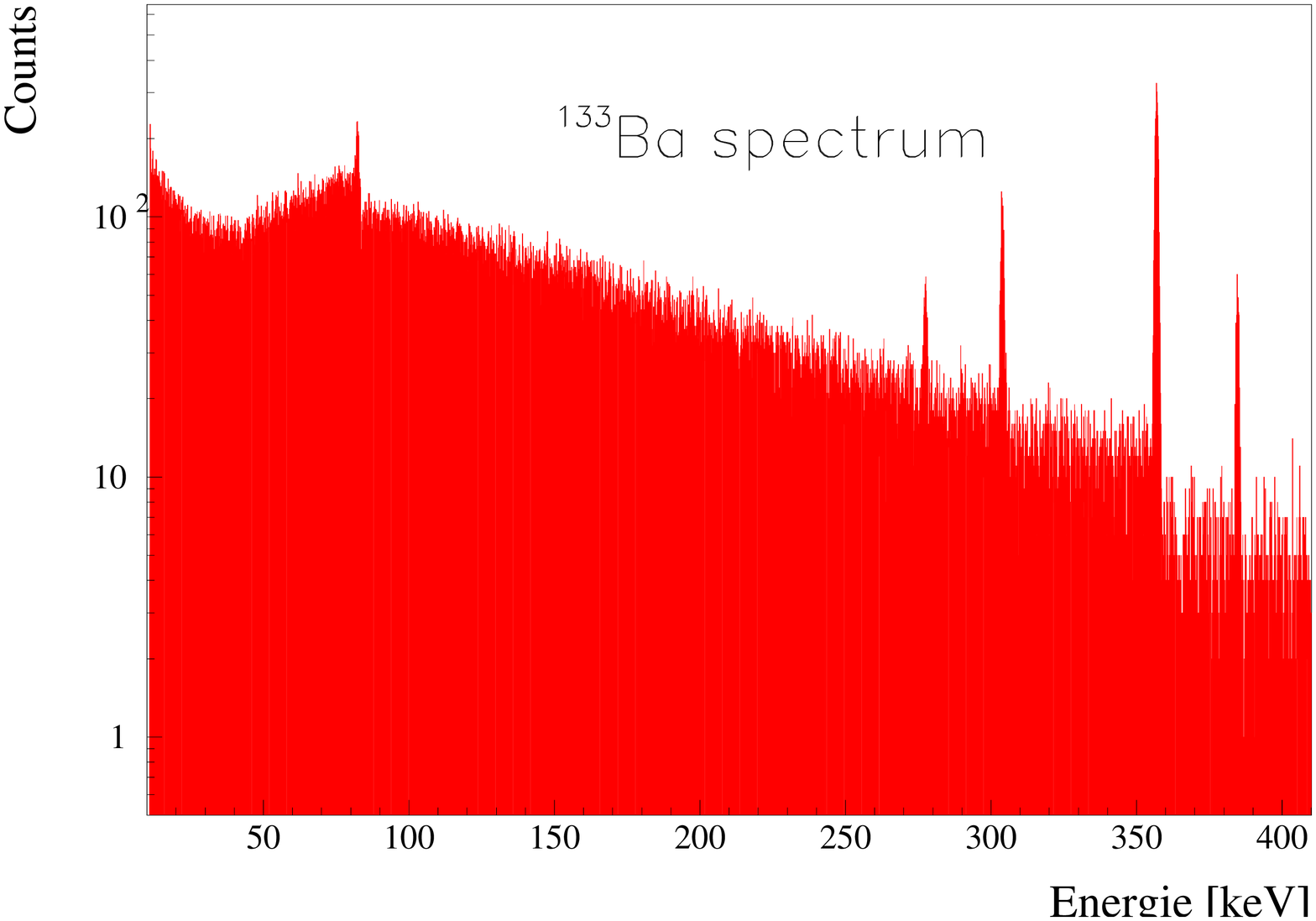}
\caption{Calibration $^{133}$Ba spectrum of a 400g naked Ge crystal in liquid nitrogen. 
The energy resolution is 1 keV at 300 keV.}
\label{ba}
\end{figure}

A third phase was dedicated to the optimization of the holder system
design (material minimization).
In figure \ref{kevlar}  a Ge crystal suspended on kevlar wires can be 
seen. The inner contact is fixed with a stainless steel spring, the 
outer contact with a thin stainless steel wire. Only 3 g of material 
in total (kevlar plus steel wires) were used. Figure \ref{back} and 
\ref{ba} show a background and a  $^{133}$Ba calibration spectrum of a 
400 g crystal in liquid nitrogen. An energy energy resolution of 1.2 keV at 300 keV and a 
threshold of 2.5 keV were achieved.

Currently we are measuring the radiopurity of kevlar and at the same time 
we are testing other possible materials for the holder system.

\section{Background simulations}

To study the expected background in the GENIUS experiment,
we performed detailed Monte Carlo
simulations and calculations of all the relevant background sources 
\cite{Bau98}.
The sources of background can be divided into external and internal ones.
External background is generated by events originating from outside
the liquid shielding, such as photons and neutrons from the Gran Sasso
rock, muon interactions and muon induced activities.
Internal background arises from residual impurities in the liquid
nitrogen, in the steel vessels, in the crystal holder system, in the Ge crystals
themselves and from activation of both liquid nitrogen and Ge crystals
at the Earths surface.

For the simulation of muon showers, the external photon flux
and the radioactive decay chains we used the GEANT3.21 package 
\cite{geant} extended for nuclear decays \cite{mueller}.
This version had already successfully been tested in establishing a
quantitative background model for the Heidelberg--Moscow experiment
\cite{HM97}.

We used the following detector geometry to perform the simulations.
The nitrogen shielding is given by a cylindrical geometry 
of variable diameter and height,
with the crystals positioned in its center. 
The vessel is surrounded by a 2 m thick polyethylene-foam isolation, which 
is held by two 2 mm thick steel layers 
(constructional data from Messer--Griesheim).
The simulated setup consists of natural Ge detectors 
integrated into a holder system of high molecular polyethylene. 


\subsection{Photon flux from the surroundings}

We simulated the influence of the photon flux with energies between
0 -- 3 MeV measured in hall C of the Gran Sasso laboratory
\cite{arpesella92}. This measurement is in good agreement with photon flux
calculations by the Borexino Collaboration \cite{borexprop}.  
The main contributions are given in table \ref{photonGS}.

\begin{table}[htb]
\begin{center}
\setcounter{mpfootnote}{0}
\vskip0.3cm
\centering
\begin{tabular}{lcc}
\hline
Isotope & Energy [keV]&Flux [m$^{-2}$d$^{-1}$]\\
\hline
$^{40}$K   & 1460    & 3.8$\times10^{7}$\\
\hline
$^{214}$Pb & 295.2   & 0.8$\times10^{7}$\\
$^{214}$Pb & 352     & 1.8$\times10^{7}$\\
$^{214}$Bi & 609.3   & 2.9$\times10^{7}$\\
$^{214}$Bi & 1120.3  & 1.4$\times10^{7}$\\
$^{214}$Bi & 1764.5  & 1.7$\times10^{7}$\\
\hline
$^{208}$Tl & 2614.5  & 1.35$\times10^{7}$\\
\hline
\end{tabular}
\caption{Simulated components of the gamma ray flux from natural
  radioactivity in the Gran Sasso Laboratory (from \cite{arpesella92}).}
\label{photonGS}
\end{center}
\end{table}

\subsubsection{Intermediate size detector (4$\times$4 m)}

The dimensions of the cylindrical tank
are dictated by the photon flux measured in the Gran Sasso 
laboratory and the radiopurity of the tank walls (made of stainless 
steel). 
With a diameter and a height of 4 m for the liquid shielding, the 
contribution from the surrounding gammas is about 70 counts/(kg y keV)
in the energy region 11 -- 100 keV. 
This is almost 4 orders of magnitude higher than the goal of 10$^{-2}$ 
events/(kg y keV) of the GENIUS experiment. 
An alternative would be to use an additional outer water shielding. 
In this case however, the
limitations of the tank dimensions are given by the radiopurity of the 
steel walls. Assuming an U/Th contamination of  
5$\times$10$^{-9}$ g/g for the steel, as measured by the BOREXINO
collaboration, a count rate of  about 1.5 counts/(kg y keV) from this
component is achieved. 
This again is too high by more than 2 orders of magnitude.
For the assumed steel radiopurity of 5$\times$10$^{-9}$ g/g, the
minimal allowed tank 
dimensions are $\sim$ 8$\times$8 m (maximal allowed countrate due to
this component was assumed to be 0.3$\times10^{-3}$ counts/(kg y keV)
in the energy 
region between 11~keV and 100~keV) or, the other way round, for a tank 
size of 4$\times$4 m, an unrealistic contamination level of the order of 
10$^{-11}$ g/g for the steel is required. 
It has been suggested to set up an intermediate size detector
in order to prove the predictions of the concept in a first step,
i.e. to prove that no unexpected background components appear, which
would dramatically decrease the sensitivity with respect to the
predictions. However, we see no sense in such an intermediate step, 
since its costs would be only
negligibly smaller than that of the full setup - increasing the total
costs of the project by about a factor of two.

This means a test of the
proposed setup by a smaller tank as a first step seems unreasonable.

\subsubsection{Full size detector}

For GENIUS as a dark matter and neutrinoless double beta decay 
detector, a  12$\times$12 m tank is suggested.
The obtained count
rate from the external gammas in the energy region 11 -- 100 keV is 
4$\times$10$^{-3}$ counts/(kg y keV).
However, to measure the solar pp- and $^7$Be neutrino flux, a tank 
size of 13$\times$13 m is needed (with a count rate from external
gammas of  
9$\times$10$^{-4}$ counts/(kg y keV) below 260 keV).

\subsection{Neutron flux from the surroundings}

We simulated the measured neutron flux \cite{arp} in the Gran Sasso
laboratory. 
The 2 m polyethylene foam isolation ($\rho$ = 0.03 g cm$^{-3}$) around 
the nitrogen tank reduces
the neutron flux for energies below 1 keV by more than 5 orders of
magnitude. Only about 3\% of neutrons with energies between 1 keV
and 2.5 MeV will pass the polyethylene isolation, whereas for energies
between 2.5 and 15 MeV the overall flux is reduced by about 40\%.
The neutron flux reaching the tank can be reduced by another
two orders of magnitude by doping the polyethylene foam isolation with
about 1.4 t of boron.  
The flux of the $^7$Li deexcitation gamma rays from
the reaction  n~+~$^{10}$B$\rightarrow~\alpha$~+~$^7$Li*, with an 
energy of 0.48 MeV,  would be too low
to reach the inner part of the liquid shielding.
After the first meter of liquid nitrogen the total neutron flux is reduced
by another 4--5 orders of magnitude, therefore we simulated the neutron capture
reactions randomly distributed in the first meter of the nitrogen shielding.   

With the conservative assumption that all neutrons reaching the nitrogen are thermalized and  
captured by the reactions $^{14}$N(n,p)$^{14}$C$^{*}$ and 
$^{14}$N(n,$\gamma$)$^{15}$N$^{*}$, 
a total of 4.4$\times$10$^{7}$ 
neutron capture reactions per year have to be taken into account. 
The relevant contribution to the background comes from the
deexcitation of the $^{14}$C$^{*}$ and $^{15}$N$^{*}$ nuclei.
The contribution of the $\beta$--decay of $^{14}$C nuclei in the liquid nitrogen 
is negligible, since only low energy electrons (E$_{\beta max}$ = 156
keV) are emitted and the decay probability is very low due to the long 
half life (10$^{-4}$ per year).

Using the assumptions of a 12$\times$12m tank and a 2m isolation
around it, 
the mean count rate in the low--energy region due to neutron capture
reactions would be about 4$\times10^{-4}$ counts/(kg y keV). 

\subsection{Activities induced by muons}

The muon flux in the Gran Sasso laboratory was measured to be
$\phi$$_{\mu}$=2.3$\times$10$^{-4}$ m$^{-2}$s$^{-1}$ with a mean
energy of $\bar{E}_{\mu}$=200 GeV \cite{arpesella92}.

We simulated the effect of muon-induced showers in the liquid
nitrogen. With the aid of a muon veto in form of scintillators or gas counters
on top of the tank, the total induced background can be drastically reduced.
Here we assumed a veto efficiency of 96\% as measured in a more
shallow laboratory \cite{heusser91}. The count rate due to muon
induced showers in the low--energy
region is about 2$\times$10$^{-3}$ counts/(kg y keV).
This can be further improved using the anticoincidence power of the
Ge detectors  among each other. For example, for 300 Ge detectors (1 
ton), the count rate reduces to 7.2$\times$10$^{-6}$ counts/(kg y 
keV) in the energy region below 260 keV.

Besides muon showers, we have to consider muon-induced nuclear 
disintegration and interactions due to secondary neutrons generated in the above
reactions. 

\subsubsection{Neutrons generated by cosmic muons}

The muon-induced production of neutrons can be approximated by
A$_{n} \sim$ 3.2$\times$10$^{-4}$ (g$^{-1}$ cm$^{2}$), due to the
$<E>^{0.75}$ dependence of the number of neutrons on the mean muon 
energy \cite{myonneutrons}. 
With the geometry of the tank h = 12 m, r = 6 m, the density of
nitrogen $\rho$ = 0.808 g/cm$^{3}$ and the cited flux, a mean
production rate of $\phi _{n\mu}$ = 2.5$\times$10$^{5}$ neutrons/year in
the whole vessel is obtained.
Table \ref{neutrons} gives the neutron-induced reactions
in the liquid nitrogen for neutron energies $<$~20~MeV (based on
all reactions found in \cite{McLane88}).

\begin{table}[htb]
\vskip0.3cm
\centering
\begin{tabular}{lcc}
\hline
Reaction & T$_{1/2}$ of the product& Decay energy\\
\hline
$^{14}$N(n,p)$^{14}$C & T$_{1/2}$=5.7$\times$10$^3$y& E$_{{\beta}^-}$=0.16 MeV\\
$^{14}$N(n,$\gamma$)$^{15}$N & stable& \\
$^{14}$N(n,2n)$^{13}$N & T$_{1/2}$= 9.96 m& E$_{{\beta}^+}$=1.2 MeV\\
$^{14}$N(n,$\alpha$)$^{11}$B  & stable&\\
$^{14}$N(n,t)$^{12}$C & stable&\\
$^{14}$N(n,2$\alpha$)$^{7}$Li &stable&\\
\hline
\end{tabular}
\caption{Neutron interactions in the liquid nitrogen for neutron
  energies $<$ 20 MeV.}
\label{neutrons}
\end{table}

All of the produced nuclides are stable or short-lived with the exception
of $^{14}$C and $^{13}$N. The contribution of gammas from the excited
$^{14}$C$^*$ nucleus corresponds to 10$^{-3}$ counts/(kg y keV) between
0 -- 100 keV. The contribution from the $\beta^{-}$-- particles
with E$_{max}$ = 156 keV is negligible due to the low decay probability
of  $^{14}$C. The production rate of $^{13}$N is 
1$\times$10$^6$ atoms per year in the whole tank. 
From 10$^6$ simulated positrons with E$_{max}$ = 1.2 MeV ($\beta
^+$-decay), corresponding to an 
exposure of about 1 year, only one event could be observed in the detectors. 
Therefore, the contribution of $^{13}$N to the background will be negligible.

In the  Germanium material, 2.3$\times$10$^2$ neutrons/(y ton) due to 
muon interactions are produced. For the low energy region the 
most significant reaction is the $^{70}$Ge(n,$\gamma$)$^{71}$Ge capture
reaction. $^{71}$Ge decays through EC (100\%) with T$_{1/2}$ = 11.43 d
and Q$_{\rm EC}$ = 229.4 keV \cite{firestone} and can not be discriminated by the
anticoincidence method.  The simulation of this decay yields 
5$\times$10$^{-4}$ counts/(kg y keV) in the energy region below 260 keV.

\subsubsection{Negative muon capture}

A negative muon stopped in the liquid shielding can be captured by a 
nitrogen nucleus, leading to one of the reactions that are listed in 
table ~\ref{myonspall}.
Estimations of the number of stopping muons in the nitrogen tank
\cite{bergamasco82,gaisser,lohman} lead to 86 stopped muons per day 
(for a 12$\times$12 m tank).
The rates of decaying and captured muons are shown in table
\ref{mcapture}.

\begin{table}[htb]
\vskip0.3cm
\centering
\begin{tabular}{lccc}
\hline
Reaction & T$_{1/2}$ & Decay energy & Rate [y$^{-1}$]\\
\hline
$^{14}$N($\mu$,$\nu_{\mu}$)$^{14}$C & T$_{1/2}$=5.7$\times$10$^4$y& E$_{{\beta}^-}$=0.16 MeV&584\\
$^{14}$N($\mu$,$\nu_{\mu}\alpha$)$^{10}$Be & T$_{1/2}$=1.6$\times$10$^{10}$y& E$_{{\beta}^-}$=0.6 MeV&29\\
$^{14}$N($\mu$,$\nu_{\mu}$p)$^{13}$B & T$_{1/2}$=17.33ms& E$_{{\beta}^-}$=13.4 MeV&116\\
$^{14}$N($\mu$,$\nu_{\mu}$n)$^{13}$C &  stable&&3798\\
$^{14}$N($\mu$,$\nu_{\mu}\alpha$n)$^{9}$Be & stable&&17\\
$^{14}$N($\mu$,$\nu_{\mu}\alpha$p)$^{9}$Li & T$_{1/2}$=178ms& E$_{{\beta}^-}$=13.6 MeV&0.6\\
$^{14}$N($\mu$,$\nu_{\mu}$2n)$^{12}$C & stable&&1168\\
$^{14}$N($\mu$,$\nu_{\mu}$3n)$^{11}$C & T$_{1/2}$=20.38m&
E$_{{\beta}^-}$=13.4 MeV&292\\
$^{14}$N($\mu$,$\nu_{\mu}$4n)$^{10}$C & T$_{1/2}$=19.3s& E$_{{\beta}^+}$=1.9 MeV&117\\
\hline
\end{tabular}
\caption{Spallation reactions from muon capture.}
\label{myonspall}
\end{table}

\begin{table}[htb]
\begin{center}
\setcounter{mpfootnote}{0}
\vskip0.3cm
\centering
\begin{tabular}{lcc}
\hline
Muon flux & 124 h$^{-1}$&\\
\hline
Stopped muons & 86 d$^{-1}$&\\
\hline
Decaying muons& $\mu ^+$ & $\mu ^-$\\
& 50  d$^{-1}$& 20 d$^{-1}$\\
\hline
Captured muons &$\mu ^+$ & $\mu ^-$\\
& 0 & 16 d$^{-1}$\\
\hline
\end{tabular}
\caption{Muon flux, stopped, captured and decaying muons in the
  nitrogen shielding of the Genius detector (for a 12$\times$12 m tank).}
\label{mcapture}
\end{center}
\end{table}

The derived production rates \cite{charalambus} for the various isotopes
are listed in table \ref{myonspall}.
Only the isotopes 
$^{14}$C, $^{10}$Be, $^{11}$C and $^{10}$C
can not be discriminated by 
muon anticoincidence, since their individual lifetime is too long.
$^{14}$C and $^{10}$Be will not be seen in our detector due to
their very low decay probabilities (10$^{-4}$ and 10$^{-10}$ per year) 
and low production rates.
The contribution of $^{10}$C and $^{11}$C,
with a production rate of 117 atoms/year and 292 atoms/year,
respectively, in the whole nitrogen tank, will be negligible.
The gamma rays from the excited nuclei produced in all the reactions can
be discriminated by anticoincidence with the muon shielding on the top 
of the tank.

\subsubsection{Inelastic muon scattering}

Another way of producing radioactive isotopes in the liquid
shielding are electromagnetic nuclear reactions of muons through
inelastic scattering off nitrogen nuclei:
$\mu$~+~N~$\rightarrow$~$\mu'~$~+~X$^*$. The only resulting isotopes
with half lifes $>$ 1s 
are $^{14}$N($\gamma$,n)$^{13}$N, with T$_{1/2}$=9.96 m and 
$^{14}$N($\gamma$,tn)$^{10}$C, with T$_{1/2}$=19.3 s. 
The production rate per day for one isotope can be written as
\cite{OConnell88}:
R(d$^{-1}$)=
6$\times$10$^{-2}\phi_{\mu}$(d$^{-1}$m$^{-2}$)N$_{T}$(kt)$\sigma_{\mu}$($\mu$b)/A,
where $\phi _{\mu}$ is the flux of muons on the detector, N$_{T}$ is the
number of target nuclei, $\sigma_{\mu}$ the reaction cross section and 
A the atomic weight of the target nucleus. For our detector this
yields R(y$^{-1}$)= 45$\times$$\sigma_{\mu}$($\mu$b).
With typical reaction cross sections of a few hundred $\mu$b \cite{OConnell88,Napoli73},
we obtain a production rate of (5--10)$\times$10$^3$ atoms per year.
A simulation of an activation time of ten years for both isotopes
yields negligible count rates in comparison to contributions
from other background components.


\subsection
[Intrinsic impurities]
{Intrinsic impurities in the nitrogen shielding, Ge crystals,
  holder system and steel vessel}

The assumed impurity levels for the liquid nitrogen 
are listed in table ~\ref{forderung}.
For the $^{238}$U and $^{232}$Th decay chains they have been
measured by the Borexino collaboration \cite{borex} for their liquid scintillator. 
Due to the very high cleaning efficiency of fractional distillation, 
it is conservative to assume
that these requirements will also be fulfilled for liquid nitrogen.
The $^{238}$U and $^{232}$Th decay chains were simulated under the assumption that 
the chains are in secular equilibrium. 
The count rate due to $^{238}$U, $^{232}$Th and $^{40}$K
contaminations of the liquid nitrogen is about 1.2$\times 10^{-3}$ in the
energy region below 100 keV.

\begin{table}[htb]
\begin{center}
\begin{tabular}{lcc}
\hline
Source & Radionuclide & Purity \\
\hline
Nitrogen & $^{238}$U  & 3.5$\times$10$^{-16}$g/g \\
           & $^{232}$Th & 4.4$\times$10$^{-16}$g/g \\
           & $^{40}$K   & 1$\times$10$^{-15}$g/g \\
\hline
Steel vessel  & U/Th       & 5$\times$10$^{-9}$g/g \\
\hline
\end{tabular}
\caption{Assumed contamination levels  for
the liquid nitrogen and steel vessel.  }
\label{forderung}
\end{center}
\end{table}

New measurements of the $^{222}$Rn contamination of freshly
produced liquid nitrogen yield 325 $\mu$Bq/m$^{3}$ \cite{rau99}.
After about a month it is reduced to about 3~$\mu$Bq/m$^{3}$ (T$_{1/2}$ =
3.8 days). 
Such a level could be maintained if the evaporated nitrogen is always
replaced by Rn--pure nitrogen, previously stored in an underground
facility or if a nitrogen cleaning system is used. 
It is planned to install a nitrogen recycling device (through condensation) 
inside the tank. This would reduce the Rn contamination to a
negligible level.
Surface emanations are reduced to a negligible level for cooled
surfaces in direct contact with the liquid nitrogen.

The mean count rate
from the contamination of $\;^{222}$Rn in the 
region below 100 keV is  3$\times$10$^{-4}$ counts/(kg y keV) assuming an
activity of 3 $\mu$Bq/m$^{3}$ in the liquid nitrogen. 

For the intrinsic impurity concentration in Ge crystals we can
give only upper limits from measurements with the detectors of the
Heidelberg--Moscow experiment. 
We see a clear $\alpha$--peak in two of the enriched detectors at 5.305
MeV, and an indication for the same peak in two other detectors. It
originates from the decay of $^{210}$Po (which decays with 99\%
through an $\alpha$--decay to $^{206}$Pb) and is a sign for a $^{210}$Pb
contamination of the detectors. However, it is very unlikely that the
contamination is located inside the Ge-crystals, most probably it is located on
the crystals surface at the inner contact.

Using three Ge detectors, we
derive an upper limit at 90\% CL (after 19 kg y counting statistics)
of 1.8$\times$10$^{-15}$g/g for $^{238}$U and 5.7$\times$10$^{-15}$g/g 
for $^{232}$Th. 
Assuming these impurity concentrations throughout the whole Ge
detector volumes, our
simulations yield a count rate of about 10$^{-2}$ counts/(kg y keV)
for both $^{238}$U and $^{232}$Th decay chains.
It is however secure to assume that these upper
limits are very conservative and that the true contamination
of HPGe is much lower. Special attention will have to be paid in order
to avoid surface contaminations of the crystals.     

An important  factor in the background spectrum 
is the effect of the holder-system. 
For the simulation we assumed the possibility to obtain a polyethylene
with an impurity concentration of 10$^{-13}$g/g for the U/Th decay
chains (this is a factor of 100 worse than the values reached at present for the
organic liquid-scintillator by the Borexino collaboration \cite{borex}). 
Encouraging are the results already achieved by the SNO experiment
\cite{sno}, which developed an acrylic with current limits on 
$^{232}$Th and $^{238}$U contamination of 10$^{-12}$g/g.
Since it is not yet sure that such a low contamination level will be 
reached for polyethylene, we are currently testing also other materials.

Assuming the above impurity level with the simulated geometry (130 g 
of material per detector) a count
rate of $\sim$ 
8$\times$10$^{-4}$ counts/(kg y keV) in  the energy region below 100
keV from this component is reached. 
This result will be further improved by using the new 
developed holder design with a minimized amount of material. 
In case of using about 10 g of material per detector, a contamination
level of 10$^{-12}$g/g for the holder system material would suffice.

For the steel vessel an impurity concentration of 5$\times$10$^{-9}$
g/g for U/Th was assumed (as measured in \cite{borexprop}). 
The contribution in the energy region 0 -- 100 keV is
1.5 $\times$ 10$^{-5}$ counts/(kg y keV). Assuming an equal
contamination for the polyethylene foam isolation as for steel, 
the contribution of both materials to the background is negligible (for a 
12$\times$12 m tank). 

\subsection{Cosmic activation of the germanium crystals}

We have estimated the cosmogenic production rates of radioisotopes 
in the germanium crystals with the $\Sigma$ programme \cite{JensB}.
The programme was developed to calculate cosmogenic activations of
natural germanium, enriched germanium and copper. 
It was demonstrated that it can  reproduce the measured cosmogenic
activity in the Heidelberg--Moscow experiment \cite{HM97} 
within about a factor of two \cite{BerndM}.  

Assuming a production plus transportation time of 10 days at sea level
for the natural Ge detectors (with the exception of $^{68}$Ge, where the 
saturation activity is assumed), and a deactivation time of three years,
we obtain the radioisotope concentrations listed in table \ref{ge_cosmo}.
All other produced radionuclides have much smaller activities due to their shorter
half lifes. The required short production time has been guaranteed by detector production companies.

\begin{table}[htb]
\begin{center}
\setcounter{mpfootnote}{0}
\vskip0.3cm
\centering
\begin{tabular}{lccc}
\hline
Isotope & Decay mode,  & Energy [keV]& A\\
        &   T$_{1/2}$                   &              & $\mu$Bq/ \\
        &              &                               &kg\\
\hline
$^{49}$V   &  EC, 330 d    & no $\gamma$, E (K$_{\alpha}$ $^{49}$Ti)=4.5 & 0.17\\
$^{54}$Mn  &  EC, 312.2 d    &E$_{\gamma}$=1377.1 E (K$_{\alpha}$ $^{54}$Cr)=5.99   & 0.20\\
$^{55}$Fe  &  EC, 2.73 a    & no $\gamma$, E (K$_{\alpha}$ $^{55}$Mn)=5.9   & 0.31\\
$^{57}$Co  &  EC, 271.3 d  & 136.5 (99.82\%)E (K$_{\alpha}$ $^{57}$Fe)=7.1  & 0.18\\
$^{60}$Co  &  $\beta ^-$, 5.27 a & 318 (99.88\%), E$_{\gamma 1,2}$=1173.24, 1332.5&0.18\\
$^{63}$Ni & $\beta^-$, 100.1 a & E$_{\beta^-}$= 66.95 no ${\gamma}$ &0.01 \\
$^{65}$Zn & EC, 244.3 d & E$_{\gamma}$=1115.55 (50.6\%),E (K$_{\alpha}$ $^{65}$Cu)=8.9 & 1.14\\
$^{68}$Ge & EC, 288 d & E (K$_{\alpha}$ $^{68}$Ga)=10.37,
Q$_{EC}$($^{68}$Ga)= 2921 &101.4\\
\hline
\end{tabular}
\caption{Cosmogenic produced isotopes in the Ge crystals for an
  exposure time at sea level of 10 days and for 3 years deactivation
  time (for $^{68}$Ge an initial saturation activity was assumed).}
\label{ge_cosmo}
\end{center}
\end{table}

The count rate below 11 keV is dominated by X--rays from 
the decays of $^{68}$Ge, $^{49}$V,     
$^{55}$Fe and $^{65}$Zn (see table \ref{ge_cosmo}).        
Due to their strong contribution, the energy threshold of GENIUS would be
at 11 keV, which is still acceptable (as can be seen from figure \ref{limits}).

Between 11 keV and 70 keV the contribution from $^{63}$Ni dominates
due to the low Q--value (66.95 keV) of the $\beta^-$--decay. 

$^{68}$Ge plays a special role. Since it can not be
extracted by zone melting like all other, non--germanium isotopes,
the starting activity would be in equilibrium with the production
rate. 
After 3 years of deactivation below ground, the activity is  about 
100 $\mu$Bq/kg. 
With a half--life of 288 d it will dominate the other background
components (with about 4$\times$10$^{-2}$ events/kg y keV below 100
keV). However, with such a background level one already would be able to 
test a significant part of the SUSY predicted parameter space for
neutralinos (see Fig. \ref{limits}).
After 5 years of deactivation the activity reduces to 14.7$\mu$Bq/kg
(5.3$\times$10$^{-3}$ events/kg y keV below 100 keV). 
At this point one can in addition reduce the contribution of $^{68}$Ga by 
applying a time analysis method. $^{68}$Ge decays through EC (100\%) 
to the ground state of  $^{68}$Ga. The resulting line at 10.37 keV 
(88\%  K-capture) would be observed in GENIUS with an energy 
resolution of about 1 keV. Thus one could use an event with an energy 
of 10.37 keV as a trigger for subsequent events in the same detector 
during a few half-lives of $^{68}$Ga (T$_{1/2}$ = 68.1 m), 
removing up to 88\%  of the $^{68}$Ga  caused events.
With about 2.5 events per day and detector one would loose only
30\% of the measuring time.

Another solution could be to process the germanium ore directly
in an underground facility or to use high purity germanium which
has already been stored for several years in an underground laboratory.

\begin{figure}[h]
\epsfxsize=12cm
\epsfbox{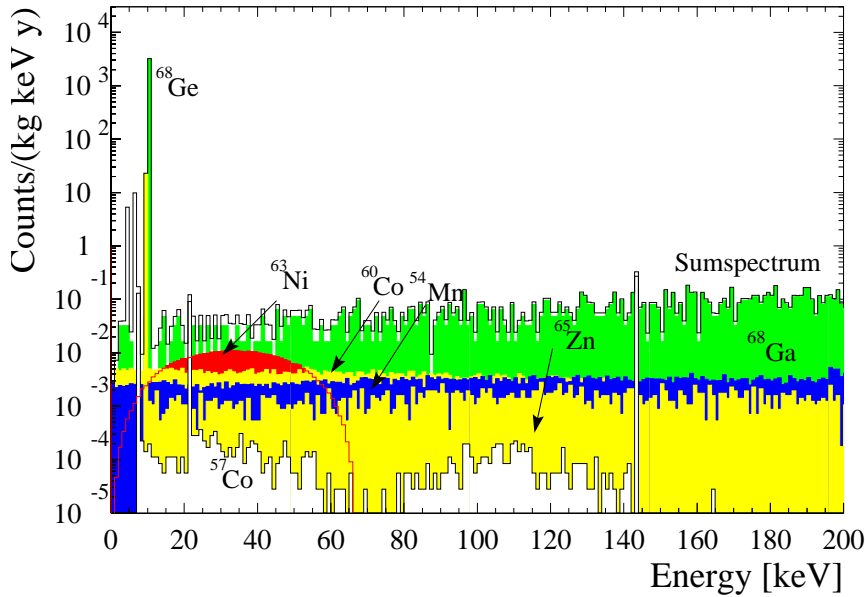}
\caption{Background originating from cosmic activation of the Ge
  crystals at sea level with 10 days exposure and 3 years
  deactivation. For 
$^{68}$Ge an initial saturation activity was assumed.} 
\label{cosmo}
\end{figure}

Figure \ref{cosmo} shows the sum and the single contributions from the
different isotopes.

The sum of all contributions from the cosmogenic activation of the Ge
crystals is 5.2$\times$10$^{-2}$ counts/(kg y keV) between 11 -- 100
keV, for an activation time of 10 days at the Earths surface and a
deactivation time of three years.  Under the mentioned conditions, after 5 years of deactivation the background would be reduced to less than 1$\times$10$^{-2}$ counts/(kg y keV), allowing for the sensitivity for dark matter shown in Fig. \ref{limits}.  
Since the cosmogenic activation of the Ge crystals  will be the dominant background component in the
low-energy region, special attention to short crystal exposure times at
sea level is essential.    
The best solution would be to produce the
detectors below ground and to apply strong shielding during the 
transportation.    

The two--neutrino accompanied double beta decay of $^{76}$Ge is not
negligible in spite of the low abundance (7.8\%) of this isotope in
natural germanium. The contribution to the background after three
years of measurement is shown in figure \ref{specall} . Due to the
already high statistics reached in the Heidelberg--Moscow experiment
\cite{HM97}, the half life and spectral form of the decay are well
known and a subtraction of this part raises no difficulties. 
The statistical error of the subtraction is not shown in fig.~\ref{limits}.

The cosmogenic activation of the Ge crystals in the Gran Sasso laboratory
is negligible in comparison to the assumed activation scenario at sea level. 
 
\subsection{Cosmic activation of the nitrogen at sea level}

An estimation of the production rates of long--lived
isotopes in the nitrogen at sea level reveals the importance
of $^7$Be, $^{10}$Be, $^{14}$C and $^3$H.
The neutron flux at sea level is  8.2$\times$10$^{-3}$cm$^{-2}$s$^{-1}$ 
for neutron energies between 80 MeV and 300 MeV \cite{allkofer}.
Since we did not find measurements of the cross sections of neutron-
induced spallation reactions in nitrogen, 
we assumed that at high neutron energies
(10$^2$--10$^4$ MeV) the cross sections are similar to the proton-induced ones.
For the reaction $^{14}$N(n,t$\alpha$n)$^7$Be the cross section 
is (9.0$\pm$2.1) mb at E$_p$ = 450 MeV, (9.3$\pm$2.1) mb at
E$_p$ = 3000 MeV by \cite{reyss} and (13.3$\pm$1.3) mb at E$_p$ = 1600 MeV
by \cite{michel}.
For the reaction $^{14}$N(n,$\alpha$p)$^{10}$Be
the cross sections are (1.5$\pm$0.4) mb at E$_p$ = 450 MeV,  
(2.6$\pm$0.6) mb at E$_p$ = 3000 MeV \cite{reyss} and (1.75$\pm$0.11)
mb at E$_p$ = 1600 MeV \cite{michel}.
 
Taking 10 mb for the $^7$Be channel we obtain a production 
rate of 3.3$\times$10$^9$d$^{-1}$ in the whole tank. 
This corresponds with a realistic 10 days sea level exposure after
production by fractional distillation to 4$\times$10$^8$ decays per day.
The simulation of this activity yields a count rate of about
10 events/(kg y keV) in the energy region between 0 -- 100 keV.
This is three orders of magnitude higher than the required level.
However, a large fraction of $^7$Be will removed
from the liquid nitrogen at the cleaning process for Rn and in
addition by underground storage (T$_{1/2}$=53.3 d) the contribution of 
$^7$Be will be  reduced to less than 10$^{-2}$ events/(kg y keV).

For $^{10}$Be, with $\sigma$= 2 mb, the production rate is
6.6$\times$10$^{8}$d$^{-1}$, which is negligible due to the long
half life of T$_{1/2}$=1.6$\times$10$^6$ y.

Tritium in nitrogen can be produced in the following reactions: 
$^{14}$N(n,t)$^{12}$C, $^{14}$N(n,t2$\alpha$)$^{4}$He,
$^{14}$N(n,t$\alpha$n)$^{7}$Be and $^{14}$N(n,tn)$^{11}$C.
The cross section for the production by $^{14}$N(n,t)$^{12}$C was
measured to be 40 mb \cite{Kincaid}. For a rough estimation, we
assumed the same cross sections for the other reactions as for the
production of $^{7}$Be to be 10 mb. The total production rate of tritium
corresponds to 2.3$\times$10$^{10}$d$^{-1}$. With T$_{1/2}$=12.33 y, the
activity after 10 days exposure at sea level would be 
3.5$\times$10$^{7}$d$^{-1}$. We simulated 10$^{10}$ decays randomly
distributed in the nitrogen tank. No events were
detected mainly due to the absorption in the dead layer of the p--type 
Ge detectors.

The muon flux at sea level is 1.6$\times$10$^7$m$^{-2}$d$^{-1}$.
The only long--lived isotopes which are produced by inelastic muon
scattering are $^{13}$N and $^{10}$C, with a production rate of about  
3.7$\times$10$^7$ atoms/days (taking $\sigma$ = 500 $\mu$b for both reactions). 
However, $^{13}$N and $^{10}$C are of no relevance due to the
short half lifes of 9.96 m and 19.3 s, respectively. 

The isotopes produced through negative muon capture with half lifes $>$
1s are  $^{14}$C, $^{10}$Be, $^{11}$C and $^{10}$C (see also table
\ref{myonspall}). Again, the number of $^{11}$C and $^{10}$C  
isotopes are soon reduced to a negligible level due to their short
half lifes. The production rate for the whole tank for $^{14}$C is
8$\times$10$^{6}$d$^{-1}$ and 4$\times$10$^{5}$d$^{-1}$ for
$^{10}$Be, which have to be added to the production rates by neutron
capture or spallation reactions.

For the production of $^{14}$C due to the $^{14}$N(n,p)$^{14}$C capture reaction,
three neutron sources at sea level are relevant.
The flux of secondary cosmic ray neutrons with energies between a few
keV and 20 MeV is 
about 2$\times$10$^{-2}$cm$^{-2}$s$^{-1}$\cite{allkofer}. These neutrons penetrate
the wall of the transportation tank and are captured in the liquid 
nitrogen. For a tank surface of 678 m$^2$, about
1.3$\times$10$^5$ s$^{-1}$ neutrons are expected.
The second component are neutrons produced in fast neutron 
spallation reactions in the liquid nitrogen. The production rate of these
neutrons is 2$\times$10$^{4}$ s$^{-1}$ in the nitrogen tank
\cite{lal}. The third component are neutrons produced in muon
reactions, which correspond to 0.85$\times$10$^{3}$ s$^{-1}$ \cite{lal}.
Thus the total flux at sea level is about 1.5$\times$10$^{4}$ s$^{-1}$.
Assuming that every neutron is captured in the nitrogen, yielding
a $^{14}$C nucleus, the production rate of $^{14}$C is about 
1.3$\times$10$^{10}$d$^{-1}$. For a production and transportation 
time of ten days, the simulation yields
less than 10$^{-4}$ counts/(kg keV y) in the relevant energy
region. Through the purification of the nitrogen this
contribution will be further reduced.

\subsection{Sum spectrum from simulations}

In table \ref{backlist} the components discussed so far are listed
and summed up. 
Not included in the table are the contributions from the intrinsic
impurities in the Ge crystals and from the $^{7}$Be activation of the
liquid nitrogen during its transportation at sea level. 
For the Ge--crystals we have only very conservative upper values for
their true contamination, which is expected to be much lower (see
Subsection 3.2.1).
Regarding the $^{7}$Be contamination of liquid nitrogen,
 the cleaning efficiency of the liquid nitrogen 
should already be high enough in order to
reduce this contribution to a negligible level. 
Furthermore it is planned to use a recycling device for the evaporated 
nitrogen, thus the activation with $^7$Be will be reduced further by storage
(T$_{1/2}$ =53 d).

\begin{table}[htb]
\begin{center}
\begin{tabular}{lcc}
\hline
Source & Component & [counts/(kg y keV)] \\
         &&11--100 keV\\
\hline
Nitrogen   & $^{238}$U   &  7$\times$10$^{-4}$ \\
intrinsic  & $^{232}$Th  &  4$\times$10$^{-4}$ \\
           & $^{40}$K    &  1$\times$10$^{-4}$ \\
           & $^{222}$Rn  &  3$\times$10$^{-4}$ \\
N activation & $^{14}$C    &  1$\times$10$^{-4}$ \\
\hline
Steel vessel   & U/Th        &  1.5$\times$10$^{-5}$ \\
\hline
Holder system  & U/Th        &  8$\times$10$^{-4}$ \\
\hline
Surrounding    & Gammas      &  4$\times$10$^{-3}$\\
               & Neutrons    &  4$\times$10$^{-4}$\\
               & Muon shower &  2$\times$10$^{-3}$ \\
               & $\mu$ $\rightarrow$ n       &  1$\times$10$^{-3}$ \\
               & $\mu$ $\rightarrow$ capture & $<<$1$\times$10$^{-4}$ \\ 
\hline
Cosmogenic     & $^{54}$Mn   & 3$\times10^{-3}$\\
activities     & $^{57}$Co   & 1$\times10^{-4}$\\
in the crystals    & $^{60}$Co   & 4$\times10^{-3}$\\
               & $^{63}$Ni   & 6$\times10^{-3}$\\
               & $^{65}$Zn   & 1.5$\times10^{-3}$\\
               & $^{68}$Ge   & 3.7$\times10^{-2}$\\
\hline

Total          & &  6.1$\times$10$^{-2}$  \\
\hline 
\end{tabular}
\end{center}
\caption{Summation of background components in the region 11 keV--100 keV.} 
\label{backlist}
\end{table}

Assuming a background as stated above, we will achieve
a mean count rate in the interesting
region for dark matter search of about 6.1$\times$10$^{-2}$ events/(kg y keV). This
means a further reduction of background in comparison to our 
best measurement (about 20 counts/(kg y keV) below 100 keV \cite{HM98}) 
by more than two orders of magnitude.
This count rate will further improve after another 2 years of running time
(decay of $^{68}$Ge), to reach a level of 1$\times$10$^{-2}$ events/(kg y keV).
These results have been confirmed by an independent simulation for the GENIUS project by the Kiev group \cite{kiev98}.

\begin{figure}[!tt]
\epsfxsize12cm
\epsfbox{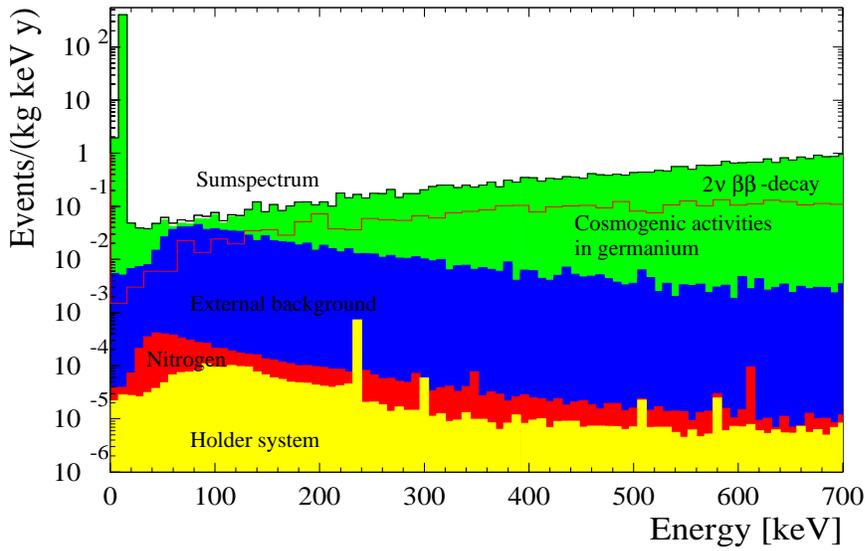}
\caption{Simulated spectra of the dominant background sources for
a nitrogen tank of 12 m diameter.
Shown are the contributions from the tank walls, the detector holder
system, from neutron capture in the nitrogen,
from natural
radioactivity and from the $^{222}$Rn contamination of the nitrogen.
The solid line represents the sum spectrum of all the simulated
components (note the different channel binning compared to figure \ref{cosmo}).}
\label{specall}
\end{figure}

In figure \ref{specall} the spectra of individual contributions 
and the summed-up total background spectrum are shown. 
As mentioned before, the low-energy spectrum is dominated by events
originating from the cosmogenic activation of the Ge crystals at the
Earths surface. Production of the detectors underground would
significantly reduce this contribution.  

\section{The GENIUS Facility: Technical Design}

\subsection{General description}

\begin{figure}[h]
\epsfxsize10cm
\epsfbox{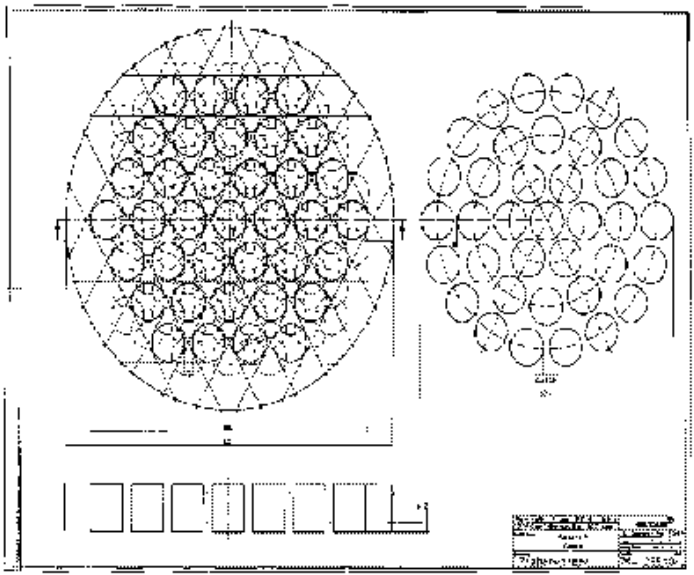}
\caption{Schematic view of possible configurations for the holder
  system of the Ge crystals.}
\label{holders}
\end{figure}

The GENIUS detector will consist of two concentric stainless steel 
tanks with the Ge crystals positioned in the centre. The inner tank 
will contain liquid nitrogen as working medium for the Ge crystals 
and as shielding from the Gran Sasso rock backgrounds. The outer vessel 
will contain the isolation material, which will be doped with Boron 
as shielding against the neutron background. The Ge crystals will be 
placed on a holder system made of teflon, which can house up to six 
layers of 40 crystals each (see Fig. \ref{holders}). 
On the top of the tank there will be a clean 
room with a lock chamber, a room for the elecronics and the data acquisition 
system and a muon veto shield.

\subsection{Technical Data}

Capacity of the inner vessel = 1400000 l liquid nitrogen $\simeq$ 
95\% of geom. vol.\\

{\bf Outer vessel:}\\
Dimension: $\oslash \times$ H $\simeq$ 14.4 m $\times$ 19 m\\
Roof type:  round roof\\
Floor type:  flat floor\\
Maximum allowed pressure: 10 mbar/ -2 mbar\\
Working pressure: 5 mbar\\
Design temperature: 77 K for the floor and 60\% for  the outer mantel, 293 K 
for the rest cylinder and roof \\

{\bf Inner vessel:}\\
Dimension: $\oslash \times$ H $\simeq$ 12 m $\times$ 13.6 m\\
Roof type:  round roof\\
Floor type:  flat floor\\
Maximum allowed pressure: 100 mbar/ -2 mbar\\
Working pressure: 50 mbar\\
Design temperature: 77 K \\
Maximum waste gas through heat flow in: 0.27\% / day of the max. 
content\\

\subsection{Design}

\begin{figure}[h!]  
\centering 
\leavevmode\epsfxsize=280pt  
\epsfbox{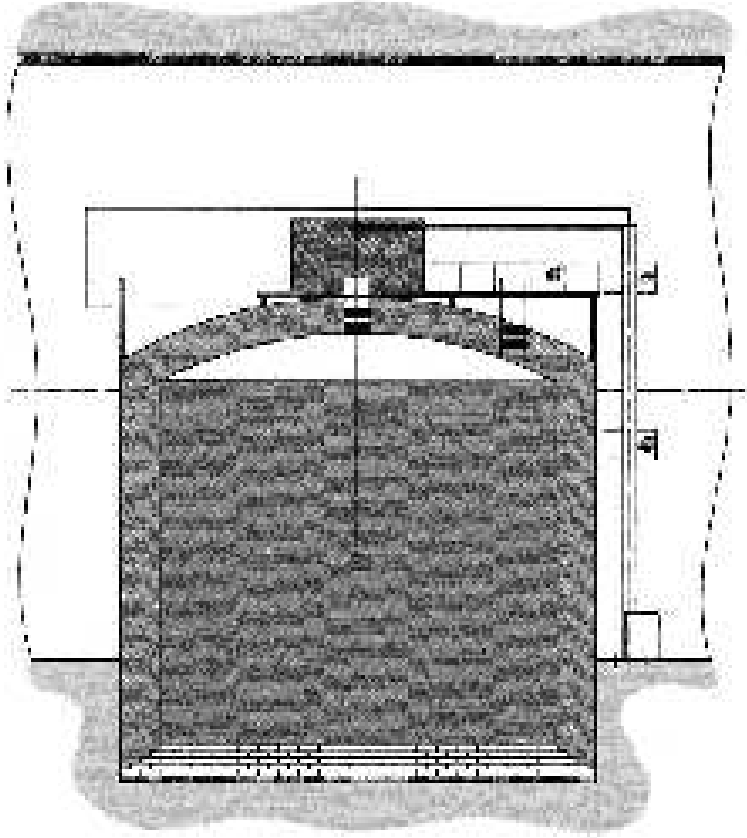}
\caption{\label{confA}Design of the GENIUS tank. The tank
is lowered 4m into the earth.}
\end{figure}

The standing, flat floor tank is made of an inner and an outer vessel in concentrical geometry.
The inner vessel is storing the liquid nitrogen, the outer vessel  the isolating material.
A removement of the isolating material is  possible. The fundament is shielded 
by polystyrol. The isolation will be flushed with gaseous nitrogen, 
thus preventing the immersion of humidity. The  vessels are gas tight 
welded. 
Both inner and outer vessel are equipped with security devices against 
over and under pressure.

Figure \ref{confA} displays a schematic view of the GENIUS tank 
(the tank is lowered 4 m into the earth).

\subsection{Time schedule of the GENIUS experiment}

\begin{figure}[h!]  
\centering 
\leavevmode\epsfxsize=350pt  
\epsfbox{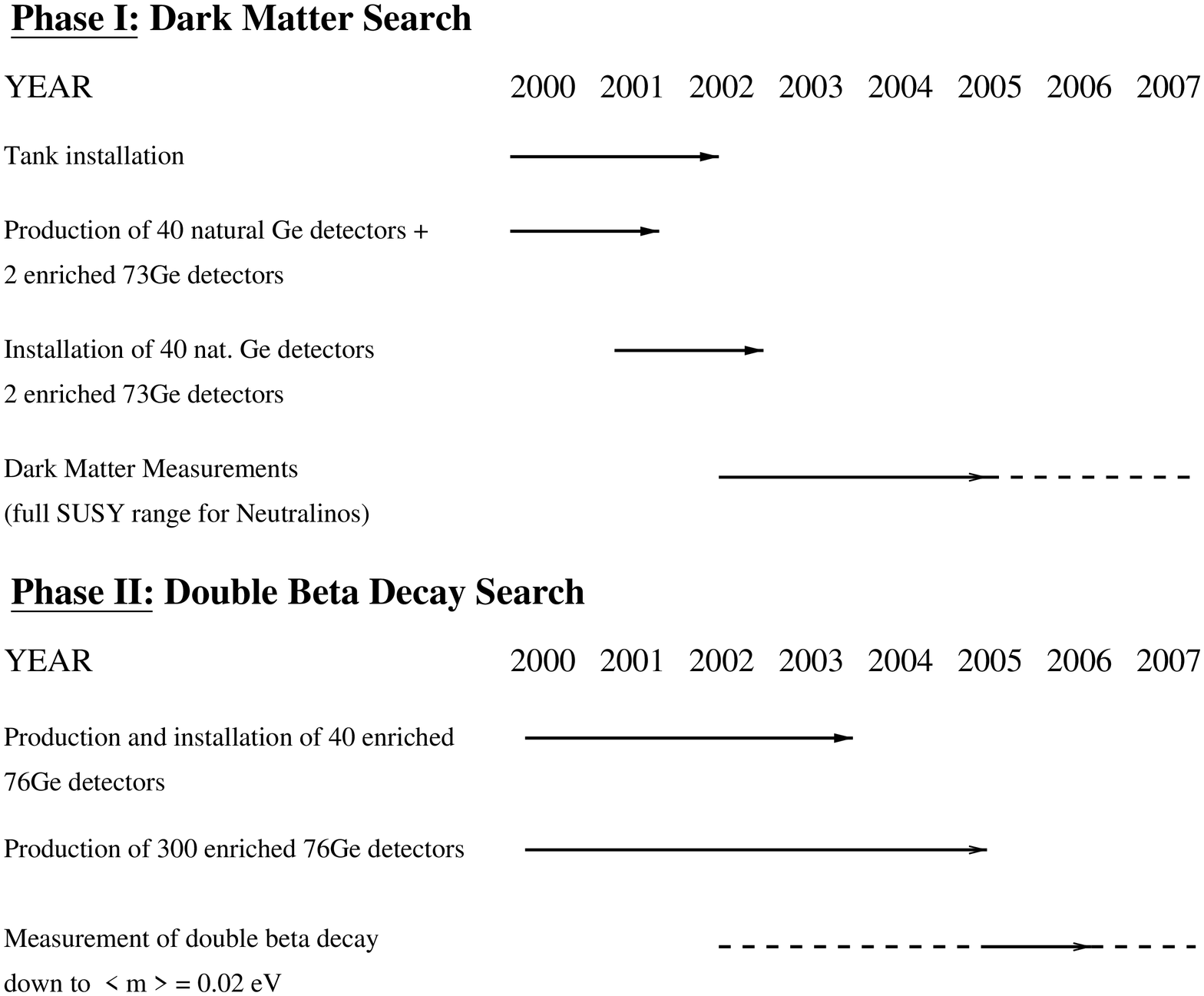}
\label{timescale}
\end{figure}

This schedule is valid under the assumptions, that the requested space is 
provided in the Gran Sasso Laboratory in near future and that detector 
production will start in the year 2000. It has been verified with potential 
producers, that technically and logistically the production of the required 
super-low level natural and enriched Ge detectors can be performed within 
the given time periods.








\end{document}